\documentclass{ar-1col-S2O}

\usepackage[comma]{natbib}
\usepackage{url}
\usepackage{amsmath,amssymb} %added
\jname{Xxxx. Xxx. Xxx. Xxx.}
\jvol{AA}
\jyear{YYYY}
\doi{10.1146/((please add article doi))}

\begin{document}

% Page header
\markboth{Aikawa et al.}{Chemistry of Dark Molecular Clouds}

% Title
\title{Chemistry of Dark Molecular Clouds}

%Authors, affiliations address.
\author{Yuri Aikawa,$^1$ Izaskun Jimenez-Serra,$^2$ and Paola Caselli$^3$
\affil{$^1$Department of Astronomy, University of Tokyo, Tokyo, Japan, 113-0033; email: aikawa@astron.s.u-okyo.ac.jp}
\affil{$^2$Centro de Astrobiología (CAB), CSIC-INTA, Ctra. de Ajalvir km 4,
28850 Torrejón de Ardoz, Spain}
\affil{$^3$Max-Planck-Institut f\"{u}r extraterrestrische Physik, Gie{\ss}enbachstrasse 1, D-85748 Garching bei M\"{u}nchen, Germany}}

%Abstract
\begin{abstract}

Recent molecular line surveys, particularly toward the starless core TMC-1 CP, have greatly expanded the inventory of interstellar molecules, revealing numerous isomers and even aromatic species. Their diverse formation pathways---from ion–molecule reactions to the possible fragmentation of carbonaceous grains---remain under debate, linking chemistry to the life cycle of the interstellar medium. Simple tracers such as carbon chains and deuterated ions are used to probe the physical conditions and evolutionary state of nearby filaments and cores, as well as in massive infrared dark clouds. Ice chemistry has also entered a new era with JWST: spatial distributions of ices indicate a connection between catastrophic freeze-out, established in the prestellar core L1544, and formation of complex organic molecules. Overall, TMC-1 CP and L1544 are not outliers but representative laboratories of molecular cloud physics and chemistry. Extending these findings across diverse environments, from the Central Molecular Zone to low-metallicity galaxies, is essential for a unified picture of how interstellar chemistry regulates the path from clouds to stars and planets.  

\noindent
%-- Deep surveys of TMC-1 CP expanded the inventory, including aromatics.

%--The origin of aromatics and the missing reservoirs of volatile elements must be placed in the context of ISM circulation and cloud formation.

%-- Molecular line and ice surveys reveal chemical processes tightly coupled with 3D cloud structure and kinematics. 

%--COMs are ubiquitous. Their spatial distribution in local cores as well as their abundances in extreme environments provide clues to elucidate their formation mechanisms. 
%\end{itemize}

\end{abstract}

%Keywords, etc.
\maketitle

%Table of Contents
\tableofcontents

% Heading 1
\section{INTRODUCTION}

Since the rise of radio astronomy, molecular line observations have led to the detection of about 350 interstellar molecules to date. Most of them have been identified in molecular clouds, making them key environments for studying chemical evolution in the universe. Molecular clouds are sites where diffuse interstellar gas becomes dense and molecule-rich, and their chemistry may ultimately be inherited by planets through star and planet formation.
Molecular line observations not only trace the kinematics of star and planet formation within these clouds, but also probe fundamental physical parameters, such as the cosmic-ray ionization rate. Furthermore, observations across various clouds, including isotopic measurements, provide insight into the history of nucleosynthesis in our Galaxy. A comprehensive understanding of molecular-cloud chemistry is therefore foundational to a broad range of astrophysical studies.

We focus on the chemistry of dark molecular clouds, defined here as dense, high-extinction regions that have not yet formed protostars. These environments provide a clean laboratory for studying chemical processes unaffected by stellar feedback. This review complements that of \citet{Jorgensen2020}, which focused on chemistry in protostellar environments, particularly warm regions such as hot cores and hot corinos\footnote{Dense hot ($T\ge 100$ K) gas rich in complex organic molecules in the vicinity of a protostar. A hot core refers to such a region around a high-mass protostar, while a hot corino is its low-mass counterpart.}. Since the review by \citet{Bergin2007}, significant progress has been made in the study of dark clouds.
Theoretical studies predict that molecular clouds form by the accumulation of gas behind interstellar shocks, such as those from supernova explosions. This scenario is supported by observations; e.g., James Webb Space Telescope (JWST) revealed bubble-like structures of the interstellar medium in nearby galaxies.%, and 3-dimensional (3D) mapping of interstellar clouds in our Galaxy.
%using a combination of extinction mapping and Gaia astrometry of background stars. 
Herschel Space Observatory (Herschel) revealed filamentary structures, which form the “skeletons” of molecular clouds. Radio observations show diffuse gas accreting onto filaments, and streamers feeding dense cores. 
These evolutionary and dynamical processes are essential for understanding dark cloud chemistry, as well as for placing it in the broader context of interstellar medium (ISM) evolution.
New high-sensitivity receivers at Yebes 40 m telescope and Green Bank Telescope (GBT) have enabled the discovery of new molecules, including aromatic species. JWST expanded the inventory of interstellar ices in the densest regions of molecular clouds through spectroscopy toward background stars. In addition, studies of giant molecular clouds across diverse environments --- the Galactic disk, the Galactic Center, the Outer Galaxy, and the Magellanic Clouds --- have begun to reveal how chemistry varies with local conditions and metallicity.

This review is organized as follows. 
Section 2 outlines the fundamental physical structures and chemical processes in molecular clouds.
Section 3 focuses on the chemical inventory, both in the gas phase and in ices, in nearby molecular clouds, and summarizes the physical and chemical properties of starless and prestellar cores. Section 4 extends the discussion to Giant Molecular Clouds, while Section 5 addresses the universality and diversity of chemistry across various environments in our Galaxy. Section 6 concludes our review. Throughout the review, we will use well-studied, representative sources to illustrate the chemical processes at play across different interstellar environments.
%by linking molecular cloud chemistry to planets and life, including our own Solar system. 

\section{GENERAL VIEW OF THE PHYSICS AND CHEMISTRY IN MOLECULAR CLOUDS}
%\textcolor{blue}{(Yuri; 10 pages)}

%Describe the basic components (i.e. gas and solid phase) and overall structure from the diffuse phase to the translucent and the dense part of the cloud. Link physical structure with chemical processes to help readers understand how physical properties are crucial for astrochemistry and vice versa. Include equations as needed; e.g. the definition of depletion.

\subsection{Physical structures}
\label{sec:physical}
%\subsubsection{Overall physical structures}
%Physical processes: CR ionization rate (Neufeld+24; Obolentseva+24); UVs (PDR) (Observation: Fuente+19, Tafalla23, Watanabe+17, Nishimura+17, Fukui+09) Theory: Gong, M.+18, Harada+19, Gomez, Walsh+22, Komichi+24)

The chemical composition of interstellar gas depends on 
%parameters such as 
column density, which determines the shielding of the interstellar radiation field (ISRF), particularly in the ultraviolet (UV) wavelength \citep{Snow06, Draine2011}. In the gas exposed to the ISRF, hydrogen is mostly atomic; the warm neutral medium ($T \sim 5000$ K)
%, $n_{\rm H}\sim ~0.6$ cm$^{-3}$) 
and the cold neutral medium ($T \sim 100$ K)
%, $n_{\rm H}\sim$ a few 10 cm$^{-3}$), 
coexist in pressure equilibrium. 
Diffuse molecular clouds ($0.2 < A_{\rm V} < 1$~mag) have $T \sim$ several tens of K and $n_{\rm H} \sim$ several tens of cm$^{-3}$, where gas density is described as the number density of hydrogen nuclei, $n_{\rm H}=n({\rm H})+2n({\rm H}_2)$, since the molecular fraction varies among regions. $A_{\rm V}=1$ mag corresponds to the hydrogen column density of $N_{\rm H}\sim 1.8 \times 10^{21}$ cm$^{-2}$.
The attenuation of short-wavelength UV allows $>10$\% of hydrogen in H$_2$, while $>50$\% of gaseous carbon remains ionized (C$^+$). In translucent clouds ($1 \lesssim A_{\rm V} \lesssim$ a few mag), hydrogen is mostly molecular (H$_2$), with C$^+$ gradually converting to neutral C and CO.
%The regions of 0.2 mag $< A_{\rm v} <$ 1 mag are diffuse molecular clouds with $T\sim$ several tens of K, and $n_{\rm H} =$ several tens cm$^{-3}$. The visual extinction of $A_{\rm V}=1$ corresponds to the hydrogen column density of N$_{\rm H}\sim 1.8 \times 10^{21}$ cm$^{-2}$.
%The short-wavelength UV radiation is attenuated so that more than 10 \% of hydrogen exists as H$_2$, while more than 50 \% of gaseous carbon is still ionized (i.e. C$^+$).
%Translucent clouds are in the range of 1 mag $\lesssim A_{\rm v} \lesssim$ a few mag; hydrogen is mostly in H$_2$, while C$^+$ is gradually converted to carbon atoms and CO. 
%The chemical composition of diffuse molecular and translucent clouds can be observed by various absorption lines. 
Molecular clouds, the focus of this review, have $A_{\rm V} \gtrsim$ a few mag. 
CO is the main carbon reservoir, and $n_{\rm H}$ exceeds the critical density of the $^{12}$CO ($J=1-0$) transition ($\sim 3 \times 10^3$~cm$^{-3}$), making CO emission observable. These thresholds are approximate, as molecular abundances also depend on local gas density.
%The main reservoir of gaseous carbon is CO, and the gas density is higher than the critical density of $J=1-0$ transition of $^{12}$CO ($\sim 2\times 10^3$ cm$^{-3}$), making the gas observable with the CO emission line. The above threshold should be considered as a rough estimate, as the molecular abundance depends not only on $A_{\rm V}$ but also on the gas density.

\begin{figure}
    \centering
    \includegraphics[width=1.0\linewidth]{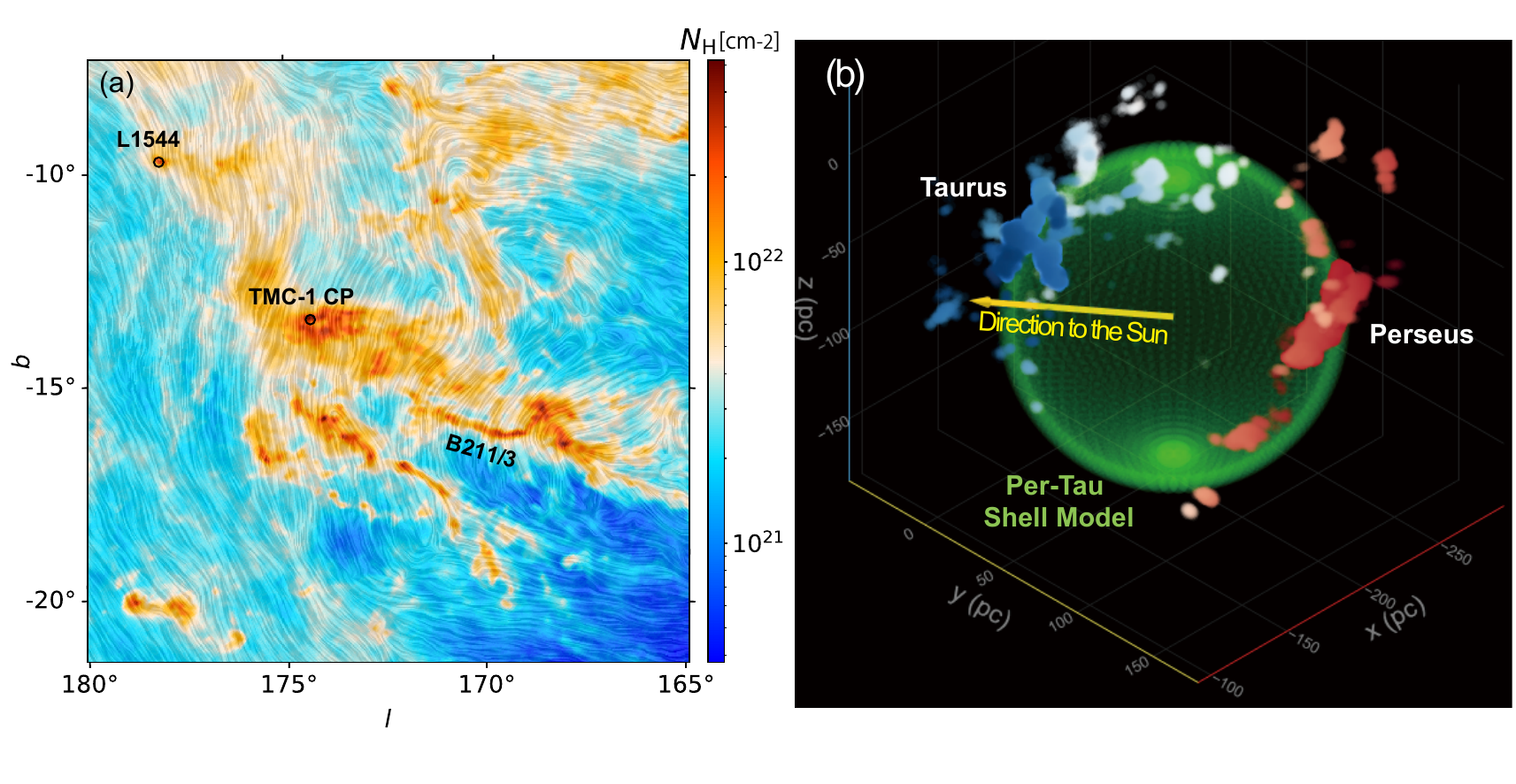}
    \caption{(a) Magnetic field orientation (drapery pattern) and column density (colour scale) measured by \citet{Planck2016} toward the Taurus Molecular Clouds. Open circles mark TMC-1 cyanopolyyne peak CP and L1544 (see Section 3). (b) Three-dimensional map of the Taurus and Perseus molecular clouds from \citet{Bialy2021}, showing their location on a 156 pc-diameter shell. Color show the density iso-surfaces of 25 cm$^{-3}$ and indicates (blue-to-red) distance from the Sun located to the left. Panel (a) adapted from Planck Collaboration et al. (2016) with permission from ESO. Panel (b) adapted from Bialy et al. (2021) with permission from AAS.}
    \label{fig:phys_structure}
\end{figure}

The densest parts of molecular clouds exhibit filamentary structures (Figure \ref{fig:phys_structure}a). Herschel dust continuum observations revealed that the radial density profile within a filament is Plummer-like:
%$ \rho(r)=\frac{\rho_{\rm c}}{1+(r/R_{\rm flat})^2}, $
$\rho(r)=\rho_{\rm c}[1+(r/R_{\rm flat})^2]^{-p/2}$,
where $p\sim 2$, $\rho_{\rm c}$ is the central density, and $R_{\rm flat}$ is the characteristic radius, within which the density profile is approximately flat. Filaments typically have width $2R_{\rm flat}\sim 0.1$ pc, and gas and dust temperatures of $\sim 10$ K. Filaments with line mass above the critical value ($\sim 16 M_{\odot}/$pc)
%at $T\sim 10$ K) 
show signs of star formation \citep{Andre2014}. The critical line mass corresponds to $A_{\rm v}\sim ~10$ mag, implying $n({\rm H}_2)\sim 3\times 10^4$ cm$^{-3}$
when divided by the filament width. 
%15*2e+33*0.7/1.67e-24/3e+18/(pi*1.5e+17*1.5e+17)/2
%High-mass star-forming regions often have hub-filament structures, where multiple filaments intersect \citep[e.g.][]{kumar2022}.

Molecular clouds have a hierarchical structure, where dense gas interacts with the surrounding lower-$A_{\rm V}$ gas via accretion and dispersal.
In the B211/3 region in Taurus molecular clouds (Figure \ref{fig:phys_structure}a), for example, an extended sheet-like gas with $n(\rm{H}_2)\sim 10^3$ cm$^{-3}$ and a thickness 0.3-0.7 pc accretes along magnetic fields toward the filament at rate 27-50 $M_{\odot}$ pc$^{-1}$ Myr$^{-1}$ \citep{Palmeirim2013, Shimajiri2019}. This implies a filament formation timescale of $1–2$ Myr, which is comparable to the chemical timescale of molecular clouds, e.g. the formation timescale of CO and N$_2$ in chemical network models (\S \ref{sec:chemicalproceeses}).
%filaments are observed in C$^{18}$O ($J=1-0$) \citep{Hacar2013}, while more extended gas is probed with $^{13}$CO and $^{12}$CO lines \citep{Goldsmith2008}. Using the latter data,  
There is even lower density gas on the periphery of the molecular cloud, which occupies a large volume fraction. 
%In the Taurus molecular cloud, the probability distribution function (N-PDF) of the total hydrogen column density has a peak around $A_{\rm v} \sim 1$ mag \citep{Schneider2022}. 
Combined HI and CO analyses indicate continuous HI gas accretion onto the cloud \citep{Fukui2009, Wang2020}.

One of the major recent advances in the study of the ISM is the reconstruction of 3D cloud structure by combining extinction mapping with Gaia astrometry of the background stars. \citet{Bialy2021} showed that the nearby low-mass star-forming clouds Perseus and Taurus are part of a shell with a diameter of 156 pc (Per-Tau Shell) (Figure \ref{fig:phys_structure}b). They estimate that the shell was formed 6-22 Myr ago by a shock wave from a supernova. The sheet-like gas around B211/3 filament could be a part of this shell. Such shell structures and the formation of filaments have also been proposed theoretically \citep{Inutsuka2015, Pineda2023}, and are supported by JWST observations of bubble structures in nearby galaxies \citep{Barnes2023a}.

\subsection{Elemental abundances and partitioning between gas and solid phases}
%\subsubsection{Elemental abundance variation (i.e. depletion) from diffuse to dense}

\begin{figure}
    \centering
    \includegraphics[width=1.0\linewidth]{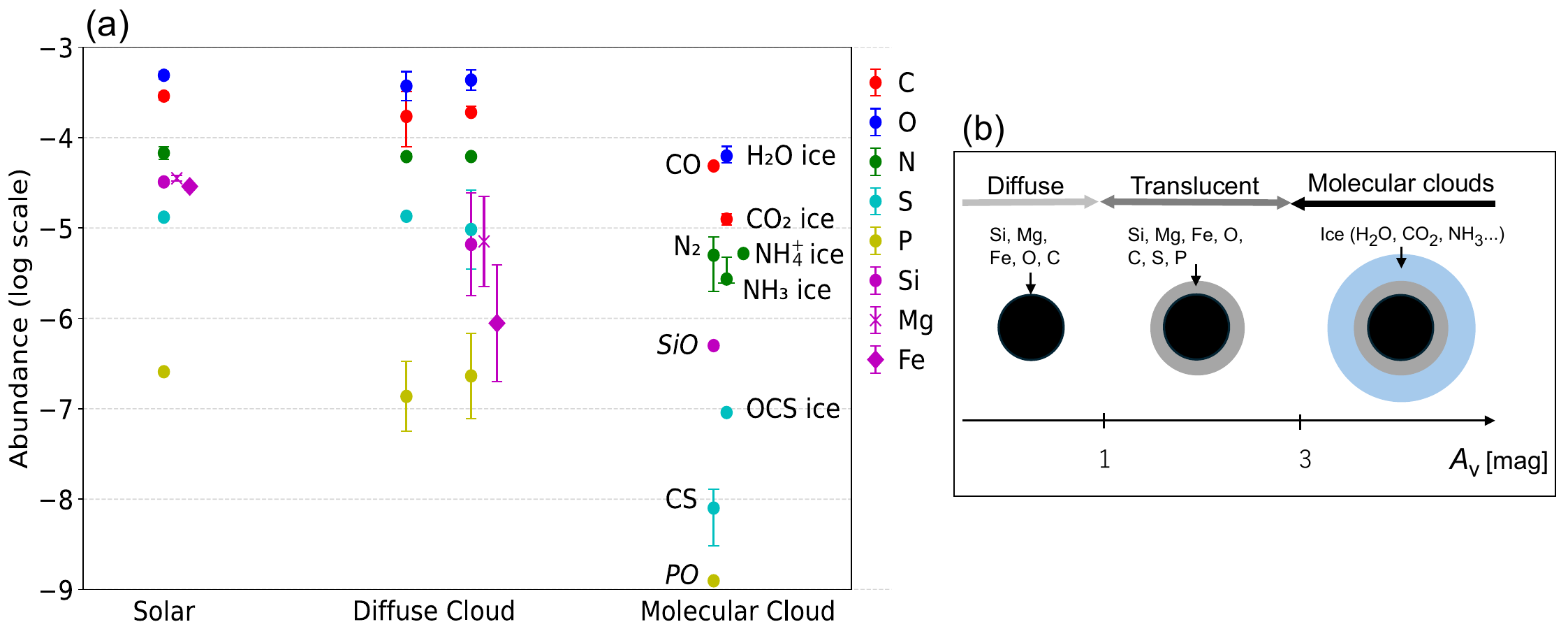}
    \caption{
    (a) Elemental and molecular abundances per hydrogen nuclei. 
    References are \citet{Asplund2021} for Solar abundance, \citet{Zuo2021} (carbon and oxygen), \citet{Meyer1997} \citep[see also][]{Jensen2007} (nitrogen), \citet{Howk2006} (sulfur), \citet{Ritchey2023} (phosphorus) for gas abundance in  diffuse clouds. 
    The right column of diffuse clouds shows the gaseous abundances in \citet{Jenkins2009}.
    %The values labeled as Jenkins09 are [X/H]0 and [X/H]1 from their Table 4. 
    The most abundant molecules for each element observed in gas and ice in molecular clouds are plotted in the right column \citep{McClure2023, JimenezSerra2025b, Xue2025}. {\it SiO} and {\it PO}. shown in italic, are observed in shocked regions \citep{Lefloch2016, Rivilla2016,Podio2017}. N$_2$ abundance is estimated from the observation of N$_2$H$^+$ \citep{Daranlot2012}.
    (b) Schematic figure showing which elements are in the solid phase in diffuse, translucent, and molecular clouds.
%    \textcolor{magenta}{[Paola: should we also cite the Hensley and Draine (2021) paper?--Some elements, e.g., N, are not listed in Hensley and Draine+21. The reference is cited now in the sentence saying that ISM could be metal rich than Solar.]}
}
    \label{fig:elements} 
\end{figure}

One of the goals of astrochemical studies of molecular clouds is to determine how elements are partitioned between gas and solids and to identify their major reservoirs.
The left column of Figure \ref{fig:elements} (a) shows the Solar elemental abundances relative to hydrogen \citep{Asplund2021}. 
Given nucleosynthesis over the 4.5 Gyr since the birth of the Sun, present-day abundances of heavy elements in the local ISM could be $\sim 10-20 \%$ higher \citep{Hensley2021}.
%Gas-phase abundances in interstellar clouds are lower than these values. Depleted elements form solids with volatilities that depend on the physical conditions of the clouds. It is thus useful to compare elemental abundances with gaseous abundances in diffuse and molecular clouds.
Gas-phase abundances in interstellar clouds are lower than these reference values because a fraction of the elements is locked into solids.
Comparison with diffuse-cloud gas-phase abundances constrains the composition of refractory dust (Figure \ref{fig:elements}b). Stronger depletion in translucent and molecular clouds implies the formation of additional solid material that is less refractory.

In diffuse clouds, gas-phase abundances can be derived from absorption lines of H, H$_2$, atoms, and atomic ions, and are shown in the middle column of Figure \ref{fig:elements}a.
%\citep{Meyer1997,Zuo2021, Jensen2007, Howk2006,Ritchey2023}. 
%The difference from the Solar abundance reflects the refractory dust composition. 
Mg, Si, and Fe are mainly locked in silicates and oxides. These refractories remain solid unless processed by strong shocks (more than several tens of km s$^{-1}$) or high temperatures ($> 1000$ K), so most are incorporated into molecular clouds. Carbon is partly incorporated into refractory dust, e.g., soot-like grains with both aromatic and aliphatic bonds \citep{Kwok2011}. About 5-10 \% of carbon could be incorporated into polycyclic aromatic hydrocarbons (PAHs) \citep[e.g.,][]{Tielens2005, Omont2025}. 

The gas/solid abundance ratios of C and O in diffuse clouds vary among the lines of sight. %\citet{Zuo2021} summarized gas-phase abundances in 10 lines of sight where the extinction and all the major dust-forming elements (i.e., C, O, Mg, Si, and Fe) were measured. The C/H and O/H ratios range from $7.9\times 10^{-5}$  to $3.2\times 10^{-4}$ and from $2.6\times 10^{-4}$ to $5.4\times 10^{-4}$, respectively. 
One of the causes of this variation is the gas volume density $n_{\rm H}$, which varies among and along the lines of sight. Assuming that the abundance of refractory grains does not vary, the rate of gas-dust collisions, and hence depletion, is proportional to the volume density. 
Indeed, \citet{Jenkins2009} showed that depletion of oxygen is weakly correlated with that of other species, e.g., Si and Mg, based on relative depletions along various lines of sight (middle right column of Figure \ref{fig:elements}) (a). \citet{Whittet2010} estimated that near the boundary between diffuse and translucent clouds, an oxygen abundance of $2.7 \times 10^{-4}$ is depleted from the gas phase. Of this, $1.1\times 10^{-4}$ is attributed to refractory dust, whereas the remaining $1.6 \times 10^{-4}$ is unaccounted for. Various potential carriers are considered, including, O-bearing carbonaceous matter \citep[see also][]{vanDishoeck2021}.
%($n_{\rm H}\sim 10$ cm$^{-3}$),
%oxygen abundance of $1.1\times 10^{-4}$ is incorporated into refractory dust, while the reservoir is unknown as much as $1.6 \times 10^{-4}$ oxygen are incorporated into refractory dust and O-bearing carbonaceous matter, respectively. 
On the other hand, no strong correlation is found between gaseous C/H ratio and $n_{\rm H}$ across various lines of sight \citep{Mishra2015}. It may indicate that carbonaceous dust is more easily formed and destroyed than silicates in the ISM.

Nitrogen is hardly incorporated into refractory dust \citep{Meyer1997, Tielens2005}. %\citet{Meyer1997} reported a nearly constant gas-phase N/H abundance of $6\times 10^{-5}$ in seven lines of sight \citep[see also][]{Jensen2007}. 
While observations %in more lines of sight 
show some variation and a tendency of lower N/H at high column density ($N_{\rm H}>10^{21}$ cm$^{-2}$), the error is large and the effect of nucleosynthesis cannot be ruled out \citep{Knauth2003, Jensen2007}.
Sulfur and phosphorus are also mostly in the gas phase in low-column-density diffuse clouds, but show depletion correlated with those of Si and Mg in the lines of sight with relatively high column densities,
%\citep{Howk2006, Jensen2007, Zuo2021, Ritchey2023}, 
suggesting incorporation into refractory dust, although the detailed mechanism is still unclear.
%{\bf{Can we omit Cl, since there are not many detections of Cl-bearing molecules; only HCl and HCl+ by Blake 1985 and Herschell?}}

In molecular clouds, volatile molecules accumulate on grain surfaces to form ice mantles (Figure \ref{fig:elements}b). In Taurus Molecular Cloud, for example, the absorption feature of H$_2$O ice at 3 $\mu$m is detected along lines of sight with $A_{\rm V}\ge 3$ mag \citep{Whittet2010, Boogert2015}.
The most abundant molecules of C, N, O, S, P, and Si observed in molecular clouds are plotted in the right column of Figure \ref{fig:elements} (a). CO gas (and ice), H$_2$O ice, and CO$_2$ ice represent the major reservoirs of oxygen and carbon. The major nitrogen reservoirs are considered to be N$_2$, NH$_3$, and atomic N, but N$_2$ and atomic N cannot be directly observed in molecular clouds. N$_2$ abundance is estimated from observations of N$_2$H$^+$ \citep[e.g.,][]{Daranlot2012}.
%While NH$_3$ is observed in both gas (rotation lines in radio) and ice (absorption bands in infrared), the latter is much more abundant in molecular clouds, except in the vicinity of protostars, where ices sublimate. 
%Ammonium hydrosulfide NH$_4$SH is found to be abundant in Comet 67P/Churyumov-Gerasimenko, which motivated and resulted in the recent detection of ammonium salt feature in interstellar ices \citep{Altwegg2022, McClure2023, Slavicinska2025}. 
Ammonium salt (e.g.,  NH$_4$SH) could be another major reservoir of nitrogen and possibly of sulfur (\S \ref{sec:iceage}).

Several S-bearing species are detected in molecular clouds, but their abundances are low; the abundance of CS, the most frequently observed among S-bearing molecules, is $\sim 10^{-8}$.
\citet{Fuente2019} observed S-bearing molecules in Taurus and estimated the gas-phase S elemental abundance by comparing the observed abundances with chemical network models at varied S/H ratios; the S/H ratio is $\sim (0.4-2.2)\times 10^{-6}$ in the region with relatively low gas density of (1-5)$\times 10^3$ cm$^{-3}$, while it further declines to $\sim 8\times 10^{-8}$ in denser regions where dust grains are covered by thick ice mantle. While the major sulfur reservoir has long been debated, e.g., volatiles such as H$_2$S ice or refractory solids such as FeS, recent observations of the Orion molecular cloud and its photodissociation region (PDR), combined with model calculations, indicate gas-phase S/H ratio close to Solar, arguing against a major refractory sulfur reservoir \citep{Fuente2025}. 
The abundances of detected S-bearing ices are however $\lesssim 10^{-7}$ (Figure \ref{fig:elements}a, see also \S \ref{sec:iceage}).
%; the upper limit of its abundance relative to H$_2$O ice is less than 1 \% ($\sim 10^{-6}$ relative to hydrogen) \citep{McClure2023}. 
Alternative postulated reservoirs include ammonium salt NH$_4$SH \citep{Vitorino2024,Slavicinska2025} and sulfur allotropes such as S$_2$ and S$_8$ \citep{JimenezEscobar2012,Shingledecker2020,Cazaux2022}.

The major reservoir of phosphorus in molecular clouds remains uncertain. PN and PO are the only neutral P-bearing molecules detected;
%\citep[e.g.,][]{Ziurys1987, Rivilla2016}. 
they are observed in the the shocked region L1157-B1 within the molecular outflow, but not toward the protostar L1157, indicating that P is sputtered from dust grains to form PO and PN via gas-phase reactions. Their low abundances ($10^{-10}-10^{-9}$)
%(2-6)$\times 10^{-10}$ and $2.5\times 10^{-9}$, respectively, 
imply depletion by a factor of 100 %\citep{Yamaguchi2011, Lefloch2016}.
\citep{Lefloch2016}.
A deep search for PH$_3$ in L1544 yields an upper limit abundance of $5\times 10^{-12}$,
%relative to H$_2$, 
implying an upper limit of $5\times 10^{-9}$ for volatile P (gas and ice more volatile than water) when combined with chemical network models \citep{Furuya2024}. For a detailed review of phosphorus in diffuse and molecular clouds, see \citet{Fontani2024a}.

In summary, the summed abundances of observed gas and ices in molecular clouds are well below Solar values. While silicates, oxides, and carbonaceous material form refractory dust already in diffuse clouds, additional gas-to-solid conversions occur in the transition phase to molecular clouds. Such ``missing'' material could contribute to gas-phase chemistry upon desorption due to energetic processing as in e.g., shocks.

% PN In Orion KL: 1e-11 - 1e-10 (Turner and Bally 1987)
%PN in Orion plateau (1-4)e-10 (Ziurys 1987)
%PN/H2 in L1157 B1 and B2, (2-6)e-10, (3-7)e-10 (Yamaguchi+11)
%L1157 B1 2.5 × 10-9 and 0.9 × 10-9 for PO and PN (Lefloch+16)

%\subsubsection{Elemental budget in dense clouds}
%C, N, O (+S, P, Cl) (Ritchey+23, Konstantopoulou+24, Furuya\&Shimonishi24, Daranlot+12)

%\subsubsection{The diffuse phase: the start of molecule formation}
%Hydrides observed by Herschel etc. (including HCl+, HS…)
%(e.g. Ebisawa+19, Gerin, Neufeld, Godard+23, Narita+24,)
%{\bf (Briefly. Refer to Gerin+16 ARAA for Hydrides/Herschel)}

\subsection{Chemical processes} \label{sec:chemicalproceeses}
%Introduction to general readers; explain the basics of chemistry, 
%\textcolor{red}{Refer Gerin+16 for hydrides review.}

The low densities and temperatures of molecular clouds prevent chemistry from reaching thermodynamic equilibrium within the dynamical timescale (e.g., filament formation timescale in \S 2.1). 
Chemical reaction network models (chemical kinetic models), comprising thousands of reactions, predict the temporal variation of molecular abundances for comparison with observations. Some reaction sets are publicly available via web-based datasets, e.g., UMIST \citep{Millar2024} and KIDA \citep{Wakelam2024}.
Despite including many estimated reactions, these models help identify possible main formation and destruction pathways of molecules, which can then be investigated by laboratory experiments and quantum chemical calculations.
Pseudo-time dependent models calculate chemical evolution under fixed physical conditions (e.g., $n_{\rm H}=10^4$ cm$^{-3}$, $T=10$ K, and $A_{\rm V}=10$ mag) typically assuming atomic initial abundances except for H$_2$. More realistic models include cloud formation dynamics from fully atomic gas \citep[e.g.,][]{Bergin2004, Furuya2015, Komichi2024}.
%\textcolor{red}{Check Tinacci+23 for heat of formation etc. It may be useful to make links to such useful database.}
%\textcolor{blue}{Maybe we could give a reference that collects the information from the most common astrochemical models: Nautilus, UCLCHEM, MAGICKAL, KMC, etc... In my 2025 Ice Age paper, I summarized all this info.}\textcolor{red}{I would name only KIDA and UMIST, which provide public data base. I added a sentence mentioning kinetic monte carlo in the subsection of grain-surface reactions, and added Ice Age model comparison paper as a reference there.}

In the following, we summarize basic reactions of C, O, and N. %, and S. 
Cold cloud chemistry ($T \le$ a few tens of K) proceeds with exothermic reactions. They are divided into gas-phase and grain-surface reactions, which are coupled via adsorption and desorption. In the gas phase, association reactions (i.e., formation of one molecule from two reactants) are often inefficient especially for small molecules, since the excess energy needs to be discarded through radiation (i.e., radiative association). For grain-surface reactions, in contrast, the excess energy can be transferred to the surface molecules. It enables efficient reactions such as H$_2$ formation, which proceeds in collisional timescale of H atoms with grains
\begin{equation}
    \left[\pi a^2\sqrt\frac{8kT}{\pi\mu m_{\rm H}}n({\rm dust})\right]^{-1}\sim 1\times 10^5\left(\frac{10 {\rm K}}{T}\right)^{1/2}\left(\frac{\mu}{1.0}\right)^{1/2}\left(\frac{10^4 ~{\rm cm}^{-3}}{n_{\rm H}}\right) {\rm yrs}, \label{eq:H2formation}
\end{equation}
where $\mu$ is molecular weight, $m_{\rm H}$ is hydrogen mass, typical grain size $a$ is assumed to be 0.1 $\mu$m, gas/dust mass ratio is 100, and dust material density is 3 g cm$^{-3}$. 
Efficient formation on grain surfaces and self-shielding against photodissociation make H$_2$ the first molecule to be abundant \citep[see review by][]{Wakelam2017}. 

\subsubsection{Gas-phase reactions}\label{sec:reactions_gas}
%Theoretical works and laboratory experiments: e.g. how to make molecules from H3+ (with reference to CRs; Oka80; Geballe\&Oka96) up to aromatics (e.g. Jones, B. M+11 PNAS, Landera+10,Cooke et al. 20; Garcia de la Concepcion+23, 24, Byrne+24)

Once H$_2$ becomes abundant, gas-phase chemistry proceeds efficiently, where ion-molecule reactions play a central role \citep{Herbst1973} \citep[see also review of hydrides by][]{Gerin2016}.
Typical reactions are:
\begin{eqnarray}
\rm{X}^+ + \rm{H}_2 &\rightarrow& \rm{XH}^+ + \rm{H} \label{reac_H2}\\
\rm{H}_3^+ + \rm{X} &\rightarrow& \rm{XH}^+ + \rm{H}_2 \label{reac_H3p}
\end{eqnarray}
H$_3^+$ is formed by cosmic-ray ionization of H$_2$ followed by H$_2^+$ + H$_2$ $\rightarrow$ H$_3^+$ + H. Ion-molecule reactions are often barrierless, and their rate coefficients remain high at low temperatures because of enhanced Coulomb attraction due to ion-induced dipole moment in the molecule.
%the Coulomb attraction caused by polarization. 
%Whether each atom or molecule undergoes reaction (\ref{reac_H2}) and (\ref{reac_H3p}), i.e. whether they are exothermic, can be checked referring to the heat of formation and proton affinity of reactants and products (e.g. Tables in UMIST95, Yamamoto 2014).
At early cloud-formation stage, i.e., low $A_{\rm V}$, oxygen and nitrogen are neutral atoms, as their ionization energies exceed 13.6 eV (the H atom ionization energy and upper limit of the ISRF), whereas carbon is ionized.

\begin{figure}
    \centering
    \includegraphics[width=0.9\linewidth]{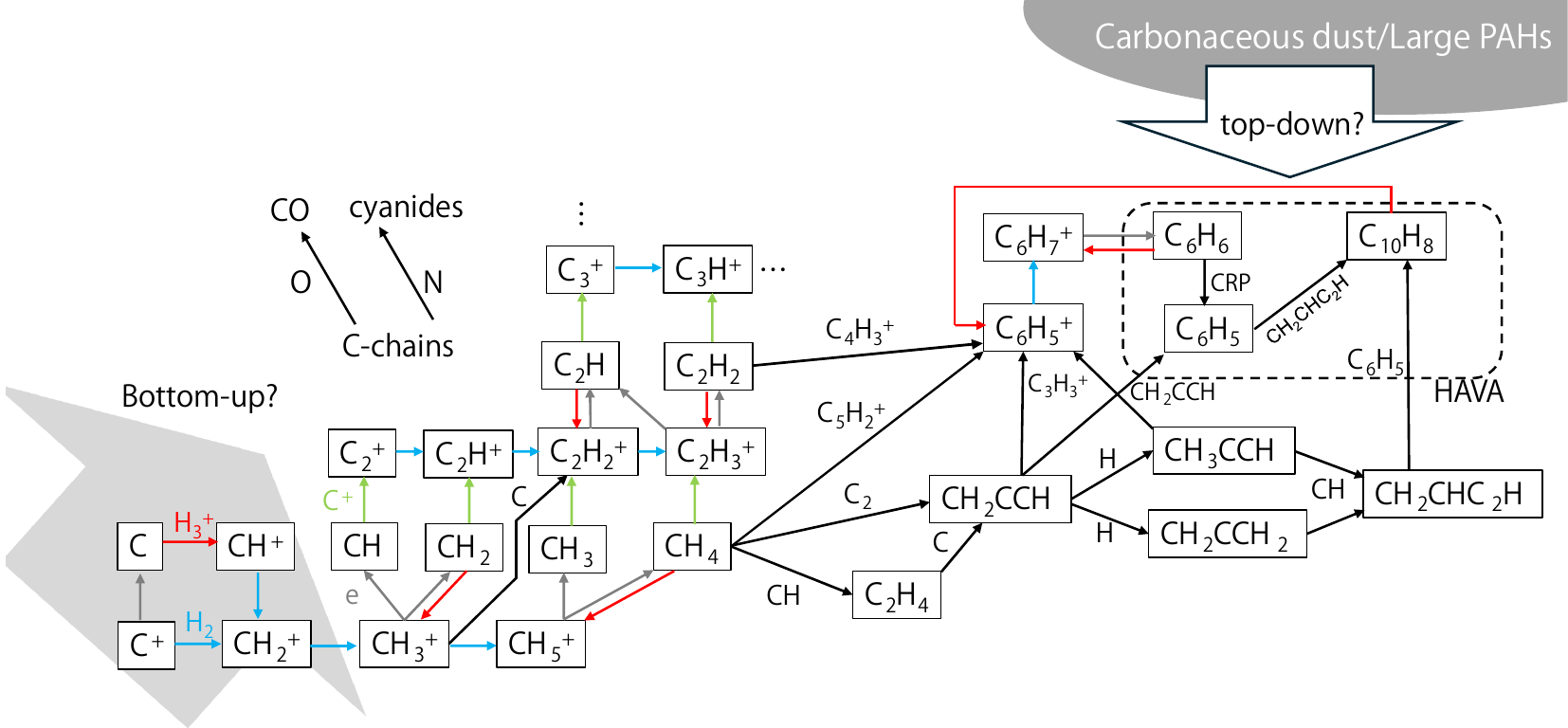}
    \caption{Representative formation paths of carbon chains \citep[left,][]{Sakai2013ChRv} and possible formation paths of aromatics
    \citep[right,][]{Byrne2024}. The red, cyan, and green arrows represent reactions with H$_3^+$, H$_2$, and C$^+$ respectively. The gray arrows represent dissociative recombination. The arrow with CRP depicts photodissociation by cosmic-ray induced UV.}
    \label{fig:network}
\end{figure}

Because reaction (\ref{reac_H2}) is endothermic for C$^+$, carbon chemistry starts with reaction (\ref{reac_H3p}) after the neutralization of C$^+$ (e.g., by  radiative recombination C$^+$ + e $\rightarrow$ C + $h\nu$). Radiative association C$^+$ + H$_2$ $\rightarrow$ CH$_2^+$ + $h\nu$
%\begin{equation}
%\rm{C}^+ + \rm{H}_2 \rightarrow \rm{CH}_2^+ + \it{h}\nu \label{reac_CH2p}
%\end{equation}
also helps. These reactions are followed by e.g., CH$_2^+$ + H$_2$ $\rightarrow$ CH$_3^+$ + H
%e.g., 
%\begin{eqnarray}
%\rm{CH}_2^+ + H_2 &\rightarrow& \rm{CH}_3^+ + H \\
%\rm{CH}_3^+ + e &\rightarrow& \rm{CH}_2 + H \\
%\rm{CH}_2 + C^+ &\rightarrow& \rm{C_2H}^+ + H \label{reac_Cplus}\\
%\rm{CH}_3^+ + C &\rightarrow& \rm{C_2H}_2^+ + H \label{reac_C}
%\end{eqnarray}
%forming hydrocarbons. 
forming unsaturated carbon chains (i.e., C$_{\rm m}$H$_{\rm n}$ with m $<$ 2n+2) (Figure \ref{fig:network} left). Hydrogenation continues by reactions (\ref{reac_H2}), but becomes endothermic before the available chemical bond is fully saturated by H. Reactions of hydrocarbons with O and N atoms yield CO and cyanides, respectively. CO is highly stable, with limited net-destruction pathways, e.g., CO + He$^+ \rightarrow$ C$^+$ + O + He, so hydrocarbon abundances decline as CO becomes the dominant carbon carrier. Thus hydrocarbons (and cyanides) are abundant in ``chemically young'' regions where carbon conversion to CO is incomplete, including the regions with inflowing atomic gas or moderate UV shielding. Their abundances are also sensitive to the gas-phase C/O ratio, which depends on H$_2$O ice formation \citep[e.g.,][]{Aikawa2003}.

A fraction of the long carbon chains is expected to transform to cyclic. For C$_3$H$_2$, both cyclic ($c$-C$_3$H$_2$) and linear (H$_2$CCC, hereafter $l$-C$_3$H$_2$) isomers are observed, with the former $1-2$ orders of magnitude more abundant in several clouds, including TMC-1 CP. \citet{Loison2017} constructed a chemical network that reproduces this ratio with reactions such as C$_3$H$^+$ + H$_2 \rightarrow c/l$-C$_3$H$_3^+$, H + $l$-C$_3$H$_2$ $\rightarrow$ H + $c$-C$_3$H$_2$. % based on literature and quantum chemical calculations. 
%In TMC-1, \citet{Cabezas2022} detected $c$-C$_5$H, which is analogous to $c$-C$_3$H but with the H atom replaced by a ethynyl (-CCH) group. In contrast to the case of C$_3$H, $c-$C$_5$H/$l-$C$_5$H is as low as 0.069.
In contrast, $c$-C$_5$H, which is analogous to $c$-C$_3$H but with the H atom replaced by a ethynyl (-CCH) group, is found to be less abundant than $l$-C$_5$H \citep{Cabezas2022g}.

Recently, aromatic molecules such as $c$-C$_6$H$_5$CN have been detected in molecular clouds (\S \ref{sec:tmc1}). 
Whether they are formed by reactions of hydrocarbons (i.e. bottom-up) or destruction of carbonaceous dust and large PAHs (i.e top-down) has been debated. 
The bottom-up mechanism may be needed, since smaller aromatics are easily destroyed by photodissociation and charge stripping due to cosmic rays 
%\citep{Montillaud2013, chabot2020, Chown2024, Goicoechea2025}.
\citep{chabot2020, Chown2024}.
Various relevant reactions have been proposed and investigated, but reaction network models under-predict the abundances of aromatic molecules by several orders of magnitude
%\citep[e.g.][]{Lee_KLK19, Burkhardt2021a, Byrne2024, Garcia2024,Yang2024,Kocheril2025}.
\citep[e.g.,][]{Burkhardt2021a, Kocheril2025}.
%\textcolor{red}{(Loison25 in arXiv disputes Kocheril25)}.
% moved from Izaskun's para on bottom-down scenario
\citet{Garcia2023} showed that neutral-neutral reactions involving the propargyl (C$_3$H$_3$) radical and the vinyl radical (C$_2$H$_3$) can form cyclic structures such as the cyclopentadienyl radical (C$_5$H$_5$) at 10 K. 
%\citet{Cabezas2025d} detected phenalene (C$_{13}$H$_{10}$) in TMC-1 and investigated its possible formation via neutral-neutral reactions of already-detected PAHs with radicals such as C$_3$H$_3$ to find that they have barriers.
%Radiative association reaction of C$_{12}$H$_8$ + CH$_3^+$ $\rightarrow$ C$_{13}$H$_{11}^+$ + $h\nu$, followed by dissociative recombination, is alternatively contemplated.
%-----------
\citet{Byrne2024} start aromatic ring formation with C$_5$H$_2^+$ + CH$_4$ $\rightarrow$ C$_6$H$_5^+$ + H (Figure \ref{fig:network} right), pointing out the need for more information on reactions of moderately-saturated hydrocarbons, including various isomers.
More recently, \citet{Loison2026} reported a barrier of $\sim 1200$ K for this reaction for the most stable C$_5$H$_2^+$ isomer (HC$_5$H$^+$), which may reduce its efficiency under cold conditions.
Non-thermal desorption is also crucial, since aromatic molecules, as well as key ingredients such as CH$_4$, accumulate on cold grains with the collisional timescale given by Equation (\ref{eq:H2formation}) with appropriate $\mu$.  
Once benzene forms, the evolution to PAHs is possible via Hydrogen Abstraction Vinylacetylene Addition (HAVA) \citep[see review by][]{Kaiser2021}, i.e., C$_6$H$_6$ + $h \nu$ $\rightarrow$ C$_6$H$_5$ + H via UV photon absorption, followed by C$_6$H$_5$ + CH$_2$CHC$_2$H $\rightarrow$ C$_{10}$H$_{8}$ + H.
%\begin{eqnarray}
%\rm{C_6H_6} + \it{h\nu} &\rightarrow& \rm{H} + C_6H_5 \\
%\rm{C_6H_5} + CH_2CHC_2H &\rightarrow& \rm{H} + C_{10}H_{8}.
%\end{eqnarray}
While the ISRF is attenuated in dark clouds, cosmic-ray induced UV photons are available
%; energetic electrons produced by cosmic-ray ionization excite H$_2$, which de-excite by emitting photons 
\citep{Gredel1989}.

% copied and modified from Izaskun's to-down paragraph
In the top-down scenario, the detected aromatic molecules are products of e.g., the UV photo-dissociation of larger PAHs and of the subsequent radiative stabilization of vibrationally hot cations. This mechanism has been shown to be viable for indene \citep{Stockett2025} and indenyl cations \citep{Bull2025}. 
%-------------------------
We discuss top-down vs bottom-up scenario referring to observations in \S \ref{sec:tmc1}.

For O-bearing species, O atoms react with H$_3^+$ (reaction \ref{reac_H3p}) and then with H$_2$ (reaction \ref{reac_H2}) to form H$_3$O$^+$, which dissociatively recombines with an electron to form OH and H$_2$O. Conversion of O to H$_2$O is, however, more efficient on grain surfaces. O atoms also react with hydrocarbons (e.g., CH$_2$) to form CO.
%which is adsorbed onto grain surfaces and partly converted to complex organic molecules (COMs) such as CH$_3$OH.
For N, however, reaction (\ref{reac_H3p}) is endothermic due to its lower proton affinity than H$_2$. N-bearing species thus tend to form by neutral-neutral reactions with rate coefficients lower than those of ion-molecule reactions by 1-2 orders of magnitude, e.g.,
CH + N $\rightarrow$ CN + H and
%\rm{OH} + N &\rightarrow& \rm{NO} + H \nonumber \\
CN + N $\rightarrow$  N$_2$ +C.
%\rm{NO} + N &\rightarrow& \rm{N_2} + O \nonumber 
Some N$_2$ is converted to N$^+$ by N$_2$ + He$^+$ $\rightarrow$ N$^+$ + N + He, where He$^+$ is produced by cosmic-ray ionization of He and plays an important role in the destruction of stable molecules such as N$_2$ and CO. N$^+$ forms NH$_3$ by repeating reactions (\ref{reac_H2}) followed by dissociative recombination. 
While the reaction of N$^+$ + H$_2$ $\rightarrow$ NH$^+$ + H is endothermic by 0.019 eV, the N$^+$ ions formed from N$_2$ have kinetic energies from the excess energy of the reaction, which can be used to overcome the endothermicity \citep{Yamamoto2017}. 
Regions with relatively bright NH$_3$ and N$_2$H$^+$ (proxy of N$_2$ that is not observable) emission are thus considered to be ``chemically old'', in contrast to the regions probed by hydrocarbons. Gas-phase formation of NH$_3$ is however much less efficient than grain-surface formation of NH$_3$ ice (\S \ref{sec:grain_surface_reactions}). 

%S$^+$ does not undergo reaction (\ref{reac_H2}), and its radiative association with H$_2$ is much slower than  (\ref{reac_CH2p}) \citep{Millar1990}.
%Major S-bearing molecules such as CS and SO form via reactions, e.g., S$^+$ + CH $\rightarrow$ CS$^+$ + H, CS$^+$ + H$_2$ $\rightarrow$ HCS$^+$ + H, HCS$^+$ + e $\rightarrow$ CS + H, and OH + S $\rightarrow$ SO + H after recombination with electrons or negatively charged PAH.

\subsubsection{Grain-surface reactions}\label{sec:grain_surface_reactions}
%- Theoretical works and laboratory experiments, e.g. CH3OH formation, COM formation {\bf refer to Cuppen+24 ARAA for surface reactions}
%- Experiments: Watanabe+02; Hidaka+09, Fuchs+09, Minissale+16,  
%                           Dartois+2018, Chuang+2018, Oba+2018, Nguyen+2020,  
%                           Kouchi+2021,Yang,..Ralf Kaiser+24, Ishibashi+24
%- Quantum chemistry: Song\&Kastner17, Wakelam+17, Shimonishi+18, Fredon+2021, Molpeceres+24                                      https://arxiv.org/abs/2502.17849
%
%- Modeling: Vasyunin+17,  Simons, Lamberts+20, Garrod+23, Riedel+23, Furuya+22, 24, Jimenez-Serra+25

\begin{figure}
    \centering
    \includegraphics[width=0.9\linewidth]{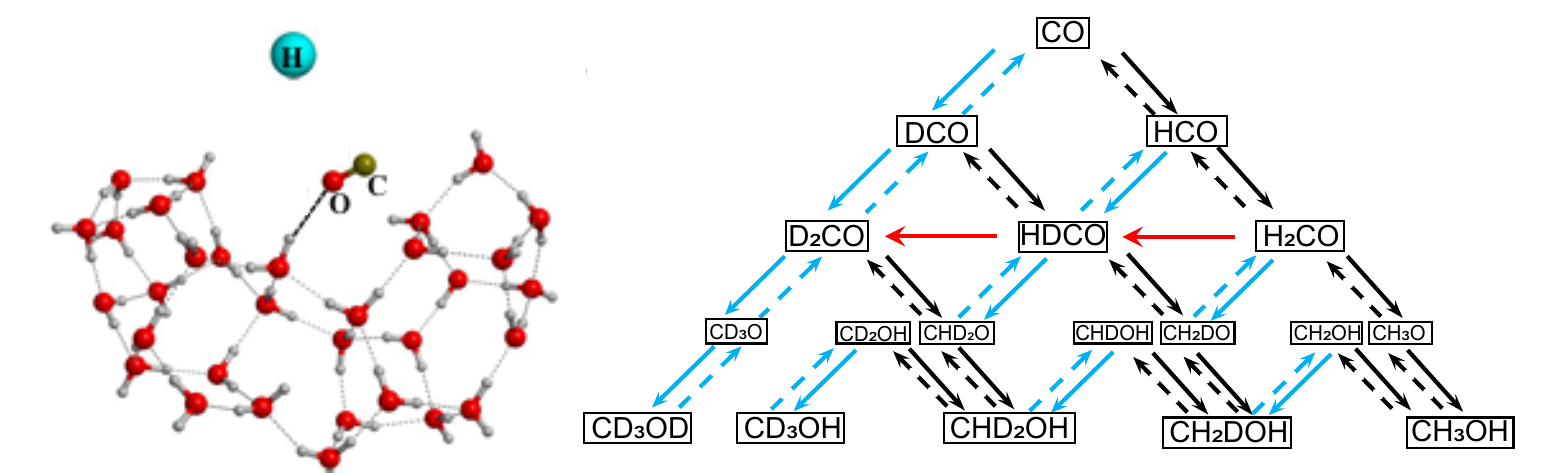}
    \caption{Left: Visualization of the quantum chemical calculation of the CO + H reaction on water ice \citep{Rimola2014}. Right: Formation pathways of CH$_3$OH and its deuterated isotopomers on amorphous water ice based on \cite{Hama2013}, \citet{Cooper2019}, and \cite{Simons2020}. The solid black and cyan arrows represent the H- and D-atom addition, respectively. The dashed black and cyan arrows represent H- and D- abstraction reactions with formation of H$_2$ and HD, respectively. Red arrows indicate direct H-D exchange processes. The left pannel adapted from \citet{Rimola2014} with permission from ESO. The right panel adapted from \citet{Hidaka2009} with permission from the American Astronomical Society.}
    \label{fig:network_ice}
\end{figure}

At $A_{\rm V}\ge$ a few mag, grain surfaces are covered by ice mantles (\S \ref{sec:iceage}). Major constituents after H$_2$O ($\ge 1\%$ relative to H$_2$O ice) are CO, CO$_2$, CH$_3$OH, NH$_3$, and CH$_4$ (Figure \ref{fig:elements}a)\citep{Boogert2015,McClure2023}. CO is adsorbed from the gas phase (Eq. \ref{eq:H2formation} with $\mu=28$), while the others and H$_2$O mainly form via grain-surface reactions. H$_2$O, NH$_3$, CH$_4$, and CH$_3$OH are saturated species, in contrast to unsaturated gas-phase molecules, and are formed primarily by H addition to O, N, C, and CO (Figure \ref{fig:network_ice})
%\citep{Watanabe2002,Fuchs2009,Hidaka2011, Qasim20}, 
\citep[e.g.,][]{Tielens1982,Watanabe2002,Qasim20}, 
although alternative pathways are also studied 
%experimentally and theoretically 
%\citep[see reviews by][]{Hama2013, Cuppen2024}.
\citep[see review by][]{Hama2013, Cuppen2024}.
Because H atoms are light and weakly bound to the grain (ice) surface, they migrate between adsorption sites mainly by thermal hopping to meet a reaction partner even at $T=10$ K
%low temperatures 
\citep{Kuwahata2015}. 
As for CH$_3$OH, there are no low-temperature gas-phase formation routes. In its formation via successive H additions to CO on grain surfaces, key steps with activation barriers (CO + H $\rightarrow$ HCO and H$_2$CO+H $\rightarrow$ CH$_3$O) can proceed through quantum tunneling \citep{Hidaka2009, Rimola2014, Song2017}.

%CO$_2$ is considered to form mainly via CO + OH $\rightarrow$ CO$_2$ + H, which is not as simple as hydrogenation for two reasons. First, the thermal hopping rate of CO and OH are much lower than that of H \citep{Kouchi2020, Miyazaki2022}. The reaction can proceed if a H atom meets an O atom adjacent to CO, a process known as non-diffusive chemistry \citep{Garrod2011}. Accurate evaluation of non-diffusive chemistry requires distribution of molecules in ice mantle and is performed using microscopic kinetic Monte Carlo codes to be compared with rate-equation methods \citep[see review and comparison in][]{Cuppen2024, JimenezSerra2025b}. Second, recent quantum calculations show that the intermediate HOCO is stable and has a lower energy on the potential energy surface than the exit channel CO$_2$ + H. The reaction is halted at HOCO, which may produce CO$_2$ or organic molecules through HOCO + X $\rightarrow$ CO$_2$ + XH or X-COOH \citep{Molpeceres2023, Molpeceres2025, Ishibashi2024}. Alternatively, CO$_2$ could form through CO + O $\rightarrow$ CO$_2$. While its reaction rate is predicted to be low \citep{Goumans2010}, recent kinetic Monte Carlo simulations show that both mechanisms (CO + OH and CO + O) are important in CO$_2$ ice formation \citep[see][]{JimenezSerra2025b}.

CO$_2$ is considered to form mainly via CO + OH $\rightarrow$ CO$_2$ + H, which is not as simple as hydrogenation. Firstly the thermal hopping rate of CO and OH are much lower than that of H \citep{Kouchi2020, Miyazaki2022}. The reaction can proceed if a H atom meets an O atom adjacent to CO, a process known as non-diffusive chemistry \citep{Garrod2011}. CO can also react with a photo-fragmented OH when a nearby H$_2$O is photolyzed \citep[e.g.,][]{Yuan2014}.
%Accurate evaluation of non-diffusive chemistry requires distribution of molecules in ice mantle and is performed using microscopic kinetic Monte Carlo codes to be compared with rate-equation methods \citep[see review and comparison in][]{Cuppen2024, JimenezSerra2025b}. 
Secondly, the reaction intermediate, HOCO, can be stabilized when the heat of reaction dissipates into the surface efficiently.
HOCO may produce CO$_2$ or organic molecules through HOCO + X $\rightarrow$ CO$_2$ + XH or X-COOH \citep{Molpeceres2025, Ishibashi2024}. Alternatively, CO$_2$ could form through CO + O $\rightarrow$ CO$_2$, while its reaction rate is predicted to be low \citep{Goumans2010}. Recent kinetic Monte Carlo simulations show that both mechanisms (CO + OH and CO + O) are important \citep{JimenezSerra2025b}.

CH$_3$OH is the most abundant complex organic molecule \citep[COMs, defined as organic molecules with 6 atoms or more;][]{Herbst2009}. Interstellar ices contain many other COMs at lower abundances, mainly formed by radical recombination (e.g., HCO + CH$_2$OH $\rightarrow$ HC(O)CH$_2$OH) \citep{Garrod2006, Chuang2016, Jorgensen2020}. Radicals are produced by H addition (e.g., CO + H $\rightarrow$ HCO), H abstraction (e.g., CH$_3$OH + H $\rightarrow$ CH$_2$OH + H$_2$) (Figure \ref{fig:network_ice}), photodissociation (e.g., CH$_3$OH $\rightarrow$ CH$_3$ + OH), and radiolysis \citep{Shingledecker2018}. They can thermally diffuse on grains in warm star-forming regions (above $\sim 30$ K), while non-diffusive chemistry should be important at lower temperatures \citep{Jin2020, Garrod2022,Iguchi2025, Borshcheva2025}. Because adsorption sites vary in potential depth, effective diffusion length is also important \citep[e.g.,][]{Kouchi2020}.
Since interstellar ices are mainly H$_2$O, OH radicals formed from photolyzed H$_2$O could drive non-diffusive chemistry by abstracting H from neighbors (e.g., OH + CH$_3$OH $\rightarrow$ CH$_3$O + H$_2$O) \citep{Ishibashi2021}. Furthermore, OH can practically be transported into ice mantle via proton-hole transfer \citep{Tsuge21}.

The parameters for calculating sublimation rate from dust to gas (e.g., adsorption energy) are obtained from laboratory and quantum chemical studies 
%\citep[e.g.][]{Molpeceres2022, Martinez-Bachs2024}
\citep{Minissale2022, Ligterink2023}. %\textcolor{magenta}{Paola: I would delete some of the previous references. For example, we could cite the review and add "and references within" and only add papers written after the reviews.} %At the sublimation temperature, the gas-phase and ice-phase abundances of a species are equal, when the adsorption rate and the sublimation rate are balanced. 
At typical densities of molecular clouds ($10^4$ cm$^{-3}$), the sublimation temperatures of molecules with heavy elements are $\sim 20$ K or higher, so thermal sublimation hardly occurs at 10 K. Instead, a small fraction of ices non-thermally desorb by mechanisms such as photodesorption, cosmic-ray heating, and reactive (chemical) desorption. For CH$_3$OH, H-abstraction followed by re-hydrogenation (Figure 4) may lead to repeated cycles of reactive desorption, analogous to the mechanism proposed for H$_2$S and PH$_3$ \citep{FuruyaOba2022}.
See \citet{Cuppen2024} for more detailed review on grain-surface processes.
%In photodesorption, molecules in the ice mantle are electronically and vibrationally excited by ultraviolet light, and dissociated fragments and surrounding molecules kicked out by vibrational excitation are desorbed (Westley+95, Andersson and van Disheck08, Yabshita+09, Arasa+15). Ultraviolet light from interstellar space does not penetrate deep into molecular clouds, but cosmic rays have a long attenuation length, and hydrogen molecules excited by cosmic rays or secondary electrons emit ultraviolet light (CR-induced UV) (Greadel+). Cosmic rays also desorb molecules by stochastic heating and sputtering dust (Hasegawa and Herbst, Dartois+18, Ivlev+).
%In reactive desorption, molecules are desorbed using part of the energy of chemical reactions in the ice (Minissale+16, Oba+2018, Nguyen+2020, Furuya+22).

\subsubsection{Cosmic-ray ionization rate}
\label{Sect:cosmic-ray}
%\textcolor{red}{copied and modified the paragraphs in Paola's subsection}
As noted above, many chemical processes in molecular clouds are driven by cosmic rays, especially low-energy ones (up to the GeV energy domain). They cannot be observed directly from Earth because the Solar wind shields them, so their spectrum and the ionization rate are estimated from molecular ion observations in interstellar clouds.
%\textcolor{red}{remove or minimize this para.ref to Paola's section}
For example, H$_3^+$ forms from H$_2^+$, which is produced by cosmic-ray ionization of H$_2$, and is mainly destroyed by reaction with CO and other atoms and molecules. In steady-state, its number density is given approximately by $n({\rm H}_3^+)=\zeta_{\rm H2}n({\rm H}_2)/\{kn({\rm CO})\}$, where $k$ is the reaction rate coefficient with CO, and $\zeta_{\rm H2}$ is the ionization rate of H$_2$. Since $n$(CO)/$n$(H$_2$) $\sim 10^{-4}$, $\zeta_{\rm H2}$ can be estimated from H$_3^+$ observations \citep{Oka2006}. 
Likewise, $\zeta_{\rm H2}$ can be estimated by the observations of HCO$^+$, DCO$^+$, and CO and their steady-state abundances in the chemical network \citep{Guelin1982}. The typical value of $\zeta_{\rm H2}$ is $10^{-17}-10^{-16}$ s$^{-1}$, yielding an ionization degree $x({\rm e})=n({\rm e})/n(\rm H_2)=10^{-8}-10^{-7}$, high enough for gas to couple with magnetic fields
\citep[see review by][]{Padovani2016,Gabici2022}
%\citet{Caselli1998} used an analytical formula and a gas-grain chemical model to take into account CO depletion onto grain surfaces. They measured ionization degree ($n({\rm e})/n_{\rm H}$) and $\zeta_{\rm H2}$ in a sample of dense cores to find a large spread of values ($\zeta_{\rm H2}$ between 10$^{-16}$ and 10$^{-17}$\,s$^{-1}$).
%The analytical method in \cite{Caselli1998} can only be applied to a limited range of deuterium fractions and CO depletion factors, as explained in \cite{Redaelli2024}, who in fact advise using the method based on the ortho-H$_2$D$^+$ column density, proposed by \cite{Bovino2020}.  

Parsec-scale maps of $\zeta_{\rm H2}$ have recently been obtained in the cluster-forming region NGC\,1333 in Perseus \citep{Pineda2024} and in the Orion Molecular Clouds OMC-2 and OMC-3 \citep{Socci2024}. Derived values are $\zeta_{\rm H2}\sim 3\times10^{-17}$\,s$^{-1}$ in NGC\,1333, with clear enhancements toward young stellar objects (YSOs), and $\sim5\times10^{-18}-10^{-16}$\,s$^{-1}$ in Orion, where a decrease of $\zeta_{\rm H2}$ with increasing H$_2$ column density is found, as predicted by cosmic-ray propagation models in dense media \citep{Padovani2009,Padovani2018}.

\subsection{Isotopes}\label{sec:D-enrichment}
In molecular clouds, molecular isotopic ratios often differ from the elemental isotope ratios, a phenomenon known as isotope fractionation. Since isotopic ratios propagate through reaction networks, they provide clues to molecular formation pathways \citep[e.g.,][]{Sakai2013}. Fractionation arises mainly from exothermic exchange reactions at low temperatures (D/H and $^{12}$C/$^{13}$C) and isotope selective photodissociation ($^{14}$N/$^{15}$N, $^{16}$O/$^{18}$O/$^{17}$O) combined with ice formation. Here we outline deuterium enrichment because it represents a powerful tool to study the initial conditions of star formation in prestellar cores; a full overview of mechanisms and observations of isotope fractionation is given in the Supplemental Text.
%\textcolor{red}{need to add a table of elemental isotope ratio of H,C,N,O,S}

%\subsubsection{Deuterium}
%\textcolor{red}{Include all referes fro Poala's 1 st paragraph. Should mention CO depletion to shorten the Paola's subsecions on ionization rate and D-enrichemnt.}
%D/H enhancement with links to CO depletion, CRs, o/p ratio of H2 (Furuya+15,19, Ueta+2016, Tsuge+2021)
While the elemental D/H ratio is $\sim 2\times 10^{-5}$ \citep{Friedman2023}, many molecules in molecular clouds show XD/XH $\gtrsim 10^{-3}$. Such deuterium enrichment (D enrichment hereinafter) is caused by several exothermic exchange reactions 
%\citep{Watson1976,Millar1989, Roueff2013, Nyman2019}. 
\citep{Millar1989, Nyman2019}. 
The most representative one is 
\begin{equation}
    \rm H_3^+ + HD \rightarrow H_2D^+ + H_2 + 230 ~ {\rm K}. \label{reac_H2D+}
\end{equation}
%The forward reaction is exothermic because of zero-point energy differences, making the reverse reaction almost negligible at 10 K 
The forward reaction is exothermic due to zero-point energy differences\footnote{The exothermicity of the reaction (\ref{reac_H2D+}), and hence the efficiency of D enrichment, depends on the ortho-to-para ratio of H$_2$, since the ground-state energy of ortho-H$_2$ is higher than that of para-H$_2$ by 170 K. See Supplemental Text.}.
Since H$_2$D$^+$ is destroyed mainly by the back reaction of (\ref{reac_H2D+}), the reaction with CO, and the electron recombination, the steady-state abundance ratio of H$_2$D$^+$ to H$_3^+$ is given by
\begin{equation}
    \frac{n({\rm H_2D^+})}{n({\rm H_3^+})}=\frac{k_{\ref{reac_H2D+}}n({\rm HD})}{k_{\ref{reac_H2D+}}\exp(-230/T)n({\rm H_2})+k_{\rm CO}n({\rm CO})+k_{\rm rec}n({\rm e})}, \label{eq:H2Dplus}
\end{equation}
where the reaction rate coefficients are $k_4 \sim k_{\rm CO}\sim 10^{-9}$ cm$^3$s$^{-1}$ and $k_{\rm rec}\sim 10^{-7}$ cm$^{3}$ s$^{-1}$. In dense cold regions ($n_{\rm H}>10^4$ cm$^{-3}$, $T\sim 10$ K), the second term dominates in the denominator, and the abundance ratio of H$_2$D$^+$ to H$_3^+$ is $\sim 0.1$. CO freeze-out accelerates the deuterium fractionation \citep{Dalgarno1984}. 
Similarly, reactions of H$_2$D$^+$ and D$_2$H$^+$ with HD produce D$_2$H$^+$ and D$_3^+$, respectively %\citep{Roberts2003,Walmsley2004}.
\citep{Walmsley2004}. 
%Reduction of abundant neutrals such as CO and O further enhances these processes, as they are major destruction partners of H$_3^+$ isotopologs \citep{Dalgarno1984}.
Since H$_3^+$ plays a central role in gas-phase chemistry (eq. \ref{reac_H3p}), its enhanced D/H ratio propagates to other molecules.

Electron recombination of deuterated H$_3^+$ produces D atoms, raising the D/H ratio of atomic hydrogen and thus of ices formed by grain-surface hydrogenation. 
The kinetic isotope effect, i.e., the higher tunneling efficiency of H atoms compared to D atoms in reactions with activation barrier, is also important.
For CH$_3$OH, H-abstraction followed by D-addition further enriches the methyl group with D \citep{Hama2013} (Figure \ref{fig:network_ice}).

\section{CHEMISTRY IN NEARBY MOLECULAR CLOUDS}

\subsection{Chemical inventory in dark clouds}
%\textcolor{blue}{(Izaskun; 5 pages)}

%In the early 2000's, it was believed that the chemistry of cold ($T=10$ K), dense [$n(H_2)\sim10^4$ cm$^{-3}$] and quiescent molecular clouds could be described by simple and small species such as CO, HCO$^+$, or N$_2$H$^+$ \citep[see e.g.][]{Caselli1999,Caselli2002a}. However, the advent of high-sensitivity instrumentation in the past decade -- especially at single-dish antennas such as the IRAM 30m, the GBT, and the Yebes 40m telescopes -- enabled detailed studies of the chemistry of molecular dark clouds to unseen complexity levels. 
In the past decade, instruments such as IRAM 30m, GBT, Yebes 40m and JWST have opened a new era in the search for interstellar molecules.
%,  with sensitivity XXX times higher than before.}
This subsection summarizes recent advances in our understanding of the chemical inventory in molecular dark clouds, in both the gas phase and ices. 

\subsubsection{Deep line surveys in TMC-1}
\label{sec:tmc1}
%GOTHAM, QUIJOTE, SANCHO: Aromatics, carbon chains, COMs, and S-bearing organics. 

%- Observations: series of papers by McGuire+, Cernicharo+, also include observations by other groups and telescopes e.g. Sakai+13)
%- Laboratory Experiments: e.g. Mebel+23
The Taurus Molecular Cloud is one of the closest molecular cloud complexes and is considered an archetype of low-mass star formation. Its distance was estimated to be $\sim140$ pc, with a median inter-cloud distance of 25 pc  \citep{Galli2019}.
As shown in Figure \ref{fig:phys_structure}, TMC-1 is part of the Taurus Molecular Cloud and exhibits a filamentary, sheet-like structure.
TMC-1, with a mass of $\sim$8 M$_\odot$, stands out among dark molecular clouds for its rich molecular repository: cyanopolyynes (HC$_{2n+1}$N), carbon chains, aromatic molecules, radicals, cations and anions, isomers, complex organics, and S-bearing molecules have been discovered toward this cloud \citep[e.g.,][and below]{Mcguire2022,Xue2025}. In addition, TMC-1 has a simple physical structure \citep{Smith2023} and, being quiescent (i.e., lacking star formation in its interior), is an excellent laboratory to study astrochemical processes.
Hence, TMC-1 has been extensively observed \citep{Kaifu2004,Gratier2016,Mcguire2020,Cernicharo2021f} and serves as a benchmark for understanding dark cloud chemistry. %\citep{Wakelam2021,Millar2024}. 

The cyanopolyyne peak within TMC-1 (hereafter TMC-1 CP) refers to the position where the emission of long carbon-chain molecules (e.g., cyanopolyynes) reach their peak values. 
It is considered as a starless core ($n$(H$_2$)$\sim$2$\times$10$^4$ cm$^{-3}$) at an earlier evolutionary stage than other cores in this cloud (e.g., TMC-1 C).
Four components with small velocity shifts are found, which could be sub-filaments moving toward the center of the filament \citep{Dobashi2018}. The overlap of multiple components in the line of sight and in velocity increases the line intensities, enabling the detection of many molecules.
The gas temperature is low ($T \sim 10$ K), and the H$_2$ column density is $\sim$10$^{22}$ cm$^{-2}$ \citep{Cernicharo1987}, equivalent to a visual extinction of $A_{\rm V}$$\sim$10 mag\footnote{Other studies give slightly higher values $\sim$(1.5-1.7)$\times$10$^{22}$ cm$^{-2}$, and hence a visual extinction closer to $\sim$20 mag 
%\citep{Navarro2021,Fuente2023,Smith2023}. 
\citep[e.g.][]{Smith2023}. 
Here, we adopt $N$(H$_2$)$\sim$10$^{22}$ cm$^{-2}$ to be consistent with the abundances reported in the QUIJOTE works.}.

The ultrasensitive spectroscopic surveys GOTHAM \citep[GBT Observations of TMC-1: Hunting Aromatic Molecules;][]{Mcguire2020} and QUIJOTE \citep[Q-band Ultrasensitive Inspection Journey to the Obscure TMC-1 Environment;][]{Cernicharo2021f}, carried out respectively with GBT and Yebes 40m telescopes toward TMC-1 CP, have played a key role in advancing our understanding of the chemical complexity of dark molecular clouds. The new molecular species discovered by these two surveys together account for $\sim$1/3 of all interstellar molecules known to date \citep{Xue2025}. These surveys exploited the fact that large molecules such as cyanopolyynes and aromatic species at 10 K present their peak emission at frequencies $\leq$80 GHz, which are less contaminated by emission from simpler species than higher frequencies \citep[][]{JimenezSerra2014a}. The FWHM of molecular lines toward TMC-1 CP is $\leq$1 km s$^{-1}$; therefore, line blending and line confusion are less severe than in
%hot corinos and massive hot molecular cores 
warm/hot cores with protostars
\citep[see e.g.,][]{Jorgensen2020}. By integrating over a thousand hours toward TMC-1 CP, these surveys have reached sub-mK rms noise levels.
%allowing the detection of molecular transitions from low-abundance species with intensities of a few mK, which were undetectable in previous surveys.
%shallower and lower-sensitivity spectral surveys \citep{Ohishi1998,Kaifu2004}. 
Furthermore, the development of new analysis techniques based on line stacking and machine learning has enabled the discovery of large carbon-chains, cyclic hydrocarbons, and aromatic molecules, even up to cyano coronene, which were undetectable in previous surveys \citep{Loomis2021,Toru2025,Wenzel2025b}.

In Supplemental Table 2, we collect all molecular species reported toward TMC-1 CP in the GOTHAM and QUIJOTE surveys. This compilation includes not only newly discovered species but also all other molecules 
previously collected in \citet{Agundez2013} and \citet{Millar2024}. Our aim is to provide an updated and as comprehensive as possible list of molecular species detected toward TMC-1 CP.
Supplemental Table 3 lists molecular isotopologs observed in the GOTHAM and QUIJOTE surveys. Below, we summarize the main findings of the GOTHAM and QUIJOTE surveys, organized by chemical family.

\paragraph{C-bearing molecules (chains and cyclic) and their CN- and CCH-derivatives}
The GOTHAM and QUIJOTE surveys have been extremely successful at discovering not only long carbon chains \citep[as e.g., $l$-H$_2$C$_5$, CH$_2$CCHC$_3$N, or CH$_2$CCHC$_4$H;][]{Cabezas2021e,Shingledecker2021,Fuentetaja2022a} but also pure hydrocarbon cyclic and aromatic molecules such as indene \citep[$c$-C$_9$H$_8$;][]{Cernicharo2021f,Burkhardt2021b} or the recently discovered phenalene \citep[$c$-C$_{13}$H$_{10}$;][]{Cabezas2025d}. These molecules, however, have a very low dipole moment and they are difficult to detect. To search for even larger aromatic molecules, the GOTHAM and QUIJOTE surveys have targeted CN- and CCH-derivatives of these cyclic molecules, which present higher dipole moments and hence, have stronger rotational emission lines. The quest for these derivatives started with the discovery of benzonitrile (c-C$_6$H$_5$CN) by \citet{Mcguire2018}, and since its detection, more than ten CN- and CCH-derivatives have been reported. Examples are the cyano derivatives of acenaphthylene, C$_{12}$H$_8$ \citep[i.e., 1- and 5-C$_{12}$H$_7$CN;][]{Cernicharo2024f}, pyrene C$_{16}$H$_{10}$ \citep[1-, 2-, and 4-C$_{16}$H$_{9}$CN;][]{Wenzel2024,Wenzel2025a}, and coronene \citep[C$_{24}$H$_{11}$CN;][]{Wenzel2025b}; or the CCH-derivatives of cyclopentadiene \citep[c-C$_5$H$_5$CCH;][]{Cernicharo2021i} or benzene \citep[C$_6$H$_5$CCH;][]{Loru2023}. 
While these derivatives are good proxies, they tend to be less abundant (by an order of magnitude or more) than the parent PAHs \citep[][Figure \ref{fig:tmc1}]{Cernicharo2022a, Wenzel2025a}. 

\paragraph{Cyanopolyynes and their Isomers}
TMC-1 CP is recognized as one of the richest repositories of cyanopolyynes \citep[HC$_n$N, with $n$ = 3, 5, 7, 9, and 11; e.g.,][]{Kaifu2004}. The detection of the largest cyanopolyyne HC$_{11}$N has been controversial \citep[][]{Cordiner2017}, but finally confirmed by stacking analysis of the GOTHAM data \citep{Loomis2021}.
The isocyanopolyyne isomers HCCNC and HNCCC \citep[][]{Cernicharo2024b} and HC$_4$NC \citep{Cernicharo2020b}, as well as protonated species \citep[e.g., HC$_3$NH$^+$, HC$_5$NH$^+$, HC$_7$NH$^+$;][]{Cabezas2022d}, have also been reported. The formation of cyanopolyynes and of their isomers is largely attributed to gas-phase reactions between atomic nitrogen and hydrocarbons (Figure \ref{fig:network}), 
and their abundances depend on the adopted elemental abundance ratio of C/O in models \citep[e.g.][]{Millar2024}.
%although chemical models tend to \textcolor{blue}{overproduce large hydrocarbon chains, including cyanopolyynes with $n>$7, for models with C/O$>$1 \citep[][]{Millar2024}.}
    
\paragraph{Molecular anions} 
%Molecular anions have been proposed to enhance the production of unsaturated carbon-chain neutrals (e.g., C$_n$H, C$_n$H$_2$, HC$_n$N), improving the agreement between the model predictions and the observations for TMC-1 CP \citep{Walsh2009}. 
Molecular anions (C$_n$H$^-$ with $n$ = 4, 6, 8, 10 and C$_m$N$^-$ with $m$ = 3 and 5) have been reported toward TMC-1 CP and other dark clouds \citep{Agundez2023d, Remijan2023}. These observations reveal that the larger the molecular size of the anion, the larger the anion-to-neutral abundance ratio. \citet{Agundez2023d} attribute this behaviour to the radiative electron attachment, which is more efficient for longer chain molecules \citep{Herbst_Osamura2008}, although other mechanisms may be possible \citep{Khamesian2016}.
%a gas-phase formation scenario that involves radiative electron attachment \citep{Agundez2023d}.
    
\paragraph{O-bearing COMs} 
%O-bearing molecules such as CH$_3$CHO, H$_2$CCO and cyclopropenone had already been reported by \citet{Soma2018} toward TMC-1 CP. \citet[][]{Agundez2021c} further expanded the inventory of O-bearing COMs in TMC-1 CP to propenal (C$_2$H$_3$CHO), vinyl alcohol (C$_2$H$_3$OH — first detection in space), methyl formate (CH$_3$OCHO), and dimethyl ether (CH$_3$OCH$_3$). Note that CH$_3$OCHO and CH$_3$OCH$_3$ had also been observed toward the methanol peak close to TMC-1 CP \citep{Soma2018} (see also Section \ref{sec:sancho}). 
O-bearing molecules such as acetaldehyde (CH$_3$CHO), cyclopropenone ($c$-C$_3$H$_2$O), and vinyl alcohol (C$_2$H$_3$OH) were reported by \citet{Soma2018} and \citet[][]{Agundez2021c} toward TMC-1 CP.
The QUIJOTE survey also discovered ethanol (C$_2$H$_5$OH), acetone (CH$_3$COCH$_3$), and propanal (C$_2$H$_5$CHO) in TMC-1 CP \citep{Agundez2023c}, revealing also a high level of chemical complexity in this core for O-bearing molecules.% in this young dark molecular clouds such as TMC-1.

\paragraph{S-bearing molecules} A wide variety of new S-bearing molecules have been reported by the QUIJOTE and GOTHAM surveys. Examples include the S-bearing carbon chains C$_n$S \citep[with $n$ = 2, 3, 4 and 5;][]{Cernicharo2021e}, all isomers of HNCS \citep{Cernicharo2024a}, $c$-C$_3$H$_2$S \citep{Remijan2025}, and the S-bearing COMs CH$_2$CHCHS \citep{Cabezas2025b} and CH$_3$CHS \citep{Agundez2025}. 
%Gas-phase chemical models \textcolor{blue}{assuming a sulfur depletion factor $>$100,} underpredict the abundances of these S-bearing species by at least one order of magnitude, indicating that key formation pathways are missing, and/or that the dissociative recombination rates of their protonated precursors need to be better constrained \citep{Agundez2025}. 
While gas-phase chemical models assuming a sulfur depletion factor $\sim$100 reproduce the observed abundance of CS in molecular clouds, they under-predict the abundances of these S-bearing species by at least one order of magnitude, indicating that key formation pathways are missing, and/or that the dissociative recombination rates of their protonated precursors need to be better constrained \citep{Agundez2025}. 
The inclusion of grain-surface chemistry and non-thermal desorption may also help to reproduce the observed abundances \citep{Cabezas2025b}. 

\paragraph{Isotopologues} 
Singly- and doubly-substituted isotopologues (D, $^{13}$C, $^{15}$N and $^{34}$S) have been detected toward TMC-1 CP. The derived isotopic ratios are: $^{12}$C/$^{13}$C$\sim$60-110, depending on whether the molecular species has 1, 2 or 3 C atoms in its structure \citep{Tercero2024,Cernicharo2024b,Xue2025}; $^{14}$N/$^{15}$N$\sim$240-330 \citep{Tercero2024,Cernicharo2024b}; H/D$\sim$40-60 \citep{Tercero2024,Xue2025}; and $^{32}$S/$^{34}$S$\sim$24.5 \citep{Cernicharo2021a}.
\citet{Fuentetaja2025} presented a comprehensive analysis of the different isotopologues of CS (including C$^{36}$S), CCS, CCCS, C$_4$S, C$_5$S, and H$_2$CS, which revealed that the measured isotopic abundance ratios are consistent with Solar values.

\subsubsection{Molecular spatial distribution in TMC-1}
\label{sec:sancho}
Molecular clouds are chemically out of equilibrium and present non-uniform compositions. Comparisons of spatial distributions of molecules are hence useful not only for identifying their formation and destruction pathways, but also for elucidating the physical structure of molecular clouds.

Up to four velocity components have been identified in molecular line profiles observed toward TMC-1 CP \citep[see e.g.,][]{Dobashi2018, Loomis2021}. 
\citet{Soma2018} found that the line profiles of carbon chains differ from those of O-bearing COMs; the former originate from gas-phase reactions, whereas the latter form on grain surfaces followed by non-thermal desorption (\S 2.3). Interestingly, H$_2$CO shows a composite line profile, reflecting both formation routes. 
Mapping around TMC-1 CP confirmed this chemical differentiation: spatial distribution of CH$_3$OH differs from that of CS (representing gas-phase production). CH$_3$OH is distributed over extended regions rather than concentrated in densest regions, as also observed in other starless cores \citep[][see also \S \ref{sec:cores}]{Soma2015}.

%By using the data from the Green Bank Ammonia Survey \citep[GAS;][]{Friesen2017}, \citet{Smith2023} identified two bright velocity components in HC$_5$N $J$ = 9–8. This dense gas tracer seems to probe two layers within one single filament: one lower density and outer layer with material flowing under gravity towards a second higher density inner layer of the filament. These maps, however, only covered a few species (NH$_3$, HC$_5$N, CCS and HC$_7$N) with limited sensitivity ($\sim$0.16 K per 0.08 km s$^{}$-channel). 

By using data from the Green Bank Ammonia Survey \citep[GAS;][]{Friesen2017}, \citet{Smith2023} identified two bright velocity components in HC$_5$N $J$ = 9–8, which seem to probe two gas components within one filament in TMC-1: (i) lower-density ``envelope'' gas is flowing under gravity toward (ii) higher density gas at the central ``spindle'' region of the filament. The HC$_5$N emission in the central region anticorrelates with NH$_3$ emission \citep[see also][]{Olano1988}, which is consistent with the expectation that HC$_5$N is a ``chemically young'' species, while NH$_3$ is ``chemically old'' \citep{Suzuki1992}. These maps, however, only covered a few species (NH$_3$, HC$_5$N, CCS and HC$_7$N) with limited sensitivity.% ($\sim$0.16 K per 0.08 km s$^{}$-channel).

More recently, the molecular line emission toward TMC-1 has been mapped with Yebes 40m telescopes at higher sensitivity in the context of the SANCHO survey \citep[Surveying the Area of the Neighbour TMC-1 Cloud through Heterodyne Observations;][]{Cernicharo2023,Tercero2024}. SANCHO has mapped a region of 320$"$$\times$320$"$ within TMC-1, with a uniform sensitivity of 1.4-4 mK and beam size of $35.6"-56.7"$ across the whole Q-band. This has allowed the simultaneous mapping of the emission of the aromatic molecule benzonitrile ($c$-C$_6$H$_5$CN) and of several carbon chains and cyanopolyynes. 
The SANCHO maps reveal that the emission peak of $c$-C$_6$H$_5$CN is shifted by $\sim$40$"$ with respect to the pointing position of the QUIJOTE and GOTHAM surveys. The spatial distribution of $c$-C$_6$H$_5$CN also closely follows that of cyanopolyynes (e.g., HC$_n$N), which indicates that the two molecular families are produced within the denser zones of the cloud \citep{Cernicharo2023}, and that their chemical synthesis is likely related \citep[see][and \S \ref{sec:others} below]{Agundez2023e}. Spatial distribution of $c$-C$_6$H$_5$CN indeed differs from that of radicals such as C$_n$H and C$_n$N, supporting the idea that cyanopolyynes are precursors of aromatic molecules. Significant differences have also been revealed in the spatial distribution between the $^{13}$C and $^{15}$N isotopologues of HC$_3$N, and DC$_3$N, with the deuterated isotopologue peaking closer to the position of TMC-1 CP. The D enhancement is likely produced by colder chemistry undergoing in this region \citep{Tercero2024}.

\subsubsection{Chemistry in other molecular dark clouds}
\label{sec:others}

After the discovery of $c$-C$_6$H$_5$CN in TMC-1 CP, \citet{Burkhardt2021a} and \citet{Agundez2023e} carried out comprehensive surveys of this molecule toward a sample of %cold clouds and star-forming clumps 
cold cores and protostellar cores rich in carbon chains. $c$-C$_6$H$_5$CN was detected across all these regions, establishing that aromatic chemistry is not unique to TMC-1 but widespread in molecular clouds, and it persists into early star formation \citep{Burkhardt2021a}. \citet{Agundez2023e} also reported some evidence for the presence of indene (C$_9$H$_8$), cyclopentadiene (C$_5$H$_6$), and 1-cyano cyclopentadiene (1-C$_5$H$_5$CN) toward L1495B, the source with the brightest cyanopolyyne emission among their sample. A trend was indeed found between the column density of $c$-C$_6$H$_5$CN and that of HC$_7$N, suggesting a possible relation in the chemistry of both species. This is also consistent with the similar spatial distribution found for $c$-C$_6$H$_5$CN and cyanpolyynes in TMC-1 CP (\S \ref{sec:sancho}). The cold cores show the highest column densities for both $c$-C$_6$H$_5$CN and HC$_7$N, which points to the idea that aromatic cycles are favored in clouds at earlier evolutionary stages where carbon chains are more abundant. 

%\textcolor{red}{MOVED AND MODIFIED FROM 3.1.2}
\subsubsection{Formation processes of carbon-chains and cyclic/aromatic molecules}
\label{sec:cyclic}

Discoveries of the GOTHAM and QUIJOTE surveys have stimulated discussion on the formation and destruction pathways of newly detected molecules (\S \ref{sec:reactions_gas}). \citet{Millar2024}, for example, updated the UMIST database, increasing the number of species from 467 to 737, and compared the predicted abundances from pseudo-time-dependent models with the observations. The best model agrees within one order of magnitude for $\sim$80 out of the 134 species considered. As the model is purely gas-phase, except for H$_2$ formation, the elemental C/O ratio was varied in the initial conditions. A model with C/O ratio of 1.4 (i.e., C-rich, which may be appropriate when water ice is frozen out; \S 2.3.1) reproduces the observations well, although a lower ratio of 1.1 reduces the overproduction of cyanopolyynes. In the model, however, aromatic molecules and O-bearing COMs such as CH$_3$CHO and CH$_3$COCH$_3$ are underestimated by several orders of magnitude. 
%The best match is obtained at 10$^6$ yr, consistent with the timescales inferred by \citet{Navarro2021} from deuterated species. 

\begin{figure}
    \centering
    \includegraphics[width=1.0\linewidth]{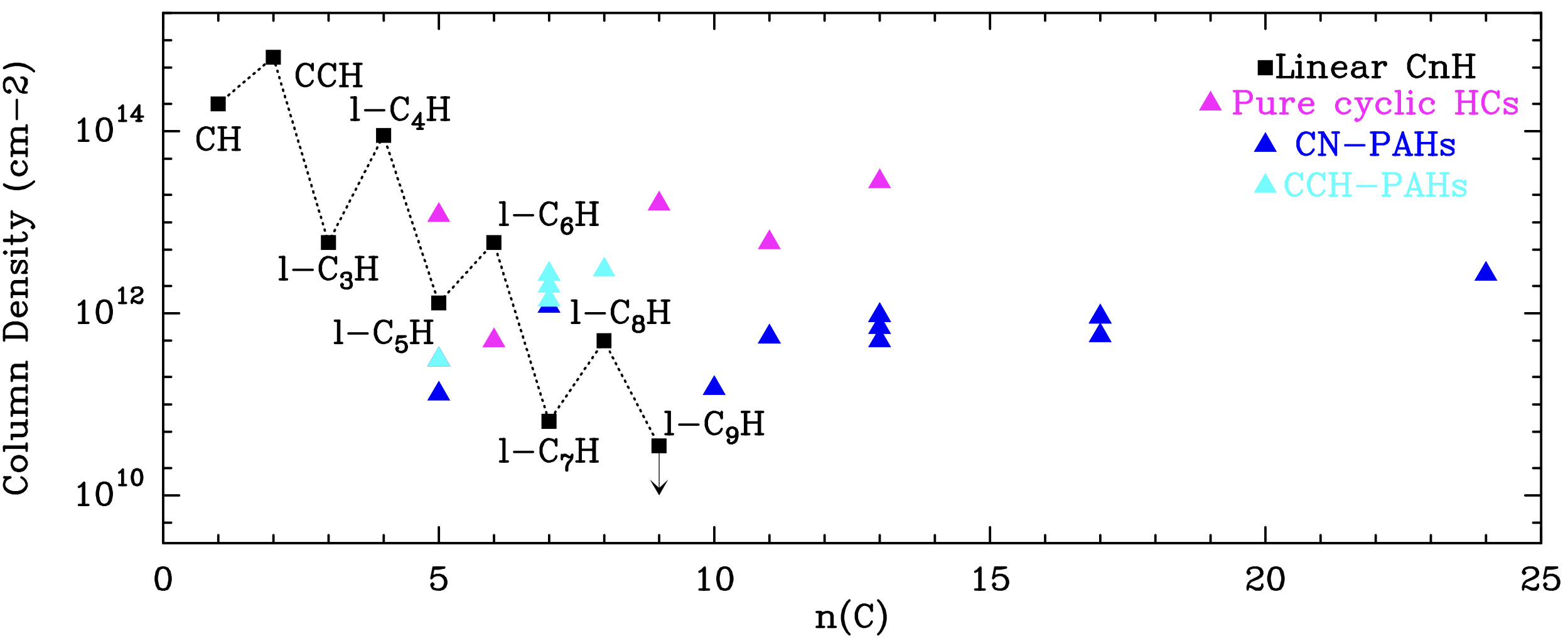}
    \caption{Column densities of carbon-chain molecules and PAHs toward TMC-1 CP as a function of the number of carbon atoms in the species. %Left panel: Blue dashed lines highlight the decreasing trend in column density with increasing C number for linear C$_n$H chains. Orange shaded area indicates the column density range within which PAH families up to 10 C atoms are found. Right panel: Orange shaded area indicates the column density range within which PAHs with more than 9 C atoms have been detected. Figure based on those reported in \citet{Cernicharo2022a} and  \citet{Wenzel2025b}.
    Linear carbon chains are shown in black. Pure cyclic hydrocarbons are shown in magenta, CN-derivatives of PAHs in dark blue, and CCH-derivatives of PAHs in light blue. Original column density data and references are listed in Supplemental Table 2.
    The data for linear C$_n$H was taken from
    \citet{Agundez2013}, \citet{Sakai2010}, \citet{Cernicharo2022c}, \citet{Agundez2023d} and \citet{Cabezas2022g}. For the pure cyclic HCs, CN-PAHs, and CCH-PAHs, the data were taken from \citet{Cernicharo2022c}, \citet{Fuentetaja2026},  \citet{Cabezas2025d}, \citet{Xue2025} and \citet{Wenzel2025b}.}   \label{fig:tmc1}
\end{figure}

While the chemical reaction networks starting from atomic conditions illustrate bottom-up chemistry, an alternative is top-down chemistry (\S \ref{sec:reactions_gas}; Figure \ref{fig:network}). While some observations favor the top-down scenario, others support the bottom-up scenario, and the issue remains unsettled.
In Figure \ref{fig:tmc1}, linear carbon-chains (e.g., C$_n$H family with $n=1-9$) show decreasing column densities with increasing $n$, consistent with the bottom-up origin dominated by ion-molecule gas-phase reactions \citep{Cernicharo2022a}. Even-numbered carbon chains are about an order of magnitude more abundant than odd numbered ones, reflecting their reactivity (i.e., destruction rate) due to differences in ground electronic state \citep[e.g.,][]{Bettens1997}. 
In contrast, cyclic molecules exhibit flatter dependence on $n$ with  column densities of $10^{12}$-10$^{13}$ cm$^{-2}$, especially above 10 C atoms, favoring a top-down origin \citep{Wenzel2025b}. 
Correlation of spatial distributions of c-C$_6$H$_5$CN and cyanopolyynes in TMC-1 (\S \ref{sec:sancho}) and other clouds  (\S \ref{sec:others}), on the other hand, support the bottom-up scenario. Isotopic ratios also provide clues: deuterated benzonitrile has  D/H $< 1.2$ \% \citep{Steber2025}.
This value lies below the range measured for large, unspecific PAHs in PDRs and external galaxies, arguing against a top-down scenario.

\subsubsection{Observations of ices by JWST} \label{sec:iceage}
%e.g. Ice Age 
% (McClure+23, Dartois+24; Noble+24; Smith, Z.+25; Jimenez-Serra+25) 

%INCLUDE PAPER DARIUS LIS ET AL. 2024 ON CHAMAELEON I IN SECTION ABOUT TMC1 AND OTHERS 

%As described in Section \ref{sec:iceage}, we are starting to unveil the chemistry of the solid phase in interstellar ices in dense molecular clouds thanks to the sensitive observations obtained in the IR with the JWST. These observations also point to the presence of COMs in ices even during the earliest and most quiescent phases of star formation in cold and dense molecular clouds. 

\begin{figure}
    \centering
    \includegraphics[width=0.9\linewidth]{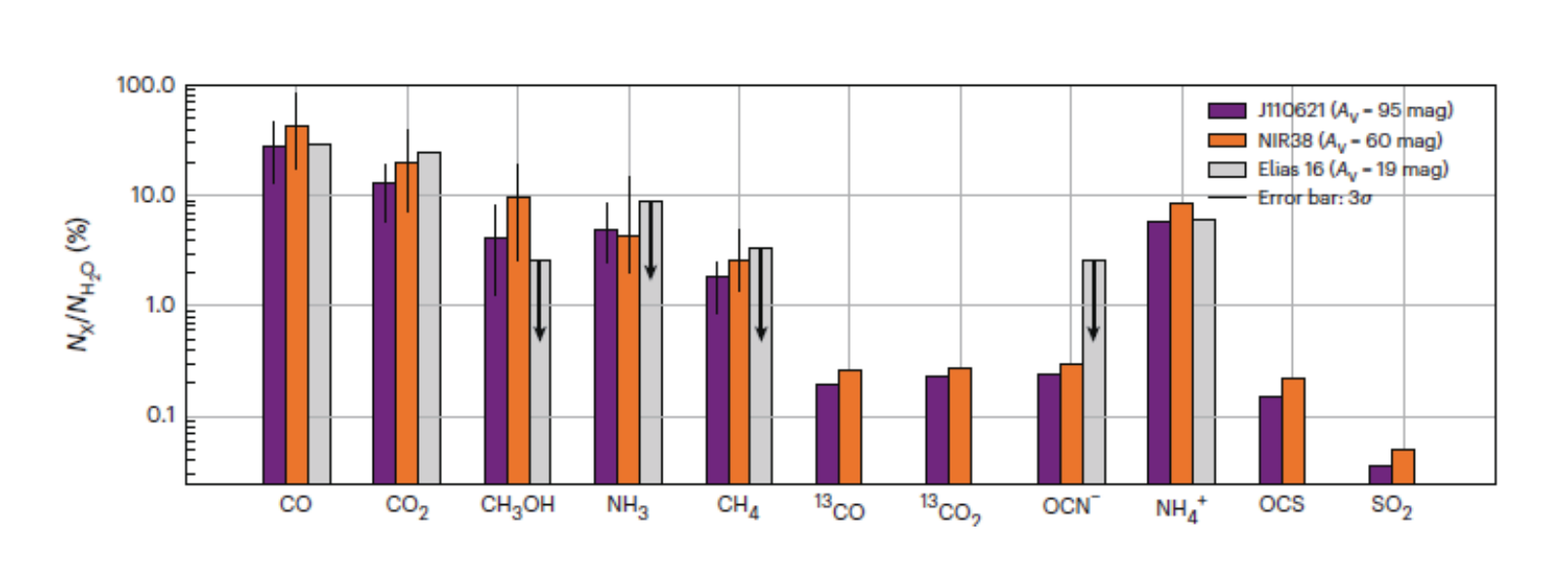}
    \caption{Abundances of icy molecules relative to H$_2$O ice toward NIR38 (K giant, $A_{\rm v}=60$ mag) and J110621 (G giant, $A_{\rm v}=95$ mag) behind the Chamaeleon I molecular cloud, and toward Elias 16 (K Giant, $A_{\rm v}=19$ mag) behind the Taurus Molecular Cloud. The H$_2$O ice abundance in the line-of-sight direction of NIR38 and J110621 is $n$(H$_2$O ice)/$n_{\rm H}\sim 8\times 10^{-5}$. Figure adapted with permission from \citet{McClure2023}. }
    \label{fig:IceAge}
\end{figure}

Ice compositions in molecular clouds are probed through infrared absorption spectroscopy against background sources. The high sensitivity and spectral resolution of JWST have enabled detailed ice observations in quiescent regions of dark molecular clouds improving greatly on pre-JWST data. Figure \ref{fig:IceAge} shows the H$_2$O-normalised ice abundances toward NIR38 ($A_{\rm V}=60$ mag) and J110621 (95 mag) behind the Chamaeleon I molecular cloud, based on JWST observations from the Ice Age project \citep{McClure2023}, as well as toward Elias 16 (19 mag) behind the Taurus Molecular Cloud \citep{Knez2005}.
%\footnote{As for the gas-phase molecular abundances in Chamaeleon I, \citet{Lis2025} performed 18–25 GHz observations of the Class 0 protostar Cha-MMS1 and prestellar core. They detected several molecules including, e.g., HC$_3$N and CCS. While the molecular column densities are an order of magnitude lower than those of TMC-1 CP, the abundances are in general in agreement with TMC-1 values, except for the enhanced abundances of $c$-C$_3$H$_2$ ($\times 25$) and NH$_3$ ($\times 125$) with respect to cyanopolyynes.}.
%at most 19 \% of the total O and C budgets are in ice, while 13 \% of the total N budget is in ice.
The abundance ratios of the dominant ices, H$_2$O, CO, CO$_2$, and CH$_3$OH, are consistent with those from previous works \citep{Oberg2011, Boogert2011, Boogert2015, Whittet2013}. Methanol is not detected toward Elias 16, despite its $A_{\rm V}$ being higher than the threshold value ($\sim 17$ mag) for CH$_3$OH detection \citep{Whittet2011}. Methanol abundance could depend on the environment, such as gas density. For example, \citet{Boogert2011} detected CH$_3$OH ice toward 7 out of 31 background stars; several non-detections had extinctions comparable to those sightlines where CH$_3$OH was detected. \citet{Pontoppidan2004} investigated the spatial distribution of CH$_3$OH ice with VLT-ISAAC L-band spectroscopy toward 10 stars in SVS~4, which is a dense cluster of pre-main
sequence stars deeply embedded in the Serpens star-forming cloud.
Some sightlines pass through the envelope of the Class~0 protostar SMM4. Methanol reaches up to 25\% relative to H$_2$O and maintains high abundance throughout SVS~4, dropping by at least an order of magnitude $\sim 0.09$ pc away from SVS~4, which is consistent with the environment-dependent CH$_3$OH abundance described above.

\citet{Chu2020} and \citet{Goto2021} also reported CH$_3$OH ice toward the prestellar cores L694-2 and L1544, with CH$_3$OH/H$_2$O ice abundance ratios of $\sim$11-14\% and a large conversion of CO into CH$_3$OH (about 50\%), consistent with the formation of CH$_3$OH ice within CO-rich ice layers in dense regions of dark molecular clouds \citep{JimenezSerra2025b}, which largely contribute to the CH$_3$OH gas abundances observed during the protostellar phase \citep{Chu2020}. Although there are indications of more complex COMs than CH$_3$OH at 6.9-7.0 $\mu$m in the Ice Age data, a firm detection would require high spectral resolution MIRI-MRS data from JWST \citep{McClure2023}.

CH$_4$, NH$_3$, and OCN$^-$ have been detected in previous studies toward protostars, and their H$_2$O-normalized abundances are comparable to the values in Figure \ref{fig:IceAge}. It suggests that the primary ice composition is set before star formation.
Absorption bands at $\sim 3.2 \mu$m and $\sim 6.8 \mu$m have been attributed to NH$_4^+$. Detection of ammonium salt NH$_4$SH in Comet 67P/C-G \citep[67P hereinafter,][]{Altwegg2019} renewed attention to these features.
\citet{Slavicinska2025} compared the laboratory measuremets of NH$_4$SH bands with observations and derived NH$_4^+$/H$_2$O ratios of $11\pm 2$ \% and $19\pm 4$ \% for background stars, Elias 16 and CK2. As for S-bearing species, \citet{McClure2023} detected OCS and possibly SO$_2$, but not H$_2$S -- expected to be the primary S-bearing molecular reservoir -- with an upper abundance limit of 0.6 \% relative to H$_2$O. If NH$_4$SH is the dominant ammonium salt, it could account for $\sim 15$ \% of elemental S.

The different shapes of the observed IR band profiles provide information about how molecules are mixed in the ice mantle, and hence, about their formation pathways. CO is dominated by a pure component, with two additional weaker components mixed with CH$_3$OH or CO$_2$. In contrast, CO$_2$ is dominated by an intimate mixture with H$_2$O, with a lesser contribution from a CO-rich mixture \citep{Whittet2013,Boogert2015,McClure2023}.
%Quantum chemical calculations show that CO + OH, one of the main formation reactions, is less efficient on CO ice than H$_2$O ice \citep{Molpeceres2023}.
%However, other formation routes such as CO + O could also be efficient in dark molecular clouds as revealed by KMC simulations \citep{JimenezSerra2025b}.}
%to suggest that this reaction could be much less efficient in CO-rich ice.
An early analysis of the Ice Age data suggested that CH$_3$OH could reside in environments containing both H$_2$O and CO. 
%While CH$_3$OH can be mainly formed by hydrogenation of CO (Fig. \ref{fig:network_ice}), 
%Alternative routes for CH$_3$OH ice formation have been proposed starting from CH$_3$ and OH and a C atom on H$_2$O ice 
However, radiative transfer modelling of the Ice Age data later showed that the observed excess of extinction on the long-wavelength side of the H$_2$O ice band at 3$\mu$m -- initially attributed to a mixture of CH$_3$OH with ammonia hydrates (NH$_3$$\cdot$H$_2$O) -- could simply be explained by grain growth to grain sizes of 0.9$\mu$m \citep{Dartois2024}. Therefore, it is important to keep in mind that the observed ice band shapes could be due not only to the different ice mixtures but also to different physical properties of dust and ice such as size and porosity.

Ice mixture and structure can also be probed via dangling-OH absorption bands. \citet{Noble2024} detected the $\sim$2.7~$\mu$m dangling-OH band toward eight background stars in Chamaeleon~I.
%in the Ice Age project. 
A narrow feature at 2.703~$\mu$m, attributed to three-bonded dangling OH, and a broader feature at 2.753~$\mu$m, likely associated with H$_2$O adjacent to CO and CO$_2$, were detected, whereas the 2.715~$\mu$m feature, expected from H$_2$O--H$_2$, was not.

Spatial distributions of ices (ice maps) suggest the importance of CO freeze-out. \citet{Smith2025} used Ice Age spectra of 44 background stars toward Chamaeleon~I to derive the spatial distributions of H$_2$O, CO, and CO$_2$ ices. 
While the correlations between these three ice species are broadly constant across the cloud, there are small but consequential spatial variations in the ice abundances.
%Ice column densities correlate well with dust column densities, but the spatial correlation is not linear. 
The CO$_2$/H$_2$O ratio is lowest ($\sim$0.15) at the region of lowest dust column densities, and rises to $\sim$0.4 toward the densest regions. The CO/H$_2$O ratio peaks in the same locations. These ice maps suggest that CO freeze-out in high-density regions promotes CO$_2$ formation (see also \citealt{Pontoppidan2006}). 
Ice mapping toward many more background stars are expected with JWST.

\subsection{Starless/Prestellar cores} \label{sec:cores}
%\textcolor{blue}{(Paola; 5 pages)}

During the process of star formation, filaments fragment and form dense cores, which have been extensively studied in the past decades with NH$_3$ inversion transitions \citep[e.g.,][]{Benson1989,Friesen2017} and dust continuum emission \citep[e.g.,][]{Andre2014}. Dense cores associated with YSOs are called protostellar cores. The rest are starless cores, considered to be in early evolutionary stage of star formation. 
In the early ‘90s, \cite{Suzuki1992} found that in nearby molecular clouds, starless cores are also chemically younger than protostellar cores, i.e., richer in carbon-chain molecules, which copiously form when carbon atoms are not yet mainly locked in CO \citep{Herbst1989,Hirota2009}. TMC-1 CP, detailed in \S \ref{sec:tmc1}, is one of such starless cores.

Starless cores can be recognized as overdensities within clouds, with average densities ($n({\rm H}_2) > 10^4$\,cm$^{-3}$) at least ten times larger than the surrounding medium and with temperatures of about 10\,K, set by the balance of cosmic-ray heating and CO line cooling \citep{Goldsmith2001}.  Turbulence is subsonic \citep[e.g.,][]{Goodman1998}, and molecular lines are so narrow that they have been used as “laboratories” for molecular spectroscopy studies \citep[e.g.,][]{Caselli1995}. When the central densities ($n_c$) of starless cores become larger than $\sim10^5$ cm$^{-3}$, dust-gas thermal coupling becomes important, and dust cooling takes over the CO line cooling. This number density value (10$^5$ cm$^{-3}$) is considered the dividing line between thermally subcritical ($n_c\leq10^5$\,cm$^{-3}$) and thermally supercritical ($n_c>10^5$\,cm$^{-3}$) starless cores \citep{Keto2008}, the latter also called more concisely prestellar cores \citep{Crapsi2005}, leading to dynamical %(or free-fall) 
collapse \citep{Tokuda2020}. The relatively high central densities of prestellar cores cause the gas and dust temperature to drop below 10\,K \citep{Crapsi2007,Pagani2007}.
Molecular freeze-out becomes catastrophic, as
non-thermal desorption rate of CO (driven by cosmic rays) can no longer counteract the adsorption rate, which increases with gas density (eq. \ref{eq:H2formation})
\citep{Caselli1999}. CO depletion accelerates deuterium fractionation (eq. \ref{eq:H2Dplus}) \citep[see][and references therein]{Ceccarelli2014}.
 These conditions immediately precede the formation of protostars and protoplanetary disks, making prestellar cores essential laboratories for studying the initial conditions of star and planet formation. 

\subsubsection{The prototypical prestellar core: L1544}
Because of the high density contrasts between their centers and outer edges, prestellar cores are dynamically unstable and contract on their way toward star formation. With free-fall times $t_{\rm ff} \lesssim 10^5$\,yr, they are  short-lived  and therefore rare. L1544 is a $\sim$8\,M$_{\odot}$ \citep{Tafalla1998} prototypical prestellar core, which lies inside the dark cloud number 1544 in the Lynds catalogue \citep{Lynds1962}, on the eastern side of the Taurus Molecular Cloud (Figure \ref{fig:phys_structure}a), at a distance of 170\,pc \citep{Galli2019}. The whole dark cloud is embedded in the Taurus Ring.
%, a 3D elliptical sub-structure within the Perseus-Taurus shell, part of the swept-up interstellar medium from a past energetic event, likely one of more supernovae that exploded between 6 and 22\,Myr ago \citep[][see Fig.\,\ref{fig:phys_structure}]{Bialy2021}. 
Because of its proximity and bright lines, the prestellar core L1544 has been studied with great detail for almost thirty years. The first detailed study of L1544 was done by \cite{Tafalla1998}, using several single dish telescopes 
%(FCRAO, Haystack, IRAM, JCMT) 
(e.g., FCRAO and IRAM) 
to investigate the physical and chemical structure of the core and its surrounding filamentary cloud. Molecular lines of abundant high-density tracers such as HCO$^+$(1-0), H$_2$CO($2_{12}-1_{11}$), CS(2-1), $c$-C$_3$H$_2$($2_{12}-1_{01}$), and, to a lesser extent, N$_2$H$^+$(1-0), present deep self-absorption, which could be explained with very subthermal excitation in the outer core layers and extended infall motions \citep[see also][who found a very low density envelope, at most a few tens of cm$^{-3}$, slowly accreting onto the core]{Redaelli2022}. 
%This work was followed by IRAM 30m observations and analysis by \cite{Caselli2002a}, who found agreement with the kinematics predicted by the magnetohydrodynamical model of \cite{Ciolek2000}, implying the important role of magnetic fields in the  dynamical evolution of these early phases of star formation. Another conclusion of \cite{Caselli2002a} was that a central “molecular hole” is needed to reproduce double-peaked line profiles of optically thin tracers. 
Subsequent IRAM 30m observations by \citet{Caselli2002a} found kinematics consistent with the magnetohydrodynamical model of \citet{Ciolek2000}, implying an important role for magnetic fields in the early dynamical evolution of star formation. They also concluded that a central “molecular hole” is required to reproduce the double-peaked profiles of optically thin tracers.
Indeed, catastrophic CO freeze-out within the central $\simeq$7000\,au \citep[using the new L1544 distance by][]{Galli2019} had already been reported by \cite{Caselli1999}, so it was no surprise that other species chemically linked to CO \citep[such as HC$^{18}$O$^+$ and HC$^{17}$O$^+$; see also][]{FerrerAsensio2022} were centrally depleted. As mentioned in Section \ref{sec:iceage}, large fractions of CH$_3$OH ice relative to H$_2$O ice (11\%) have been found along the line of sight of a background star about 20000\,au northeast from the L1544 center, suggesting significant CO freeze-out also in the surrounding molecular cloud and efficient CH$_3$OH production on grain surfaces \citep{Goto2021}.

Thanks to the high sensitivity of Herschel and the high-resolution HIFI spectrometer \citep{Pilbratt2010,deGraauw2010}, the ground state transition of ortho-H$_2$O ($o$-H$_2$O) ($1_{10}-1_{01}$) was detected in L1544, marking the first detection of water vapor toward a prestellar core \citep{Caselli2012b}. The line exhibits an inverse-P Cygni profile, which can be reproduced only if central infall is ongoing, cosmic-ray induced UV photons drive some photodesorption of water ice, and at least 2.6 Jupiter masses of water ice is present within the whole core, thus showing that a large amount of water is already present {\em before} star formation. The $o$-H$_2$O ($1_{10}-1_{01}$) line, together with previously observed N$_2$H$^+$(1-0) and C$^{18}$O(1-0), enabled \citet{Keto2015} to prove that L1544 is contracting quasi-statically. The central density predicted by these models, $\sim 10^7$\,cm$^{-3}$ \citep{Keto2010}, was later confirmed by ALMA \citep{Caselli2019,Caselli2022}, revealing a dense central region “kernel” (radius $\simeq$1800\,au and mass $\simeq$0.16\,M$_{\odot}$) resembling the flattened structure predicted by theory, within which the future protoplanetary disk will form  \citep[the ``pseudodisk'';][]{Galli1993a,Galli1993b}. 

Within the “kernel”, nearly all (99.99\%) species heavier than helium appear to be frozen onto dust grains \citep{Caselli2022}, implying catastrophic freeze-out also predicted by theory \citep{Roberts2004,Walmsley2004} after the discovery of a strong $o$-H$_2$D$^+$ ($1_{10}-1_{11}$) line in L1544 \citep{Caselli2003} with the Caltech Submillimeter Observatory. Bright $o$-H$_2$D$^+$ ($1_{10}-1_{11}$) lines imply significant depletion of species like CO and O, the major destroyers of H$_3^+$ and its deuterated isotopologues (\S \ref{sec:D-enrichment}). As the “kernel” is the precursor of the future stellar system, these findings show that prestellar chemistry is stored in thick icy mantles on dust grains and may be inherited to the next stages of planet formation. This conclusion has also been reached to explain the excess of deuterated water in our Solar System \citep{Cleeves2014} and the high abundance of deuterated methanol in comet 67P \citep{Drozdovskaya2021}, as discussed in the Supplemental Text. Figure\,\ref{fig:L1544} summarizes the main observational findings achieved in the past 20 years toward the prototypical prestellar core L1544. 

\begin{figure}
    \centering
    \includegraphics[width=0.9\linewidth]{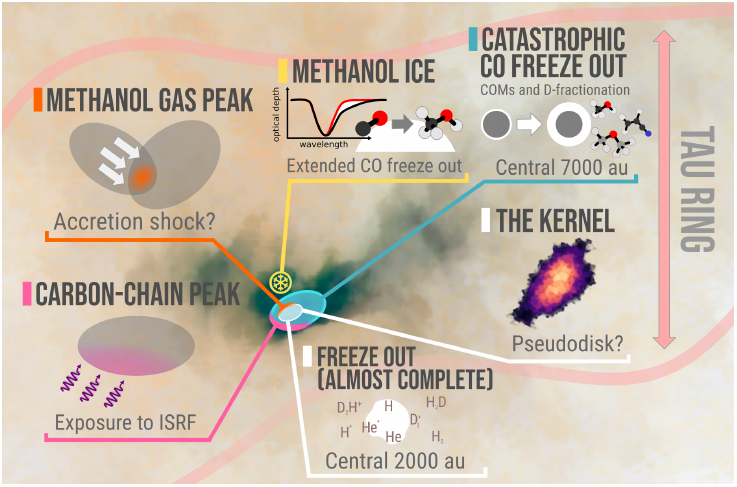}
    \caption{Summary sketch of the prototypical prestellar core L1544 featuring: (i) in the background, Herschel image of the Taurus Molecular Cloud containing the L1544 filament and core (green color: 500\,$\mu$m dust continuum emission). The pink thick curves enclose the Taurus ring \citep{Bialy2021}; (ii) the carbon-chain peak, toward the sharp drop in column density \citep{Spezzano2016}; (iii) the methanol gas peak \citep{Bizzocchi2014}, where large COMs have been detected in the gas phase \citep{JimenezSerra2016}, and coincident with a density enhancement caused by a slow accretion shock \citep{Lin2022}; (iv) the methanol ice detection \citep{Goto2021}, witnessed by the broad long wavelength wing of the 3\,$\mu$m water ice feature (see black curve). This shows that CO freeze out, first step toward methanol formation, is extended beyond (v) the catastrophic CO freeze out zone, within which $>$90\% of the CO molecules are adsorbed onto dust grains \citep{Caselli1999}, and where production of COMs \citep{Borshcheva2025} and deuterated species \citep{Caselli2002b} are enhanced; (vi) the {\it kernel}, discovered with ALMA, whose characteristics resemble the elongated structure (or pseudodisk) formed during the collapse of magnetized clouds \citep{Caselli2019}, and where the freeze out becomes almost complete \citep[$>$99\% of species heavier than He are locked in ice, leaving the gas phase rich inlight species;][] {Caselli2022}. The inset is the NH$_2$D ALMA image from  \cite{Caselli2022}.} 
%    \caption{Summary sketch of the prototypical prestellar core L1544 featuring: (i) in the background, Herschel image of the Taurus Molecular Cloud containing the L1544 filament and core (green color: 500\,$\mu$m dust continuum emission). The pink thick curves enclose the Taurus ring; (ii) the carbon-chain peak, toward the sharp drop in column density; (iii) the methanol gas peak, where large COMs have been detected in the gas phase, and coincident with a density enhancement caused by a slow accretion shock; (iv) the methanol ice detection, witnessed by the broad long wavelength wing of the 3\,$\mu$m water ice feature (see black curve). This shows that CO freeze out, first step toward methanol formation, is extended beyond (v) the catastrophic CO freeze out zone, within which $>$90\% of the CO molecules are adsorbed onto dust grains, and where production of COMs and deuterated species are enhanced; (vi) the {\it kernel}, discovered with ALMA, whose characteristics resemble the elongated structure (or pseudodisk) formed during the collapse of magnetized clouds, and where the freeze out becomes almost complete ($>$99\% of species heavier than He are locked in ice, leaving the gas phase rich in light species). The inset is the NH$_2$D ALMA image from  \cite{Caselli2022}.}     
    \label{fig:L1544}
\end{figure}

\subsubsection{Comparison among cores}
\label{Sect:Comparison}

A first sample of six prestellar cores\footnote{One of the seven prestellar cores identified by \cite{Crapsi2005}, L1521F, was soon after recognized as hosting a very low luminosity object, or VeLLO \citep{Bourke2006}.} was identified by \cite{Crapsi2005}, using level of deuterium fraction (\S \ref{Sec:D-fractionation}) in N$_2$H$^+$ as the criterion. Besides L1544, they identified the prestellar core L183, embedded in an isolated high-latitude cloud; its structure and chemical composition have been studied by, e.g., \cite{Swade1989}, \cite{Pagani2007}, and \cite{Lattanzi2020}. Like L1544, L183 presents significant molecular freeze-out (with $>$90\% of CO in the solid phase) and low temperatures ($\sim7$\,K) toward its central regions. However, the two cores show chemical differences; carbon chain molecules are more abundant in L1544, while O-bearing molecules, including SO and SO$_2$, are more abundant in L183. This can be understood considering the different environments in which the two prestellar cores are embedded. L1544 is more exposed to the ISRF than L183, allowing more carbon chains to thrive \citep{Spezzano2016,Lattanzi2020}. Differences in the dynamical evolution can also affect the chemical structure of dense cores \citep[e.g.,][]{Lee2003,Aikawa2005,Sipila2018}.

\cite{Spezzano2016,Spezzano2017} found a clear chemical segregation in L1544, with CH$_3$OH and other O-bearing molecules toward the north of the core \citep[][]{Bizzocchi2014,Vastel2014,JimenezSerra2016}, while C-rich molecules, such as c-C$_3$H$_2$ \citep[and cyanopolyynes;][]{Bianchi2023}, peak toward the southern region, more exposed to the ISRF. The impinging ISRF maintains a significant fraction of carbon in atomic form, enabling the efficient formation of C-rich molecules. A similar chemical differentiation has also been found toward other nearby prestellar cores in different environments (HMM-1 and Oph D in Ophiuchus, and the isolated L694-2) and the starless core L1521E in Taurus. The Bok Globule B68 (another starless core), where CO and N$_2$H$^+$ depletion has been measured \citep{Bergin2002} shows no chemical differentiation, as the ISRF is uniformly distributed around the core \citep{Spezzano2020}. This shows that the core chemical composition depends on the structure of the surrounding cloud. It is noteworthy that L1521E, with a lower central density than L1544 and embedded in the same molecular cloud (Taurus), is richer in S-bearing molecules, suggesting that sulfur depletion increases during the evolution from the starless to the prestellar phase \citep{Nagy2019}. This is consistent with the findings of \cite{Fuente2019} and \cite{HilyBlant2022} in Taurus and theoretical predictions from \cite{Laas2019}. In general, freeze-out is expected to increase with the dynamical evolution of dense cores, as also found by \cite{Lin2023}, who studied one starless core (L1517B) and two prestellar cores (L429 and L649-2) using high-angular-resolution VLA+GBT NH$_3$ observations. They found a clear trend of increasing NH$_3$ depletion with increasing central density \citep[see also][who measured significant NH$_3$ depletion in the prestellar core HMM-1]{Pineda2022}. 

Considering the importance of prestellar cores as precursors of stellar systems, an unbiased search has started to identify them in nearby molecular clouds and to study how their physical and chemical properties depend on environment. \cite{Caselli2025} used dust continuum emission from the {\em Herschel Gould Belt Survey} \citep{Andre2010} archive to select all cores with central densities (within the Herschel beam) $\ge 3\times10^5$\,cm$^{-3}$, a threshold provided by L1544. They found 40 prestellar core candidates (out of 1746 starless cores), which were then studied with APEX observations of N$_2$H$^+$(3-2) and N$_2$D$^+$(4-3). The first results from detailed radiative transfer modeling is reported for IRAS\,16293E and CrA 151\citep{Spezzano2025, Redaelli2025}. IRAS16293E is a prestellar core next to the multiple Class 0 source (IRAS16293-2422) in Ophiuchus and affected by interaction with one of its outflows \citep[see also][]{Lis2016,Pagani2024}. Interestingly, no depletion of N$_2$H$^+$ and N$_2$D$^+$ is present in IRAS16293E \citep[unlike in L1544;][]{Redaelli2019}, while N$_2$D$^+$/N$_2$H$^+\sim 44$\% is almost twice higher than in L1544.
%one of the largest D-fractions was measured ($\sim$44\%, almost two times larger than in L1544 
%\textcolor{red}{specify }\textcolor{cyan}{PAOLA: for me it is obvious that we are refering to N2H+ and N2D+; do we need to repeat this?}). 
CrA\,151 is a prestellar core with a high central density ($\sim10^7$\,cm$^{-3}$), thus among the most dynamically evolved of the sample. In fact, it shows moderate central heating, which could be caused by gravitational energy release or by the presence of a very young protostar or a VeLLO. Also in this case, the D-fraction measured in N$_2$H$^+$ is very high ($\sim$50\%). These works demonstrate how chemistry and molecules are excellent probes of the physical conditions and processes taking place in dark molecular clouds and their cores (\S\ref{sec:physics}).

\subsubsection{Molecules as probes of physics}
\label{sec:physics}

Molecules are important diagnostic tools for probing the physical properties of interstellar matter. Here, we focus on the  use of molecules to measure the cosmic-ray ionization rate (which regulates the ionization fraction), the deuteration fractionation (formation of useful tracers to study the densest and coldest gas which are precursors of stellar systems), and the accretion of material onto dense cores (important for the dynamical evolution of star-forming regions).

\paragraph{CR ionization rate}  
%Molecular clouds maintain relatively low levels of electron fractions (fractional abundances of electrons with respect to H$_2$ molecules, $x(e)$ between $10^{-8}$ and $10^{-6}$), thanks to the presence of low energy cosmic rays, which are accelerated in remnants of supernovae explosions and, locally, by shocks along jets driven by young stellar objects \citep[][for a comprehensive review on low energy cosmic rays in the dense interstellar medium]{Padovani2016,Gabici2022}. Being partially ionized, the clouds and cores dynamically evolve under the influence of magnetic fields \citep[e.g.,][and references therein]{Pattle2023}. It is therefore important to measure the cosmic-ray ionization rate per H$_2$ molecule, $\zeta_{H2}$, which regulates the electron fraction in dense clouds and, consequently, their dynamical and chemical evolution. 

%How to measure $\zeta_{H2}$ in dense clouds has been originally discussed by \cite{Guelin1982} and \cite{Wootten1982}, using the DCO$^+$/HCO$^+$ and HCO$^+$/CO column density ratios, sensitive to $\zeta_{H2}$. Following this original work, and taking into account the freeze-out of CO onto dust grains, 
The ionization rate $\zeta_{\rm H2}$ in cores can be estimated by observing CO, HCO$^+$ and DCO$^+$ and considering the effect of CO depletion (\S,\ref{Sect:cosmic-ray}).
\cite{Caselli1998} used an analytical formula and a gas-grain chemical model to estimate $x$(e) and $\zeta_{\rm H2}$ in a sample of dense cores, finding a large spread of values ($\zeta_{\rm H2}$ $\sim 10^{-16}-10^{-17}$\,s$^{-1}$). This analytical formula can only be applied to a limited range of deuterium fractions and CO depletion factors, as explained in \cite{Redaelli2024}, who advise using the method based on the $o$-H$_2$D$^+$ column density, proposed by \cite{Bovino2020}. 
%--moved to section 2
%Large (parsec) scale maps of $\zeta_{H2}$ in dense molecular clouds have been recently presented by \cite{Pineda2024} in the cluster-forming region NGC\,1333 in Perseus and by \cite{Socci2024} in the Orion Molecular Clouds OMC-2 and OMC-3. Values derived are $\zeta_{H2}\sim 3\times10^{-17}$\,s$^{-1}$, with clear enhancements toward young stellar objects, in NGC1333, and $\sim5\times10^{-18}-10^{-16}$\,s$^{-1}$ in Orion, where a decrease of $\zeta_{H2}$ with increasing H$_2$ column density is also found, as predicted by models of cosmic ray propagation in dense media \citep{Padovani2009,Padovani2018}. 
%----

\cite{Redaelli2021} used high-sensitivity, high-spectral-resolution IRAM-30m observations of four ions (N$_2$H$^+$, N$_2$D$^+$, HC$^{18}$O$^+$, and DCO$^+$), coupled with non-local thermodynamic equilibrium radiative transfer calculation, to constrain $\zeta_{\rm H2}$ in L1544. They derived an average value of $\zeta_{\rm H2}= 3\times10^{-17}$\,s$^{-1}$, in agreement with the attenuation model based on Voyager measurements of the cosmic-ray spectrum and similar to the reevaluation of $\zeta_{\rm H2}$ in diffuse clouds by \cite{Obolentseva2024}. A lower value of $\zeta_{\rm H2}$ $(5\times 10^{-18}-1\times 10^{-17}$ s$^{-1}$) was found by \cite{Harju2024} in the prestellar core HMM-1 in Ophiuchus, with APEX observations of $o$-H$_2$D$^+$, N$_2$D$^+$ and DCO$^+$. Interestingly, \cite{Indriolo2012} found low upper limits toward diffuse sight lines in the Ophiuchus-Scorpius region, in harmony with the result of \cite{Harju2024}, suggesting that the cosmic-ray flux across the Ophiuchus Molecular Cloud is overall reduced compared to other regions. In contrast, higher $\zeta_{\rm H2}$ values have been found toward warmest cores near protostars, which contribute to the re-acceleration of local cosmic rays \citep{Redaelli2025b}.

\paragraph{Deuterium fractionation} \label{Sec:D-fractionation}
The catastrophic freeze out of CO in prestellar cores enhances the D enrichment (\S \ref{sec:D-enrichment}). The atomic D/H ratio in prestellar cores, for example, can approach unity \citep{Roberts2003,Walmsley2004}, i.e., more than four orders of magnitude higher than the elemental D/H ratio. 
%---------added  by Yuri after we decided to move isotopes to supplemental material-----------
While CO freeze-out and D enrichment become more pronounced as prestellar core densities increase, CO sublimates at $\sim 20$ K, and exchange reactions (e.g., reaction \ref{reac_H2D+}) proceed efficient in the backward direction once the central temperature rises due to star formation. Deuterated molecules therefore serve as valuable tracers of core evolution. Molecular ions such as N$_2$D$^+$ and H$_2$D$^+$ are especially useful, promptly reflecting the physical condition due to their short chemical timescales \citep[e.g.,][]{Crapsi2005, Tokuda2025}. 
The deuterated molecules N$_2$D$^+$ and para-NH$_2$D have recently been used by \citet{Arzoumanian2026} to probe the ion-neutral drift velocity in L1544, providing evidence of ambipolar diffusion.
Further overview is in Supplemental Text.

\paragraph{Accretion of material onto dense cores} \label{Accretion}
The distribution of gas-phase CH$_3$OH in starless and prestellar cores often appears asymmetric. Although this is partly due to the non-uniform environments within which cores are embedded as in the case of L1544 and other starless cores (\S 3.2.2), additional processes may contribute to local enhancements of the CH$_3$OH column density. In particular, asymmetric accretion flows can generate local, slow shocks at the interface between the infalling material and the dense core 
%\citep{Punanova2018,Lin2022,Hsu2025,Choudhury2025}, 
\citep[e.g.,][]{Choudhury2025}, 
leading to enhanced gas density. Because CH$_3$OH is not efficiently formed in the gas phase but is primarily produced on dust-grain surfaces (\S 2.3.2), it is especially sensitive to such processes. Increased density promotes CO freeze out and CH$_3$OH formation, while weak shocks may trigger grain-grain collisions, which could help desorption of surface species \citep{Kalvans2022}. Sheared flows inducing Kelvin-Helmholtz instability in a surface layer of the prestellar core H-MM1 have been inferred from ALMA CH$_3$OH observations by \cite{Harju2020}, who found a maximum layer thickness of about 1000\,au. In this case, velocity fluctuations caused by the two-dimensional turbulence can induce grain-grain collisions with the subsequent sublimation of icy species. 

Molecules abundant at the interface between the core and the cloud thus provide key information on the ongoing gas accretion onto starless and prestellar cores, which continue to grow in mass. Such accretion continues to later evolutionary stages, as demonstrated by molecular streamers connecting protostars to their envelope and surrounding clouds. Recent discoveries of streamers from cloud or core scales to protostars \citep[e.g.,][]
{Tokuda2014,Pineda2020,Pineda2023,Flores2023,ValdiviaMena2024,Gieser2024,Podio2024} 
and from clouds to planet-forming disks \citep[e.g.,][]{Alves2020,Ginski2022,Harada2023,Gupta2024,Winter2024} reveal a continuous connection between molecular clouds and various stages of star and planetary-system formation. These streamers, predicted by hydrodynamical simulations of star-forming regions \citep{Kuffmeier2017,Dullemond2019,Hanawa2022,Yano2024}, transport gas, dust, and ice from the cloud to planet-forming regions, enriching the local chemistry with pristine (“fresh”) material. Following the discovery of a streamer feeding the young protostar Per-emb-2 by \citet{Pineda2020}, \citet{Taniguchi2024} found its reservoir, a cold, dense, chemically young starless core $\sim 20,000$ au north of the protostar. Streamers thus enable cloud material to influence the chemical \citep[and dynamical, see][]{Mauxion2024,Morbidelli2024} evolution of planet formation over millions of years. Although research on this topic is still in its infancy, with further observations needed to test theories, these results underscore the crucial role of molecular clouds and prestellar cores in shaping stellar systems and their chemistry such as in our own.

\subsubsection{COMs: detection and formation} 
COMs beyond CH$_3$OH were traditionally detected in the hot ($T \gtrsim$ 100\,K) region around YSOs, where ices sublimate (i.e., hot cores and hot corinos).
These COMs were threfore thought to form on icy dust grains in warm ($\sim 30$ K) region surrounding the hot core or corino, where thermal diffusion and reaction of radicals are efficient on grain surfaces \citep[e.g.,][]{Garrod2008}. %\citep[e.g.,][]{Charnley1992,Caselli1993,Viti1999,Garrod2008}.
 
Detection of COMs beyond CH$_3$OH in starless and prestellar cores started with the discovery of propylene (CH$_2$CHCH$_3$) toward TMC-1 CP \citep{Marcelino2007} and the observations of CH$_3$CHO, methyl formate (HCOOCH$_3$), dimethyl ether (CH$_3$OCH$_3$), and ketene (CH$_2$CO) toward L1689B \citep{Bacmann2012}. While their abundances are low ($10^{-11}-10^{-9}$ relative to H$_2$) compared to those in hot cores and corinos, these detections motivated theoretical studies on the formation and desorption of COMs at 10 K. \cite{Vasyunin2013} proposed reactive desorption of precursor species (such as H$_2$CO and CH$_3$OH) followed by gas-phase reactions \citep[see also][and references therein]{Balucani2015,Ceccarelli2023}. 
Reactive desorption is important also for COMs formed by grain-surface reactions (see \S \ref{sec:grain_surface_reactions}). 
%It is based on the assumption that part of the formation energy is transformed to kinetic energy of the product icy molecule, which can then break the bond with the neighboring molecules and lift off into the gas phase \citep[thus the name “reactive desorption”;][]{Garrod2006FaDi,Garrod2007}. The rest of the energy can be dissipated into the dust/ice lattice via coupling to phonon modes.

\cite{Vastel2014} then realized that COMs appear to trace the outer layers of prestellar cores. In L1544, the CH$_3$OH peak is offset by about 5000\,au to the northeast of the dust peak. (\S 3.2.2).
%Similarly, \cite{Bizzocchi2014} mapped methanol across L1544, finding that its emission has an asymmetric ring morphology around the dust peak. The CH$_3$OH peak was shifted by about 5000\,au northeast of the dust peak. 
\cite{JimenezSerra2016} carried out high-sensitivity IRAM\,30m observations toward the dust peak and the CH$_3$OH peak positions of L1544, detecting several COMs, including HCOOCH$_3$ and CH$_3$OCH$_3$. The O-bearing COMs show higher abundances toward the CH$_3$OH peak than toward the dust peak (by factors between a few and 10), while the N-bearing COMs only show a modest (factors $\leq$3) increase toward the CH$_3$OH peak. 

The increase of O-bearing COMs toward the methanol peak is reproduced by the gas-grain model of \cite{Vasyunin2017}. 
By incorporating the physical structure of L1544, the model shows a sharp increase in gas-phase CH$_3$OH at the radius of catastrophic CO freeze-out, about 5000 au from the core center.
%This model, which takes into account the physical structure of L1544, clearly shows that at the location of the catastrophic CO freeze-out (about 5000\,au from the center), there is a sharp increase in the gas phase abundance of CH$_3$OH. 
Catastrophic CO freeze-out enhances the gas-phase abundance of methanol for two reasons: (i) CO ice is promptly hydrogenated to form CH$_3$OH (Figure \ref{fig:network_ice}); and (ii) reactive desorption is more efficient from apolar (CO-rich) ice than polar (H$_2$O-rich) ice. Although the reactive desorption efficiency is low \citep[$\lesssim$\,1\%; see][]{Minissale2016,Ligterink2018}, it is sufficient to reproduce the observed CH$_3$OH abundance at the emission peak.  Other O-bearing COMs are also predicted to peak near the CH$_3$OH maximum \citep[see also][]{Vazart2020, Molpeceres2024}.

N-bearing COMs, in particular CH$_2$CHCN, appear to form earlier than O-bearing COMs during the chemical evolution of dense cores. Molecules such as CH$_2$CHCN (and its saturated form CH$_3$CH$_2$CN) can efficiently form on the surface of cold (10\,K) dust grains from adsorbed precursors such as HC$_3$N \citep{Raaphorst2025}, which is abundant in the less dense envelope of molecular cloud cores \citep{Bianchi2023}.

The abundances and spatial distributions of COMs vary among cores. O-bearing COMs have also been detected with IRAM 30m in the prestellar core L183 \citep{Lattanzi2020}, where the abundances of CH$_3$OH and CH$_3$CHO are higher than in L1544, probably because L183 is almost uniformly surrounded by a molecular cloud without direct exposure to the ISRF (unlike L1544). Observations in L1498 and L1517B \citep{JimenezSerra2021, Megias2023} show a general tendency for COMs to be more abundant toward the CH$_3$OH peak than toward the dust peak, similar to L1544. However, the chemical complexity in these two cores is lower than in L1544, as indicated by the non-detection of CH$_3$CHO, CH$_3$OCH$_3$, and CH$_3$OCHO, and the detection of N-bearing COMs such as CH$_2$CHCN and CH$_3$CN. The different abundance behavior between O- and N-bearing species in young starless cores (L1498 and L1517B) compared to more evolved core L1544 may reflect an evolutionary effect, in which N-bearing COMs form first, while O-bearing COMs are produced in ices after the catastrophic CO depletion sets in. \cite{Scibelli2021} detected O-bearing COMs and CH$_2$CHCN toward the starless core L1521E using the ARO 12m telescope (2.5 times larger beam compared to IRAM 30m), suggesting that COM emission is spatially extended. This was later confirmed by \cite{Punanova2022}, who mapped CH$_3$OH emission in seven cores embedded in the L1495 filament in Taurus. 

A statistically significant sample of starless and prestellar cores ($\sim 60$ objects) in the Taurus and Perseus molecular clouds were observed with ARO 12m, finding CH$_3$OH in all sources (100\% of the sample), and CH$_3$CHO in 70\% (50\%) of the objects in Taurus (Perseus). 
%\citep{Scibelli2020,Scibelli2024}. 
A subset of the Perseus cores was also searched for larger COMs with Yebes 40m, and 20\% show CH$_2$CHCN, HCOOCH$_3$, and CH$_3$OCH$_3$ \citep{Scibelli2020, Scibelli2024}. There are problems in reproducing the observed abundances using models \citep[such as][]{Vasyunin2017}, which require the reactive desorption of ice followed by gas-phase reactions. As discussed in \cite{Scibelli2021}, additional mechanisms are required to enhance COM formation and desorption in cold environments, such as non-diffusive chemistry (\S 2.3.2),
%\citep{Jin2020,Borshcheva2025}, 
radiolysis by cosmic-rays \citep{Shingledecker2018}, and cosmic-ray sputtering \citep{Wakelam2021}. However, detailed physical structure of the cores needs to be taken into account for quantitative comparison between observations and models. 
%Formation of COMs in icy dust mantles could also help to explain the presence of COM features in dark cloud ices observed with JWST along the line of sight of background stars \citep{McClure2023} \textcolor{red}{(--COMs features are tentative and not clearly detected towards gackground stars--)} and in the cold envelope of young stellar objects \citep{Yang2022,Chen2024,Nazari2024,Rocha2024,Rocha2025}. 

\section{CHEMISTRY IN GIANT MOLECULAR CLOUDS}
\label{sec:gmcs}
%\textcolor{blue}{(Izaskun + Paola)}
%\textcolor{red}{Yuri modified some sentences (e.g. removed repetition) without changing the contents. Please check. The modifications are not necessarily based on comments from the editor and colleagues. So please check their comments as well. Yuri did not remove any refs. Please reduce them.}

%Review chemistry in Giant Molecular Clouds. IRDCs correspond to the densest parts of Giant Molecular Clouds, where future massive stars and star clusters will form. They are the equivalent of dark dense clouds in nearby low-mass star-forming regions described in Section 2. a-b. We describe similarities and differences between GMC and nearby clouds. Section 3-a covers the chemistry observed in IRDC, while chemistry induced by stellar feedback (PDRs, cloud-cloud collisions, enhanced CR ionization rate) is described in Section 3-b.  The latter is essential in evaluating e.g. the elemental budget in gas and solid phases, and also shows that chemistry can be used as probes of physical processes.

In this Section, we review the chemistry of Giant Molecular Clouds (GMCs), the massive counterparts of nearby low-mass molecular clouds. We first present the chemistry observed in Galactic GMCs, which have higher levels of turbulence (linewidths of a few km s$^{-1}$) and slightly higher gas kinetic temperatures ($T_{\rm kin}$=15-20 K) than in nearby molecular clouds (\S \ref{sec:gmcs-galactic}). We also discuss the chemistry in GMCs whose physical and chemical conditions differ significantly from those in the Galactic disk: (i) the GMCs in the Galactic Center, characterized by high metallicity and energetic processing (\S \ref{galacticcenter}), and (ii) the GMCs in the outer Galaxy and Large and Small Magellanic clouds (LMC and SMC), which are characterized by low metallicity and lower levels of extinction resulting in significant UV photo-processing (\S \ref{sec:low-metallicity}). 

\subsection{GMCs in the Galactic disk }
\label{sec:gmcs-galactic}

Giant Molecular Clouds (GMCs) are massive, %($\ge 10^5 M_{\odot}$), 
cold molecular complexes in the Milky Way’s disk, predominantly composed of H$_2$, but traced observationally by CO. Their masses typically range from 10$^4$ to 10$^6$ M$_\odot$ with average H$_2$ gas densities of $\sim$10$^2$-10$^3$ cm$^{-3}$, although in dense clumps where stars form the densities can reach values $\sim$10$^4$-10$^5$ cm$^{-3}$. Their spatial scales ($\sim 100$ pc) are much larger than that of nearby clouds (a few tens of pc). 
%\textcolor{magenta}{The main difference with nearby clouds is the total mass, as GMCs have masses of about 10$^5$ cm-3 or higher. We should include the mass at the beginning of the paragraph (soon after "massive").}\textcolor{blue}{OK. Added by Yuri.}
%They are much larger than nearby clouds since they present sizes between 10 and 100 pc across. 
GMCs contain filaments, clumps, and cores, spanning scales from tens of parsecs down to $\leq$0.1 pc. GMCs are the nurseries of stars, especially massive ones, and hence, the study of the physical properties, kinematics and chemistry of these clouds has attracted a lot of attention to understand how massive stars form \citep{Beuther2025}. 

A key discriminant for the different scenarios of massive star formation is the initial mass of dense clumps yielding massive stars and star clusters, and their level of fragmentation  \citep{Beuther2025}. Infrared-Dark Clouds (IRDCs) are the densest and coldest regions in GMCs, observed as dark silhouettes against the bright Galactic mid-infrared background \citep[][]{Egan1998,Peretto2009}. IRDCs have masses ranging between 10$^2$ and 10$^5$ M$_\odot$ and they are cold \citep[$T\leq$25 K;][]{Pillai2007,Pillai2012} and dense \citep[$n_{\rm H}\geq$10$^4$-10$^5$ cm$^{-3}$ in their clumps of sizes $\leq$0.1 pc; see e.g.,][]{Butler2009}, representing the massive counterparts of nearby dark clouds. Their column densities ($\geq$10$^{22}$-10$^{24}$ cm$^{-2}$) and mass surface densities ($\geq$0.1 g cm$^{-2}$) are similar to those found in well-known massive star-forming regions \citep[e.g.,][]{Rathborne2006} and hence, they are considered as the initial conditions of massive star and star cluster formation. The chemistry of IRDCs has been studied both at global scales and at small scales, not only to constrain the physical and chemical properties of IRDCs (and of their clumps) and their evolutionary stage, but also to establish how massive stars form in these clouds and whether their chemistry resembles that observed in nearby clouds \citep{Fontani2011,RedaelliBovino2021}. %\textcolor{red}{Modified since the original sentence have two "to understand".}

\subsubsection{Chemical properties of IRDCs}
%\textcolor{blue}{(Izaskun)}
%Use molecules to understand the physics, e.g. extended CO freeze-out, higher densities than low-mass clouds as seen by extended N2H+, average dense core densities, which are counterparts of prestellar cores.

%Tatematsu+08, Fontani+11, Hernandez+11, Sanhueza+12, Henshaw+13, Ragan+14, Jimenez-Serra+14,  Fontani 16, 19, Sokolov+18, Tatematsu+20, Redaelli+21; Sabatini+24, Law+25, refer Beuther+ARAA for physical structure

%The Chemical Clock of High-mass Star-forming Regions: N2H+/CCS by Chen, J. L.; Zhang, J. S.; Ge, J. X.; Wang, Y. X.; Yu, H. Z.; et al. Astronomical Journal

The study of IRDCs provides a baseline for high-mass star formation chemistry, i.e., the ``starting conditions'' before protostars alter the surrounding gas. The low temperature and high density in IRDCs lead to the freeze-out of molecules onto dust grains, especially of CO. Through a comprehensive chemical study of the filamentary IRDC G+035.39-00.33, an IRDC with very low levels of star-formation activity, \citet{Hernandez2012} found widespread CO depletion across this cloud. This contrasts with low-mass prestellar cores, where the CO freeze-out zone is restricted to the central regions of the core (\S \ref{Sec:D-fractionation}). The widespread CO depletion is attributed to the higher overall gas densities in IRDCs than in nearby clouds \citep{Butler2009,Peretto2009}, and is supported by the emission of the $J$=1 $\rightarrow$0 line of N$_2$H$^+$ extended across parsec-scales in IRDCs \citep{Henshaw2014}. N$_2$H$^+$ is not only a high-density tracer but also a sensitive tracer of CO depletion, because CO has a higher proton affinity than N$_2$; when CO is abundant, the proton of N$_2$H$^+$ is transferred to CO. Widespread deuteration is also found as seen in N$_2$D$^+$ $J$=2$\rightarrow$1 \citep{Barnes2016}. The emission from higher-$J$ transitions of these molecules, as well as of N$_2$H$^+$, is well suited to identifying prestellar cores in IRDCs since it is observed to be more compact and to peak toward the densest regions \citep[][]{Zhang2015,Barnes2023b}. Ortho-H$_2$D$^+$ has also been proposed to be an excellent probe of prestellar cores in IRDCs and in high-mass star-forming regions \citep{Miettinen2020,RedaelliBovino2021}.  

Large-scale chemical studies of IRDCs require large amounts of telescope time even with the most sensitive telescopes (e.g., the IRAM 30m telescope). Therefore, IRDC chemistry surveys have mainly focused on the chemical composition of large samples of IRDC clumps \citep[see, e.g.,][]{Sanhueza2012}. When protostellar feedback begins, the chemical composition of the surrounding gas is expected to change, enhancing the abundances of other molecules. In this way, depending on the observed chemistry and evolutionary stage, massive IRDC clumps can be classified as \citep{Sanhueza2012,Rathborne2016}:

\begin{itemize}
    \item Quiescent clumps: These are IR-dark clumps at 70$\mu$m. Their chemistry is characterized  by simple species such as NH$_3$, N$_2$H$^+$, CCH, and CCS \citep{Miettinen2014,Worthen2025}. For clumps at the prestellar stage, CO depletion and high deuteration are common \citep{Hernandez2012,Barnes2016,Sabatini2022}
%,Sabatini2024,Cosentino2025b}. 
    \item Intermediate clumps: These host young protostellar objects and present either enhanced 4.5 $\mu$m emission \citep[also known as Extended Green Objects -- EGOs;][]{Cyganowski2008} or a 24$\mu$m source (but not both). They show regions with enhanced CH$_3$OH and emission from large COMs, and sometimes SiO shocks \citep{Liu2020}. 
    \item Active clumps: These are cluster-forming clumps. They are associated with an EGO and an embedded 24$\mu$m source. They could also show bright 8$\mu$m emission, likely arising from an HII region \citep{Liu2025}. Chemically, they resemble young massive hot cores as feedback heats the gas and ices evaporate \citep{Rathborne2011}.
\end{itemize}
Because chemistry evolves systematically with evolutionary stage, several abundance ratios have been proposed as chemical clocks, such as N$_2$H$^+$/HCO$^+$, N$_2$H$^+$/HNC, and N$_2$H$^+$/CCS, to constrain the evolutionary stage and timescale of these IRDC clumps \citep[see e.g.,][]{Sanhueza2012,Sabatini2021,Chen2025}.  

%\subsection{Chemistry induced by stellar feedback}

%\subsubsection{PDRs}

%Sulfur and COMs in PDRS - PDRs4ALL - Berne+22, Goicoechea+06, 21; Cuadrado+17; Fuente+24 (atomic S in Orion); Fuente+17 (S2H); Berne et al. 2023 (CH3+ in disk) 
%{\bf Refer to Wolfire+22 ARAA for PDR/XDR}

\subsubsection{Chemical probes of the physical processes involved in IRDC formation}

The detection of various gas density tracers across parsec scales enables detailed studies of molecular gas kinematics from the low-density envelope of IRDCs, traced by CO and its isotopologues, to the densest regions traced by N$_2$H$^+$ \citep[][]{Henshaw2014,Jimenezserra2014b,Barnes2021}. These studies have enhanced our understanding of the internal physical structure of IRDCs -- with multiple filamentary structures physically connected in space and velocity \citep{Henshaw2014} -- and of the global gravitational collapse these clouds are undergoing \citep{Peretto2013,Jimenezserra2014b,Rathborne2016}. When observed at higher angular resolution with interferometers, the observed filaments split further into a collection of elongated structures, analogous to the so-called fibers in low-mass clouds \citep[][]{Hacar2013}, with widths narrower than 0.1$\,$pc \citep{Henshaw2017}.

\begin{figure}
\centering
\includegraphics[angle=0,width=0.22\textwidth]{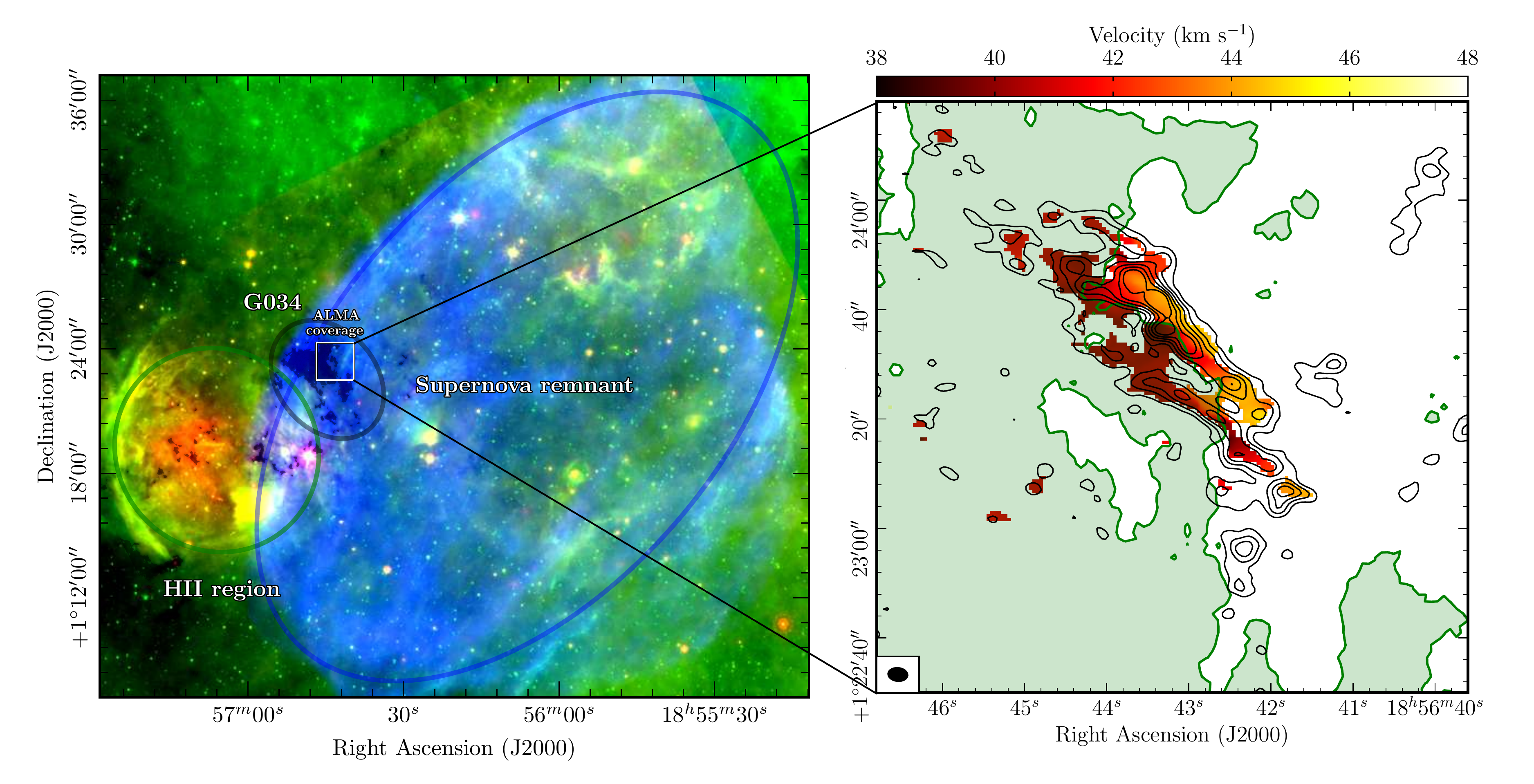}
\caption{Left panel: Three-color image of G034.77-00.55 showing its location (black circle) between SNR W44 (blue circle) and the HII region G034.77-00.55 (green circle). Red is 24$\mu$m emission \citep[Spitzer MIPSGAL;][]{Carey2009}, green is 8$\mu$m emission \citep[Spitzer GLIMPSE;][]{Churchwell2009}, and blue shows a combined JVLA+GBT 21 cm continuum map \citep[THOR survey;][]{Beuther2016}. The white square indicates the ALMA SiO mosaic area and the gray shadow corresponds to the A$_{\rm V}$ = 20 mag level within the IRDC. Right panel: SiO integrated intensity map (black contours) toward G034.77-00.55 superimposed on its moment 1 velocity map (red scale). Contours are from 3$\sigma$ ($\sigma$ = 0.016 Jy km s$^{-1}$) by steps of 3$\sigma$. Green contour and shadow indicates the A$_{\rm V}$ = 20 mag dense material in the IRDC \citep{Kainulainen2013}. Figure based on material published by \citet{Cosentino2019}.}
\label{fig:gmcs}
\end{figure}

Sensitive mapping toward IRDCs has also revealed widespread emission of shock tracers such as SiO and CH$_3$OH \citep{Jimenezserra2010,Cosentino2018,Liu2025}. Two velocity components are identified in these shock tracers: (i) a broad and compact component (linewidths of tens of km s$^{-1}$), red-/blue-shifted relative to the ambient cloud velocity; and (ii) a narrow and spatially-extended component (linewidths $\leq$3 km s$^{-1}$) arising from the ambient gas. While the former is clearly associated with molecular outflows in star-forming clumps, the origin of the latter remains uncertain. It could be associated with a widespread population of low-mass protostars that remain unresolved in the single-dish observations, or it could be a fossil record of a cloud-cloud collision responsible for the formation of the IRDC itself \citep{Jimenezserra2010,Cosentino2018}. Cloud-cloud collisions have been invoked as the dominant mechanism in the formation of massive stars not only in well-known high-mass star-forming regions in the Milky Way, but also in nearby interacting galaxies, including the Magellanic system and the Antennae Galaxies \citep[see review by][]{Fukui2021}. The cloud-cloud collision scenario has been confirmed toward IRDC G034.77-00.55 (at 2.9 kpc), where ALMA SiO $J$=2$\rightarrow$1 observations resolved, for the first time, the internal structure of a large-scale ($\geq$46000 au or $\geq$0.22 pc), highly-magnetized ($\sim$0.9 mG) CJ-type MHD shock, induced by the expansion of the nearby SNR W44 \citep[Figure \ref{fig:gmcs};][]{Cosentino2019}. Similar configurations have been found toward other IRDCs located in the vicinity of SNRs or HII regions \citep{Liu2025}. In G034.77-00.55, the shock compresses the gas in the interaction layer (the dense gas ridge in Figure \ref{fig:gmcs}, right panel), forming dense cores with enhanced D/H ratios \citep{Cosentino2023,Cosentino2025a}, which could be the population of future star-forming cores triggered by the expansion of the SNR.  

%\subsection{Cloud-cloud collision chemistry}

%Widespread SiO, D/H enhancement as probes of prestellar clusters induced by cloud-cloud collisions; (induced by supernovae) Jimenez-Serra+10; Cosentino+18, 19, 20, 23, Barnes+16

%\subsubsection{Enhanced CR ionization rate}
%Yamagishi+23
  
%\subsection{CHEMISTRY IN GIANT MOLECULAR CLOUDS IN EXTREME ENVIRONMENTS}
%\textcolor{blue}{(Izaskun+Yuri; 6 pages)}

\subsection{GMCs in the Galactic Center}
\label{galacticcenter}

The Galactic Center spans the inner 600 pc of the Milky Way and contains 5\% of the Galaxy's total molecular gas reservoir \citep[$\sim$3-5$\times$10$^7$ M$_{\odot}$;][]{Henshaw2023}. The majority of this gas is concentrated in a chain of Giant Molecular Clouds, the so-called Central Molecular Zone (CMZ), which represents nearly 80\% of the dense gas in the Milky Way. Despite this, its star formation rate (SFR) is only $\sim$0.1 M$_\odot$ yr$^{-1}$, well below expectations from star-formation vs. molecular-gas relations \citep{Longmore2013}. The physical properties of CMZ clouds differ markedly from those of typical Galactic disk GMCs: the average H$_2$ gas densities ($\sim$10$^{4}$ cm$^{-3}$) and kinetic temperatures ($\sim$70-150 K) are significantly higher \citep[][]{Guesten1983,Krieger2017}. 
In contrast, dust temperatures are low \citep[$\leq$20-25 K;][]{Battersby2025}, indicating that gas and dust are thermally decoupled. 

\subsubsection{Chemistry in the Central Molecular Zone}
CMZ GMCs are located in a high-metallicity environment \citep[by a factor of $\sim$2 with respect to Solar;][]{Najarro2009} and their gas and dust are affected by energetic processing induced by: (i) widespread low-velocity shocks ($v_s \sim$ 10-50 km s$^{-1}$) produced by cloud shearing and compression by the central bar potential \citep[][]{Martin-pintado1997,Santamaria2021}; (ii) intense UV background radiation produced by the central massive stellar clusters \citep[$G_0 \sim$ 10$^3$-10$^4$ in Habing units,][]{Goicoechea2004}; (iii) X-ray flares coming from the central black hole in Sgr A$^*$ \citep[][]{Terrier2018}; (iv) intense magnetic fields \citep{Pillai2015}; and (v) enhanced cosmic-ray ionization rates \citep[$\zeta_{\rm H2} \sim$10$^{-15}$-10$^{-14}$ s$^{-1}$;][]{lepetit2016,Oka2019}. Among these processes, low-velocity shocks and enhanced cosmic-ray ionization rates seem to be the mechanisms dominating the chemistry in GMCs in the Galactic Center \citep[][]{Harada2015,Zeng2018}.

While the basic chemical processes should be common to those in our local molecular clouds, the energetic processing of the gas and dust in CMZ GMCs induces a peculiar chemistry on large scales, especially for COMs, which is also found in starburst galactic nuclei \citep[as in NGC253;][]{Martin2021}. Besides the compact COM emission associated with hot cores and molecular outflows in Galactic Center massive star-forming regions \citep[see e.g.,][]{Jorgensen2020,Busch2024}, which is outside the scope of this review, COM emission in the CMZ is widespread across several tens of parsecs \citep[e.g.,][]{Martin-Pintado2001,Zheng2024}. Most of these clouds show no signs of massive star formation, indicating that the widespread COM emission is not associated with stellar feedback \citep{Zeng2020}. Observations toward tens of GMCs across the CMZ revealed a common COM chemical composition \citep{Requena2006}. Our current understanding is that COMs in the Galactic Center mainly form on grain surfaces\footnote{Exceptions exist, e.g. vinyl cyanide (C$_2$H$_3$CN), which likely form in the gas \citep{Zeng2018}.} and are injected into the gas phase through grain sputtering induced by recurrent, low-velocity shocks \citep{Requena2006,Zeng2018}. Frequent ($\leq$10$^5$ yrs) shocks with velocities $> 6$ km s$^{-1}$ are needed to explain the uniform COM/CH$_3$OH abundance ratios observed in Galactic Center GMCs \citep{Requena2006}. Although COMs could freeze back onto grains at the low $T_{\rm dust}$ ($\leq$25 K), the recurring shocks would maintain COMs gas-phase abundances high. 

Due to their high COM abundances, Galactic Center GMCs have been very prolific in the discovery of new interstellar COMs \citep[see][and references therein]{McGuire2016,Jorgensen2020}. Many COMs have been detected in absorption against the bright Sgr B2 (N) and (M) hot cores as COMs are present in the colder, extended molecular shell around the Sgr B2 star-forming clusters. COM emission is indeed extended across the Sgr B2 cloud \citep{Li2017}, representing one of the largest COM repositories in the Galaxy.

\subsubsection{A prototypical quiescent Galactic Center cloud: G+0.693}
\label{sec:g0693}

%Due to their high COM abundances, the GMCs in the Galactic Center have been very prolific in the discovery of new COMs in the ISM (see e.g. the detections of glycolaldehyde, ethylene glycol, acetic acid or propylene oxide; Hollis et al. 2000, 2002; Remijan et al. 2002; McGuire et al. 2016). These detections were achieved in absorption against the bright continuum of the Sgr B2 (N) and (M) hot cores. The COM emission appears in absorption because it arises from the cooler molecular envelope of the Sgr B2 molecular cloud (Goicoechea et al. 2004), which presents lower H$_2$ volume densities than those of the Sgr B2 (N) and (M) star-forming regions. 

%\textcolor{magenta}{Paola: can we say upfront, maybe even in this title, that this is a starless GMC or a giant molecular dark cloud - to show that this can be considered a dark cloud, the theme of the review?} \textcolor{blue}{I have added the word "quiescent"}
The quiescent cloud G+0.693-0.027, hereafter G+0.693, located within the Sgr B2 molecular complex, stands out for its chemical complexity \citep[][]{Requena2006,WidicusWeaver2017,Zeng2018}. It lies $\sim$100$"$ northeast of Sgr B2 (M) and appears as the peak of HNCO \citep[][]{Henshaw2016} and CH$_3$OH Class I 36 GHz maser emission in Sgr B2 cloud, indicating shocked gas \citep{Liechti1996,JimenezSerra2025a}.
%HNCO and CH$_3$OH Class I masers are known to trace shocks and hence, since G+0.693 is associated with their peak emission, a strong shock interaction is likely undergoing in the region. 
Interferometric observations revealed two molecular flows that seem to be colliding at this location, causing a large-scale, low-velocity shock that enhances the gas-phase molecular abundances \citep{Zeng2020,Colzi2024}.
%Since no signs of massive star formation are found in this cloud, the chemical richness of G+0.693 is attributed to a cloud-cloud collision (Zeng et al. 2020). 

A wide variety of molecular species spanning all chemical families has been detected toward G+0.693: O-, N-, S-, and P-bearing species, carbon chains, and even cyclic molecules. %\citep[including the first S-bearing cyclic hydrocarbon;][]{araki2025}. 
A list of the molecules detected in G+0.693 is given in Supplemental Table 4. 
Compared to the Sgr B2 (N2) hot molecular core, G+0.693 is rich in unsaturated species and in N-bearing species, such as methyl amine (CH$_3$NH$_2$) and cyanamide \citep[NH$_2$CN;][]{JimenezSerra2025c}, which is likely due to the combination of shock chemistry and an enhanced cosmic-ray ionization rate (by a factor of $\geq$100; see below). 
%To date, more than 25 new molecules have been detected for the first time toward this cloud, the majority of which are COMs of prebiotic interest (see recent review by Jimenez-Serra et al. 2025a; see also Section \ref{prebiotic}). These detections have been possible thanks to broadband, ultrasensitive spectral surveys carried out at 7mm, 3mm, 2mm and 1mm with the Yebes 40m and IRAM 30m radiotelescopes. Like for the case of TMC-1, the ultrasensitive 7mm Q-band observations have played a critical role in these detections. 
Molecular emission toward G+0.693 is largely sub-thermally excited, with typical excitation temperatures of $T_{\rm ex}$$\leq$15 K. This reflects the relatively low H$_2$ volume densities (a few 10$^4$ cm$^{-3}$), which are below the critical densities of COMs. As a result, COM spectra show their peak intensity at lower frequencies, less confused by emission from simpler molecules. Despite the broad linewidths (FWHM$\sim$20 km s$^{-1}$), the levels of line blending and line confusion are low, making G+0.693 an excellent laboratory for discovering new interstellar molecules. 

%\subsubsection{Chemistry in G+0.693}
%\label{prebiotic}

Nearly 30 new molecular species have been detected toward this cloud to date. These discoveries were enabled by broadband, ultrasensitive spectral surveys at 7mm, 3mm, 2mm and 1mm conducted with the Yebes 40m, \textcolor{blue}{IRAM 30m and APEX} radio telescopes. 
%\textcolor{red}{Figure \ref{fig:prebiotic}} summarizes the prebiotic molecules detected toward G+0.693.  
As in the case of TMC-1, ultrasensitive 7mm Q-band observations obtained with Yebes 40m telescope have played a critical role in these discoveries. Several S-bearing and P-bearing molecules have been detected, including mono-thioformic acid \citep[HC(O)SH;][]{Rodriguez2021a}, PO$^{+}$ \citep[][]{Rivilla2022b}, HOCS$^{+}$ \citep[][]{Sanznovo2024Aa}, MgS and NaS \citep[][]{Reymontejo2024}, and 2,5-cyclohexadien-1-thione \citep[a structural isomer of thiophenol, c-C$_6$H$_6$S;][]{Araki2025}. Chemical modelling shows that a cosmic-ray ionization rate $\geq$100 times higher than the standard value is required to reproduce abundances of ions such as PO$^{+}$ and HOCS$^{+}$ \citep{Rivilla2022b,Sanznovo2024Aa}. From these models, atomic sulfur seems to be undepleted \citep{Sanznovo2024Aa}, while 99\% of P remains in grains \citep{Rivilla2022b}. 

Interestingly, the majority of the detected species are COMs of prebiotic interest \citep[see recent review by][]{JimenezSerra2025a}.
They include precursors of ribonucleotides, lipids, and amino acids, which are essential for the synthesis of RNA/DNA, proteins and cell membranes, the three pillars of life \citep[see e.g.][]{patel2015}. Examples are ethanolamine \citep[NH$_2$CH$_2$CH$_2$OH;][]{rivilla2021a}, n-propanol \citep[n-C$_3$H$_7$OH;][]{Jimenezserra2022}, and vinyl amine and ethyl amine \citep[C$_2$H$_3$NH$_2$ and C$_2$H$_5$NH$_2$, respectively;][]{Zeng2021}. 
The discovery of these prebiotic molecules has triggered the interest in the astrochemistry community as little is known about their formation and survivability under interstellar conditions. Quantum chemical calculations have been performed to investigate the formation processes of species such as urea (NH$_2$CONH$_2$), HC(O)SH, or H$_2$CNCN, both on grain surfaces and in the gas phase \citep{Slate2020,Molpeceres2021,Fortenberry2024}. Laboratory experiments exploring their formation and resilience in interstellar ice analogs exposed to UV and cosmic-ray irradiation have also been performed \citep[e.g.][]{Mate2021,Herrero2022,Zhang2023},
%Biancalani2024,Suhasaria2025,Quitian2025,Mate2025}, 
revealing different levels of molecular stability. A clear example is the comparison between urea and 2-aminooxazole, two proposed precursors of ribonucletides \citep{patel2015, Menor2020}, 
where urea is found to be more resistant to energetic processing in ices \citep{Herrero2022}. 

The detection of different isomers from the same molecular species has also raised interest in understanding isomerization processes in the ISM \citep{Garcia2021,Garcia2022,Molpeceres2022}. Studies on the stability and detectability of isomers within the same chemical family, carried out both theoretically and experimentally, have provided insights into which molecular species are likely to be detectable in the ISM \citep{Singh2022,Watrous2024,Noriega2025}. %\citep{Singh2022,Marks2023,Watrous2024,Noriega2025}.
All these studies suggest that the basic ingredients needed for early replicative and metabolic processes could be synthesized in the ISM \citep{Zhang2024,Mcanally2025}.

\subsection{GMCs in the outer Galaxy and the Magellanic Clouds}
\label{sec:low-metallicity}
%Title change sujested by Shimonishi

%Nishimura+17, Fontani+24, Shimonishi+10,16,18,20,21,23, Sewilo+18, 22a,22b

Another factor that may influence the chemistry in molecular clouds is metallicity. Low-metallicity environments are found in LMC and SMC, whose metallicities are lower than the Solar value by factors of $\sim$2 and $\sim$4, respectively \citep{DeCia2024}. Low metallicity affects not only molecular abundances through an overall scaling, but also molecular formation processes, owing to the reduced dust abundance and consequent decrease in UV shielding and total grain-surface area available for reactions.
In evaluating molecular abundances in these regions, hydrogen column densities $N_{\rm H}$ are estimated from dust continuum emission by assuming a gas-to-dust mass ratio that scales with metallicity. As a result, molecular abundances relative to hydrogen carry systematic uncertainties associated with the adopted gas-to-dust ratio, whereas abundance ratios between molecules (e.g., HCN/CCH) are largely unaffected.

At molecular cloud scales ($>$10 pc), single-dish observations reveal distinct behaviours among different molecules \citep[][]{Shimonishi2024}: while HCN shows a decreasing trend with a decreasing metallicity, other species such as CCH and HCO$^+$ do not show any variation. 
Importantly, this trend appears to be independent of the star-formation activity within the observed clouds, indicating that it is not affected by star formation \citep{Nishimura2016a}.
The decrease in HCN abundance clearly follows the reduced elemental nitrogen abundance in low-metallicity environments. In contrast, CCH is a well-known PDR tracer \citep{Jansen1995}, and HCO$^+$ also traces dense gas in PDRs in Galactic disk molecular clouds \citep{Trevino2016}; their relative enhancement in low-metallicity environments is therefore attributed to the increased volume of PDR gas. 

For more complex molecules such as CH$_3$OH, low gas-phase abundances have been found in low-metallicity nearby galaxies such as LMC and IC10  \citep{Nishimura2016a,Nishimura2016b}. 
%This trend is also confirmed in the ice, as shown by IR ice absorption spectra obtained against bright high-mass YSOs in the LMC and SMC \citep{Shimonishi2016}. 
This trend is further confirmed by ice observations toward bright high-mass YSOs in LMC: the abundance of CH$_3$OH ice relative to H$_2$O ice is low, whereas CO$_2$ ice abundance is comparable or higher than that in Galactic molecular clouds \citep{Shimonishi2016}. 
Warm dust temperatures in LMC and SMC suppress CO hydrogenation, while thermal diffusion of CO and OH could promote CO$_2$ formation \citep{Acharyya2015,Shimonishi2016, Furuya2024_Ebpop}.
%However, note that the decreasing abundances for species formed on dust grains such as CH$_3$OH, do not depend exclusively on the metallicity \citep{Shimonishi2020,Shimonishi2024} but also on the initial physical conditions of the source \citep[e.g. dust temperature, degree of shielding, local radiation field strength;][]{Acharyya2015,Acharyya2016,Acharyya2018,Vermarien2025}. 
Recent JWST observations reported the detection of various COM ices larger than CH$_3$OH in the envelope of an embedded protostar in LMC \citep{Sweilo2025}. High-resolution ALMA observations towards high-mass YSOs (i.e. hot cores) indeed revealed the presence of gaseous large COMs, such as dimetyl ether (CH$_3$OCH$_3$) and methyl formate (CH$_3$OCHO), as well as N-bearing COMs such as methyl cyanide (CH$_3$CN) \citep[e.g.][]{Sewilo2022, Golshan2024, Shimonishi2025}. Further studies, e.g., combined ice and gas observations toward the same objects, are required to clarify the relationship between the COM abundances in gas and ice.
%Interestingly, high-resolution ALMA observations toward high-mass YSOs (i.e., hot cores) reveal that the low CH$_3$OH abundance does not prevent the formation of larger COMs such as dimethyl ether (CH$_3$OCH$_3$), methyl formate (CH$_3$OCHO) or methyl cyanide (CH$_3$CN) \citep{Shimonishi2018,Shimonishi2020,Sewilo2018,Sewilo2022,Broadmeadow2025}. 

The Outer Galaxy also has metallicity values lower by a factor of $2-10$ than the Solar value \citep[e.g.][]{Wenger2019} and has become the target of active chemical studies in recent years.
% \citep{Bernal2021}. 
The CHEMOUT (CHEMical complexity in star-forming regions of the OUTer Galaxy) project characterized the chemical composition of 35 sources at Galactocentric distances of $9- 24$ kpc  \citep{Fontani2022a,Fontani2022b,Colzi2022b}. COMs such as CH$_3$OH are detected in $\sim$40\% of the sources, while CH$_3$CCH is found in $\sim$17\%. The detected emission, with beam size $\sim 2.3$ pc at a distance of $\sim 10$ kpc, indicates that the COM emission likely arises from the colder envelopes around protostellar objects, at temperatures below the CH$_3$OH sublimation threshold, rather than from hot cores. The derived CH$_3$OH abundances, (0.6-7.4)$\times$10$^{-9}$, are consistent with those in local and inner Galaxy star-forming regions observed at similar spatial scales \citep{Fontani2022b}. This trend is confirmed by ALMA for the star-forming cloud WB89-671 at Galactocentric radius $\sim 23$ kpc \citep{Fontani2024b}.
\citet{Shimonishi2021}, on the other hand, derived the COM/CH$_3$OH column density ratio in the ice-sublimating region of the Outer Galaxy hot core WB89-789 SMM1. The value is similar to that in the inner Galaxy hot core NGC7192 FIRS2, implying that large COM formation efficiency does not strongly depend on metallicity. Therefore, although CH$_3$OH formation may be inhibited in low-metallicity environments, once formed it leads to the production of larger COMs. This idea is further explored in Section \ref{sec:comparison-metallicity}.% where we will compare the COM/CH$_3$OH abundance ratios across different Galactic environments (Solar neighbourhood, inner Galaxy, outer Galaxy) and low-metallicity external galaxies.

\section{COMPARISON BETWEEN MOLECULAR CLOUD ENVIRONMENTS}
\label{sec:comparison}
%\textcolor{blue}{(It may be moved to the summary and outlook section)}

\subsection{Comparison between TMC-1 CP and G+0.693}
\label{sec:comparison-clouds}

\begin{figure}
\centering
\includegraphics[angle=0,width=0.27\textwidth]{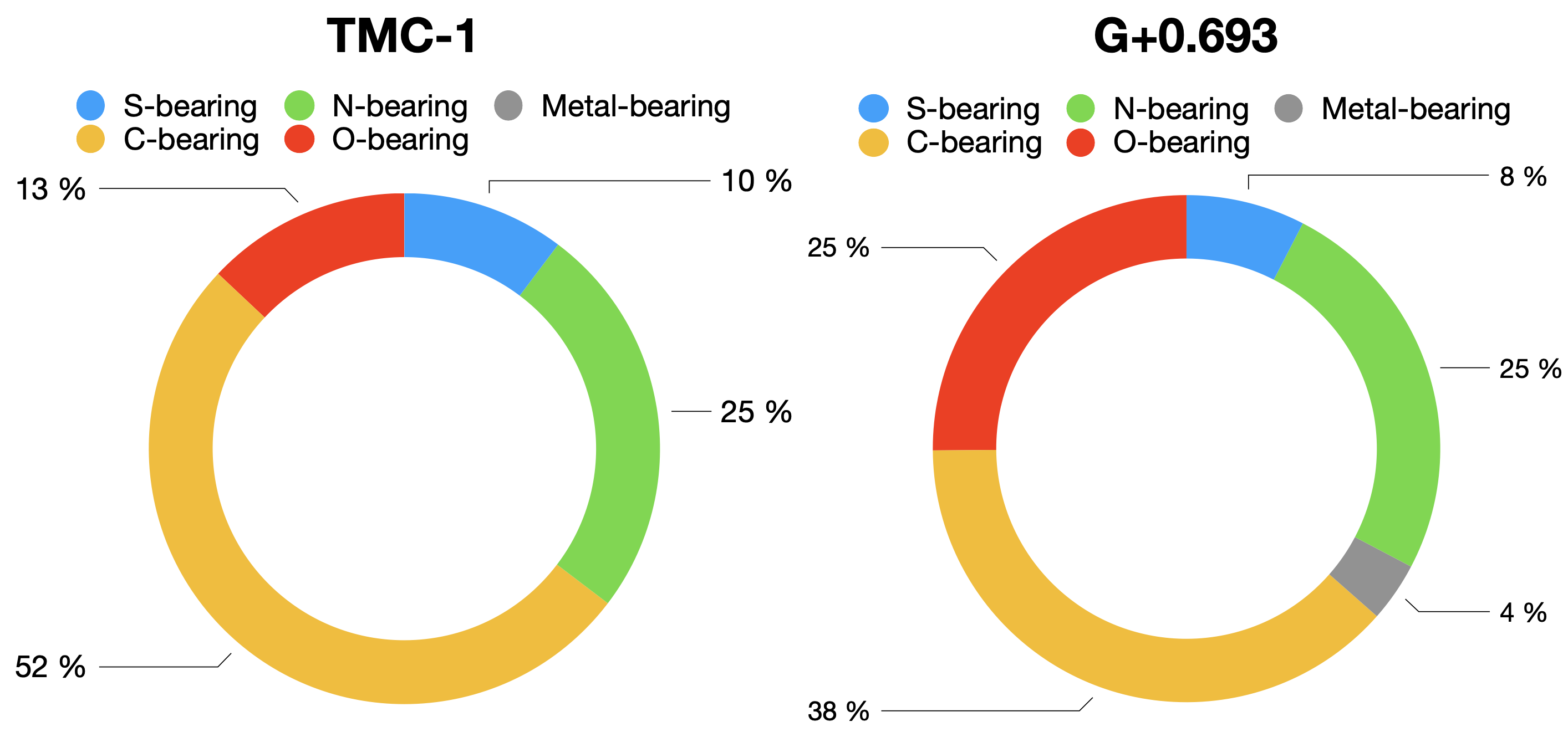}
\caption{Pie-charts with the distribution of molecular species detected toward TMC-1 CP and G+0.693 across different chemical families. The fractions of C-, N-, O-, S-, and metal-bearing species have been calculated by counting all species in Supplemental Tables 2 and 4, with at least one C, N, O, S, and metal atom (Si, Mg, Na, Ca, and P), respectively.}
\label{fig:statistics}
\end{figure}

The detailed analysis of spectroscopic surveys toward TMC-1 CP and G+0.693 allows a comparison of their chemical inventories. Figure \ref{fig:statistics} presents a statistical comparison of the two clouds for different chemical families. %The fractions are calculated by counting all species with at least one C, N, O, S and metal atom. 
%\textcolor{red}{can we move the following sentences to the Table (or Fig) caption?}
%Figure \ref{fig:statistics} shows that 
The production of C-bearing molecules in TMC-1 CP is efficient, consistent with its early evolutionary stage and the high abundance of atomic C not converted to CO yet \citep{Maezawa1999}. In contrast, in G+0.693, the production of O-bearing molecules is favoured with respect to TMC-1 CP. The fraction of O-bearing molecules in G+0.693 is a factor of $\sim 1.8$ larger (25\% vs. 13\%) than that in TMC-1 CP, possibly because more oxygen (either in atomic form or in radicals such as OH) is available to react on grain surfaces in G+0.693. Galactic Center GMCs are known to have an enhanced cosmic-ray ionization rate (\S \ref{sec:g0693}), which heavily processes the ices. Enhanced ionization rate results into higher abundance of atomic H (\S \ref{sec:D-enrichment}), which enhances the hydrogen addition/abstraction reaction cycle in the ices. For N- and S-bearing species, the derived fractions are similar in both clouds. No metal-bearing molecules are detected in TMC-1 CP, whereas 4\% of the species reported in G+0.693 contain elements heavier than O, possibly due to dust sputtering in the large-scale shock in G+0.693 \citep{Zeng2020}.  

%\begin{table}[h]
%\tabcolsep5.5pt
%\caption{Statistical comparison between TMC-1 CP and G+0.693 for the different chemical families}
%\label{tab:comp-clouds}
%\begin{center}
%\begin{tabular}{@{}l|c|c|c|c|c|c@{}}
%\hline
%& Total number of & \multicolumn{5}{c}{Fraction (\%)} \\
%& species detected & C-bearing & N-bearing & S-bearing & O-bearing & Metal-bearing \\
%\hline
%TMC-1 CP & 188 & 26 & 34 & 19 & 22 & 0 \\
%G+0.693 & 99 & 4 & 32 & 13 & 42 & 8 \\
%\hline
%\end{tabular}
%\end{center}
%\begin{tabnote}
%$^{\rm a}$ For C-bearing molecules, only molecules with C and H atoms are considered. For N-bearing species, molecules with C, H and N are included. For S-bearing species, any molecule with a S atom is considered except those with elements heavier than O (e.g. SiS, NaS, MgS). For O-bearing molecules, any species with C, H, O are considered (they may contain N as well in some cases, as e.g. HNCO). For metal-bearing molecules, we consider those species that contain an atom heavier than O (e.g. PO). 
%\end{tabnote}
%\end{table}

\subsection{Exploring chemical variations as a function of metallicity}
\label{sec:comparison-metallicity}

%- Comparison between the abundance ratios wrt CH3OH for TMC1, prestellar cores, and G+0.693

\begin{figure}
\centering
\includegraphics[angle=0,width=0.5\textwidth]{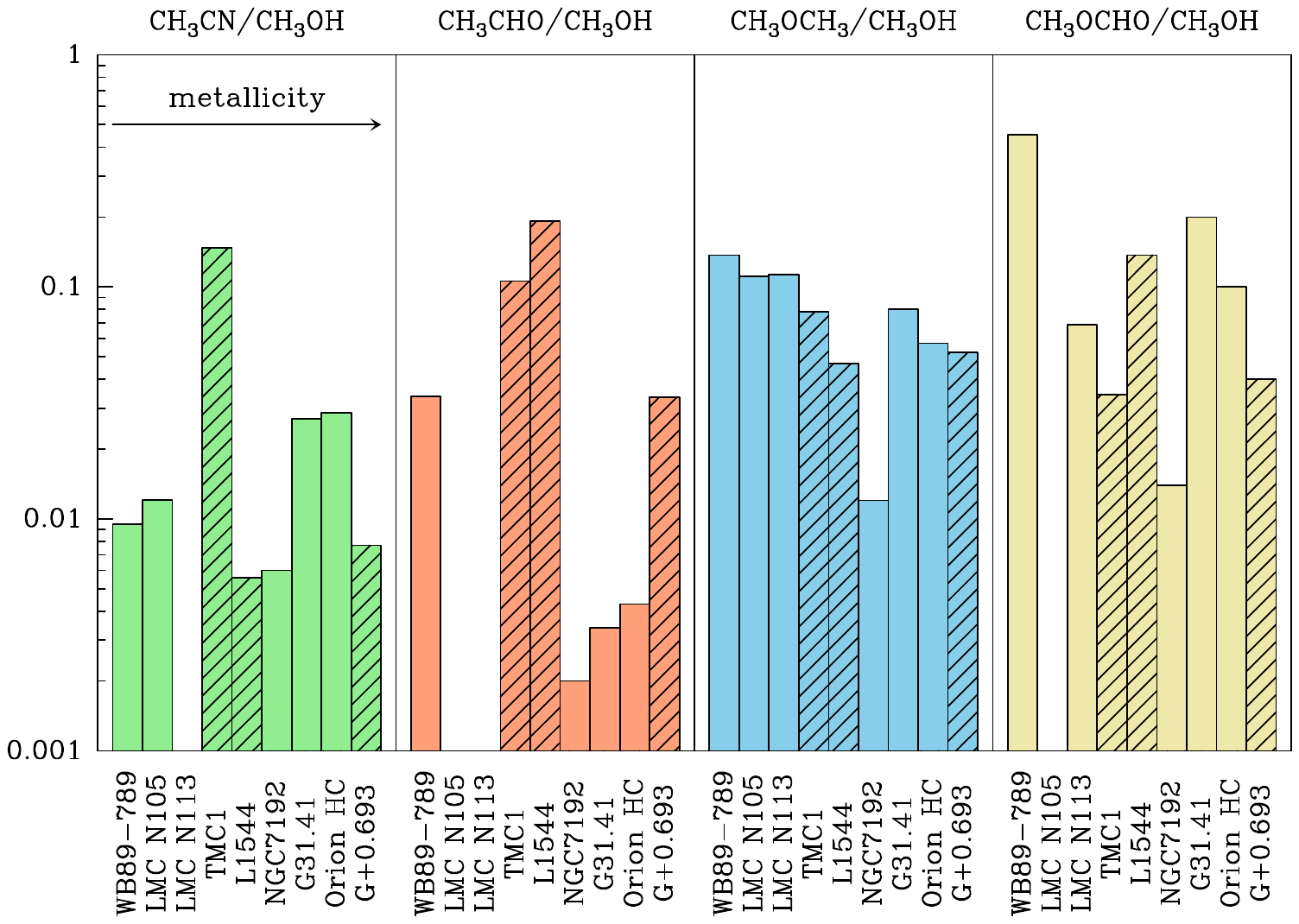}
\caption{Column density ratios for several COMs with respect to CH$_3$OH measured toward sources with different metallicity values. Metallicity increases from left to right as: 0.25 for WB89-789; 0.5 for the LMC hot cores N105 2-A and N113 A1; 1 for TMC1 CP, L1544 CH$_3$OH peak, NGC7192 FIRS2, G31.41+0.31, and the Orion HC; and 2 for G+0.693. Data are taken from \citet{Shimonishi2021} for WB89-789; \citet{Sewilo2022} for LMC N105 2-A; \citet{Sewilo2018} for LMC N113 A1; \citet{Agundez2013}, \citet{Agundez2021c}, \citet{Cabezas2021a,Cabezas2025b} for TMC-1 CP; \citet{JimenezSerra2016} and \citet{JimenezSerra2021} for L1544 CH$_3$OH peak; \citet{Fuente2014} for NGC7192 FIRS2; \citet{Mininni2020} and \citet{Mininni2023} for G31.41+0.31; \citet{Sutton1995} and \citet{Ikeda2001} for the Orion Hot Core; and \citet{Rodriguez2021a}, \citet{Zeng2018} \citet{SanzNovo2022}, \citet{SanzNovo2025a}, \citet{SanzNovo2025b} for G+0.693.
Hatched histograms indicate the objects that are $"$starless$"$.%\textcolor{red}{How about putting shades for "starless" objects to distinguish from hot cores?}
}
\label{fig:metallicity}
\end{figure}

As described in \S \ref{sec:low-metallicity}, observations and astrochemical simulations suggest that the abundance of COMs in low-metallicity environments do not simply scale with metallicity. Instead, other physical parameters such as the initial dust temperature during the prestellar phase could play a key role in COM formation. To test this idea, in Figure \ref{fig:metallicity} we compare the COM/CH$_3$OH column density ratios across different environments characterized by distinct metallicity values ranging from 0.25 for the Outer Galaxy hot core WB89-789 to 2 in the Galactic Center  for G+0.693. Figure \ref{fig:metallicity} shows that there is no trend for the COM/CH$_3$OH column density ratios as a function of metallicity. This was already noted by \citet{Sewilo2022} and \citet{Shimonishi2024} using a sample of Galactic hot cores. However, Figure \ref{fig:metallicity} shows that this behaviour can be extended to quiescent molecular dark clouds, both in our Solar neighbourhood (TMC-1 and L1544) and in the Galactic Center (G+0.693).

We note that the dominant desorption mechanisms of COMs differ among the objects shown in Figure \ref{fig:metallicity}. While COMs are thermally sublimated in hot cores, including those in distant low-metallicity environments, non-thermal desorption (e.g., chemical reactive desorption) is at play in TMC-1 and L1544. We assume that desorption efficiencies are broadly similar among COMs, supported by the absence of systematic differences in COM/CH$_3$OH ratios between hot cores and starless objects.

Figure \ref{fig:metallicity} shows that the majority of the ratios lie within a factor of 10, which suggests that the COM production efficiency is not significantly affected by metallicity. However, a few exceptions exist; the anomalously high CH$_3$CN/CH$_3$OH ratio in TMC-1 CP and the low CH$_3$CHO/CH$_3$OH ratios in intermediate- and high-mass hot cores (NGC7192 FIRS2, G31.41+0.31 and Orion HC hot core). Note that the latter cannot be due to the use of interferometric data for the hot cores: the values of Orion HC were obtained using single-dish observations but yet probe the hot component \citep{Sutton1995,Ikeda2001}. 
The discrepancies observed for CH$_3$CN and CH$_3$CHO could be due to efficient processing in the gas phase at the high temperatures in hot cores \citep{Blazquez2020}.

\section{SUMMARY AND OUTLOOK}
%\textcolor{blue}{(1 page)}

\label{Sect:Inheritance} 

Molecular clouds form through the accumulation of diffuse interstellar gas, driven by interstellar shock waves. Once H$_2$ forms via grain-surface reactions, gas-phase reactions produce unsaturated hydrocarbons, which are subsequently converted into CO, while fully hydrogenated molecules such as H$_2$O and CH$_3$OH predominantly form as ices on dust grains. 
Recent deep line surveys toward TMC-1 CP and GMCs have greatly expanded the chemical inventory of the ISM. 
%The detection of numerous isomers and COMs is driving active theoretical and laboratory efforts to elucidate their formation and destruction pathways.
Detections of aromatic molecules, in particular, have fueled discussion on bottom-up versus top-down scenarios. In the latter, aromatic species may originate as fragments of large PAHs and carbonaceous grains, linking molecular cloud chemistry to the large-scale circulation of the ISM in the Galaxy. 
Ice observations and mapping indicate that the major ice compositions are established in molecular clouds, while CO freeze-out in high-density regions enhances the formation of CH$_3$OH, a key parent species of many COMs, and CO$_2$.
Although COM features are only tentatively detected in ices toward background stars, the detection of gaseous COMs in prestellar cores suggests that their formation is initiated prior to star formation. 
%The detection of COMs in the gas phase along the bipolar outflows driven by protostars, at thousands of au from the central source, further supports their formation and storage in ices, since in these outer regions of dense cores it would take much longer than the shock timescale to synthesize COMs in the gas phase starting from simpler species \citep{Arce2008,Yamaguchi2012,Burkhardt2016,Lefloch2017}, with the exception of formamide \citep[NH$_2$CHO;][]{LopezSepulcre2024} and possibly acetaldehyde \citep[CH$_3$CHO;][]{Codella2020}.
%Interestingly, \cite{Chahine2024} measured the D$_2$CO/HDCO abundance ratio along the outflow cavities of the Class 0 source NGC\,1333 IRAS\,4A, finding evidence of a decreasing D-fraction with distance from the central protostar. This trend is consistent with the radial decrease in D-fraction observed in prestellar cores \citep{Redaelli2019} and predicted by chemical modeling \citep{Aikawa2012,Sipila2015}. Thus, \cite{Chahine2024} provide strong evidence that prestellar ices are released in the gas phase via sputtering along outflows, reinforcing the hypothesis that many COMs observed in these regions are directly injected into the gas by the shock processing of dust. 

Cores with central densities $n({\rm H}_2)\ge 10^5$ cm$^{-3}$ tend to be thermally supercritical and are classified as prestellar cores. Significant freeze-out and deuteration proceed in the central regions of these cores. Chemical differentiation of ambient gas around cores suggests that the composition of infalling material, and thus of the cores themselves, depend on the location within the filaments, which regulates, for example, exposure to the ISRF.
Core chemistry also depends on evolutionary stage; compared with L1544, young starless cores harbor fewer O-bearing COMs but more N-bearing COMs, which may formed at earlier stages.

Chemistry in IRDCs, the densest and coldest regions in GMCs, is characterized by extended CO freeze-out due to their higher gas densities compared to nearby clouds. Widespread emission of shock tracers, such as SiO, is also observed. Its narrow velocity component could be a fossil record of a cloud-cloud collision responsible for the formation of the IRDC. Observations of IRDC clumps show that characteristic molecular species vary with evolutionary stage; for example, carbon chains are observed in clumps at the prestellar stage.
GMCs in CMZ exhibit peculiar chemistry, likely driven by a high cosmic-ray ionization rate and widespread low-velocity shocks induced by cloud shearing and compression by the central bar potential. A variety of prebiotic molecules, including precursors of RNA and DNA, has been detected in the quiescent GMC G+0.693 in the CMZ. 
In GMCs in LMC and SMC, low metallicity affects molecular formation processes through reduction of dust abundance and UV shielding. The relatively low abundance of CH$_3$OH ice and the high abundance of CO$_2$ ice may result from slightly high dust temperatures compared with those in Galactic molecular clouds.
%\textcolor{cyan}{PAOLA: can we say here "warm dust" instead of "warm dust temperature"? Temperature cannot be cold or hot, but low or high. It is the gas or dust that is cold or hot depending on their temperature}. 
In the Outer Galaxy, another low-metalicity environment, the gaseous CH$_3$OH abundance in cold envelopes and the COM/CH$_3$OH abundance ratios toward hot cores are similar to those in the inner Galaxy and the Solar neighborhood. These observations suggest that, in low-metallicity environments, CH$_3$OH formation may be inhibited, whereas the subsequent formation of larger COMs may not.

%-----
Various challenges remain to be addressed through a combined effort of observations, laboratory experiments, quantum chemical calculations and astrochemical simulations. 
Deep line surveys toward TMC-1 CP have stimulated active discussions on the formation and destruction pathways of newly detected gas-phase molecules, including various isomers, isotopologues, and aromatic species.
Elemental reaction processes are being investigated through laboratory experiment and quantum chemical calculations, on the basis of which chemical reaction networks are being extended. For aromatic molecules, the top-down scenario, i.e., destruction processes of carbonaceous dust and large PAHs, also needs to be quantitatively constrained, and observations of their spatial distributions are particularly valuable.
In prestellar cores, while the detection of gaseous COMs has highlighted the importance of non-diffusive chemistry, icy COMs beyond CH$_3$OH have not yet been clearly detected. JWST observations toward background stars are expected to quantify their icy abundances in quiescent regions, as well as their spatial distributions.
The relative contributions of diffusive and non-diffusive processes to COM formation likely vary with dust temperature and ice surface composition, and thus along evolution from molecular clouds to prestellar and protostellar cores. The formation and destruction pathways of individual COMs, including prebiotic species detected in CMZ clouds, therefore require substantial further study. Laboratory experiments and quantum chemical calculations have significantly advanced our understanding of grain-surface processes such as diffusion and non-thermal desorption, providing a solid foundation for further progress. The major reservoirs of sulfur and phosphorous in molecular clouds are still unknown. The detection of ammonium salt in comet 67P, and a signature of its presence in interstellar ices, have attracted considerable attention. How and when such salts form in the ISM remains an open question.

Molecular clouds are the birthplace of stars and planets, and the chemistry reviewed here is at least partially inherited by planetary systems. In the central region of prestellar cores, volatile species heavier than He, including COMs, are largely stored in thick icy mantles of dust grains.
When protostars form, some cores are observed as hot corinos, %\citep[i.e., with size $10-100$ au,][]{Ceccarelli2007},
%where dust temperatures exceed the ice sublimation threshold ($\sim$100 K) owing to e.g., stellar irradiation and accretion shock.
%, and mechanical heating in forming disks.
which are rich in COMs, by definition, and also in deuterated molecules \citep{Ceccarelli2007, Jorgensen2020}. The relative abundance of COMs and molecular D/H ratios in hot corinos correlate well with those in prestellar cores and comet 67P \citep{Drozdovskaya2019, Scibelli2025new}. 
JWST observations revealed icy COMs beyond CH$_3$OH along protostellar sightlines, with relative abundances comparable, within a factor of a few, to those measured in comet 67P \citep{Rocha2024}. Additional similarities between cometary and interstellar compositions, including PO, PN and CH$_3$SCH$_3$ \citep{Rivilla2020a, SanzNovo2025a}, further support the inheritance of cometary ices from the pre-Solar molecular cloud.  
Although gas and solids undergo substantial chemical and physical processing in protoplanetary disks \citep{Obserg2023, aikawa2024}, isotopic ratios serve as key tracers of inheritance from molecular clouds. In this context, samples from the near-Earth asteroid Bennu reveal volatile-rich material, abundant ammonia, and N-rich soluble organic matter with enhanced $^{15}$N, pointing to an origin in cold molecular clouds or the outer proto-Solar disk \citep{Glavin2025} (see Supplemental Text on N isotope fractionation).
Beyond observational evidence for inheritance, ongoing studies of the formation and destruction of aromatic molecules in molecular clouds, as well as of prebiotic molecules in CMZ, offer crucial insights into molecular evolution around protostars and in protoplanetary disks. Quantifying the survival, destruction, and re-formation efficiencies of these species under diverse physical conditions is essential for understanding the carbon budget and the potential habitability of forming planetary systems.

\begin{summary}[SUMMARY POINTS]
\begin{enumerate}

\item Deep line surveys toward TMC-1 CP have greatly expanded the interstellar chemical inventory and stimulated new laboratory and theoretical studies of the formation and destruction pathways of newly detected molecules, including numerous isomers. In particular, the origin of aromatic molecules, via bottom-up versus top-down routes, remains a major challenge. %Their nearly flat abundance with carbon number indicates that bottom-up chemistry alone is probably insufficient.

\item Observations and astrophysical simulations are revealing the three-dimensional structure and kinematics of molecular clouds, which constrain important parameters for chemistry such as UV shielding and the dynamical timescales over which time-dependent chemistry proceeds.

\item In terms of both gas dynamics and chemistry, TMC-1 CP appears to be a representative starless core rather than a special case, implying that many of the molecules detected there are widespread in molecular clouds. 

\item Elemental budget in molecular clouds is not fully understood. For example, ammonium salt, allotropes, and refractories could be important reservoir of S. But how, when, and under which conditions they form remains unclear.

\item As starless cores become supercritical, they evolve into prestellar cores such as L1544, characterized by a compact high-density kernel and catastrophic freeze-out of heavy-element species. The key role of this freeze-out in COM formation, predicted by theory and experiments, is supported by deep molecular line observations toward the methanol peak in L1544 and by spatial distribution of interstellar ices.

\item Detection of gaseous COMs in prestellar cores has highlighted the importance of non-diffusive chemistry in COM formation; however, their observable abundances are limited mainly by non-thermal desorption. JWST observations toward background stars will be crucial for determining whether most COMs form in ice mantles during the prestellar phase or in warmer protostellar environments.

\item Observations of GMCs under various physical and metallicity conditions pose new puzzles, but also provide powerful constraints on COM formation mechanisms. This includes, for example, the detection of diverse prebiotic molecules in G+0.693 in CMZ and the similarity of hot core COM/CH$_3$OH abundance ratios between the Outer Galaxy and in the Solar neighborhood.

%\item It is now clear that COMs are ubiquitous, but our understanding of their formation remains incomplete; e.g., why O-bearing and N-bearing COMs are often spatially differentiated in prestellar (and protostellar) cores, and why N-bearing COMs are enhanced in the CMZ GMCs. Accurate treatment of non-diffusive surface chemistry requires Monte Carlo simulations, but such calculations remain computationally challenging for complex species. The accessible diffusion length set by the potential energy distribution could also be crucial. The tentative nature of COM features measured in the ice against background stars further suggests that COM abundances increase substantially in the warm ($\sim$10–50 K) regions around protostars.

\end{enumerate}
\end{summary}

%Disclosure
\section*{DISCLOSURE STATEMENT}
%If the authors have noting to disclose, the following statement will be used: 
The authors are not aware of any affiliations, memberships, funding, or financial holdings that
might be perceived as affecting the objectivity of this review. 

% Acknowledgements
\section*{ACKNOWLEDGMENTS}
%Acknowledgements, general annotations, funding.
We thank Ewine van Dishoeck, Tom Millar, Asunci\'on Fuente, Mario Tafalla, Brett McGuire, Marcelino Ag{\'u}ndez, Silvia Spezzano, Takashi Shimonishi, Kenji Furuya, Naoki Watanabe, Germ\'an Molpeceres, Laura Colzi, Miguel Sanz-Novo, Olli Sipil{\"a}, Elena Redaelli, Tommaso Grassi, Ashley Barnes, Samantha Scibelli, Thomas Speak, Reace Willis, Nami Sakai, Rob Garrod, Yoshito Shimajiri, and the GOTHAM team for their input on text.
Y.A. acknowledge support by Grant-in-Aid for Transformative Research Areas (A) grant Nos. 20H05844 and 20H05847, and JSPS KAKENHI grant No. 24K00674.
I.J.-S. acknowledges funding from the ERC grant OPENS, GA No. 101125858, funded by the European Union. I.J.-S. also acknowledges partial support from grant PID2022-136814NB-I00 funded by the Spanish Ministry of Science, Innovation and Universities/State Agency of Research MICIU/AEI/ 10.13039/501100011033 and by “ERDF/EU”, and from the CSIC ILINK project SENTINEL (ILINK23017). 
P.C. would like to acknowledge Tommaso Grassi for the help with the L1544 Figure and Shuro Takano for providing material about $^{13}$C chemistry - although this has been moved to Supplemental Text. I. J.-S. would like to thank Giuliana Cosentino for providing Figure 8.

\bibliography{ref}

\end{document}

\renewcommand{\tablename}{Supplemental Table} 
% Page header
\markboth{Aikawa et al. - Supplemental Text}{Supplemental Text: Isotopes and Chemical Inventory}

% Title
\title{Supplemental Text: Isotopes and Chemical Inventory in Dark Molecular Clouds}

%Authors, affiliations address.
\author{Yuri Aikawa,$^1$ Izaskun Jimenez-Serra,$^2$ and Paola Caselli$^3$
\affil{$^1$Department of Astronomy, University of Tokyo, Tokyo, Japan, 113-0033; email: aikawa@astron.s.u-okyo.ac.jp}
\affil{$^2$Centro de Astrobiología (CAB), CSIC-INTA, Ctra. de Ajalvir km 4,
28850 Torrejón de Ardoz, Spain}
\affil{$^3$Max-Planck-Institut f\"{u}r extraterrestrische Physik, Gie{\ss}enbachstrasse 1, D-85748 Garching bei M\"{u}nchen, Germany}}

%Abstract
\begin{abstract}
In molecular clouds, the isotopic ratios of molecules often differ from the elemental isotope ratios, a phenomenon known as isotope fractionation. Molecular isotope ratios have been used to investigate the formation pathways of molecules, as well as whether and how interstellar material evolves through star and planetary system formation. Conversely, molecular isotope ratios can also be used to infer the elemental isotope ratios, providing diagnostics of stellar nucleosynthesis across our Galaxy. In this Supplemental Text, we review the basic mechanisms of isotope fractionation and observational studies of molecular isotope ratios in dense cores of nearby molecular clouds and in giant molecular clouds in the Central Molecular Zone.
We also summarize the column densities and abundances of all
molecules detected toward TMC-1 CP and G+0.693$-$0.027.

%Abstract text, approximately 225 words and 
%inclusive of 3--5 bullet items describing 
%the findings of the current research:

\end{abstract}

%Keywords, etc.
\begin{keywords}
keywords, separated by comma, no full stop, lowercase
\end{keywords}
\maketitle

%Table of Contents
\tableofcontents

% Heading 1
\section{Basic mechanisms of isotope fractionation}
The elemental abundance ratios of the local interstellar medium (ISM) and the Sun are summarized in Supplemental Table \ref{tab:isotope}. Differences between the ISM and Solar values may arise from stellar nucleosynthesis in our Galaxy after the birth of the Sun.
Nitrogen isotopic ratios in the ISM have significant uncertainty. While \citet{Ritchey2015} derived a $^{14}$N/$^{15}$N ratio of $\sim 270$ from CN absorption lines in diffuse clouds, \citet{Colzi2018} reported a higher value of $\sim 400$ based on observations of HCN and HNC in high-mass star forming regions. 
Isotope fractionation is primarily driven by exothermic isotope-exchange reactions at low temperatures and by isotope-selective photodissicoation.
%\textcolor{red}{need to add a table of elemental isotope ratio of H,C,N,O,S}

\begin{table}[b]
\tabcolsep7.5pt
\caption{Isotope ratios of elements}
\label{tab:isotope}
\begin{center}
\begin{tabular}{@{}l|c|c|c@{}}
\hline
Element & ISM & Solar & reference\\
\hline
%D/H & ($1.5\pm 0.04)\times 10^{-5}$ & $1.94\times 10^{-5}$ & \citet{Linsky2006, Lodders2003}\\
D/H & ($2\pm 1)\times 10^{-5}$ & $1.94\times 10^{-5}$ & \citet{Friedman2023, Lodders2003}\\
$^{12}$C/$^{13}$C  & $69\pm 6$ &   & \citet{Wilson1999, Milam2005}\\
$^{14}$N/$^{15}$N & $274\pm 18$ & $458.7\pm 4.2$ & \citet{Ritchey2015,Marty2011}\\
                  & $\sim 400$ & & \citet{Colzi2018}\\
$^{16}$O/$^{18}$O & $557\pm 30$ & $530\pm2$  & \citet{Wilson1999,McKeegan2011}\\
$^{16}$O/$^{17}$O & & $2776\pm 14$ &\citet{McKeegan2011}\\
%$^{32}$S/$^{34}$S & $\approx 22$ & &   \citet{Wilson1999}  \\
\hline
\end{tabular}
\end{center}
%\begin{tabnote}
%$^{\rm a}$ While \citet{Ritchey2015} derived $^{14}$N/$^{15}$N using absorption lines of CN in diffuse clouds, \citet{Colzi2018} obtained $^{14}$N/$^{15}$N$\sim 400$ by observing HCN and HNC in high-mass star forming regions. Difference between the ISM and Solar value could be due to the stellar nucleosynthesis in our Galaxy after the birth of Sun.
%\end{tabnote}
\end{table}

\subsection{Deuterium}\label{sec:D-enrichment}
%\textcolor{red}{Include all referes fro Poala's 1 st paragraph. Shoudl mention CO depletion to shorten the Paola's subsecions on ionization rate and D-enrichemnt.}
%D/H enhancement with links to CO depletion, CRs, o/p ratio of H2 (Furuya+15,19, Ueta+2016, Tsuge+2021)
While the elemental D/H ratio is $\sim 2\times 10^{-5}$ (Supplemental Table \ref{tab:isotope}), various molecules have high XD/XH ratios of $\gtrsim 10^{-3}$ in molecular clouds. Such deuterium enrichment (hereinafter referred to as D enrichment) is caused by several exothermic exchange reactions \citep{Watson1976,Millar1989, Roueff2013, Nyman2019}. The most representative one is 
\begin{equation}
    \rm H_3^+ + HD \rightarrow H_2D^+ + H_2 + \Delta E. \label{H2D+}
\end{equation}
The forward reaction is exothermic by $\Delta E=230$ K (but see below) due to the difference in zero-point energies between the reactants and products, so that the reverse reaction is almost negligible at 10 K.
A similar reaction of H$_2$D$^+$ with HD produces D$_2$H$^+$, and  D$_2$H$^+$ + HD produces D$_3^+$\citep{Roberts2003, Walmsley2004}. %Depletion of abundant neutrals, such as CO and O, further enhances the D enrichment, as they are major destruction partners of all the H$_3^+$ forms \citep{Dalgarno1984}.
Since H$_3^+$ plays a central role in gas-phase reactions (H$_3^+$ + X $\rightarrow$ XH$^+$ + H$_2$), its enhanced D/H ratio propagates to other molecules.
Recombination of deuterated H$_3^+$ produces D atoms, which enhances the D/H ratio of atomic hydrogen and icy molecules formed by hydrogenation on grain surfaces. 
The kinetic isotope effect, i.e., the higher tunneling efficiency of H atoms compared to D atoms in reactions with activation barrier, is also important.
In the case of CH$_3$OH, H-abstraction and D-addition further enrich the methyl group with D \citep{Tielens1983,Hidaka2009,Hama2013} (Figure 4 in the main text). Reviews such as \citet{Caselli2012} and \citet{Ceccarelli2014} detail the deuteration process from clouds to planets.

The exothermicity of the reaction (\ref{H2D+}), and thus the efficiency of D enrichment, depend on the ortho-to-para ratio (OPR) of H$_2$, since the ground-state energy of ortho-H$_2$ ($o$-H$_2$) is higher than that of para-H$_2$ ($p$-H$_2$) by 170 K \citep{Pagani1992,Flower2006, Hugo2009}. While the branching ratio for forming $o$-H$_2$ is 75 \% in grain-surface H$_2$ formation \citep{Watanabe2010}, \citet{Furuya2015} calculated OPR of H$_2$ and deuterium chemistry during molecular cloud formation behind an interstellar shock and found that the OPR of H$_2$ can be lowered to $\lesssim 10^{-3}$ through gas-phase reactions with H$^+$ and H$_3^+$ by the time $A_{\rm V}$ reaches $\sim 3$ mag, such that D enrichment via reaction (\ref{H2D+}) becomes efficient enough to account for the observed molecular D/H ratios. The decline of H$_2$ OPR can be further accelerated by ortho-to-para conversion on grain surfaces \citep{Ueta2016, Bovino2017, Furuya2019, Tsuge2021_NSC}.
% Brunken+14 is not cited here as it is an observation towards a warm target, IRAS16293

In high-density low-temperature regions ($n_{\rm H}>10^4$ cm$^{-3}$, $T\sim 10$ K), the abundance ratio of H$_2$D$^+$ to H$_3^+$ is proportional to $n({\rm HD})/n({\rm CO)}$, since CO is the main reactant with H$_2$D$^+$ (\S 2.4 in the main text). CO freeze-out then accelerates deuterium fractionation \citep{Dalgarno1984}. Even multiply deuterated molecules become abundant and are detected in prestellar cores and the central regions of protostellar cores, where ices are sublimated (\S \ref{sec:multiD}).

The effectiveness of isotope exchange reactions for deuteration depends on the species involved and on their exothermicity. The rate of the backward reaction of (\ref{H2D+}) is proportional to $n({\rm H}_2)\exp(-230/T)$. Considering the canonical CO abundance in molecular clouds, $n$(CO)/$n$(H$_2$)$\sim 10^{-4}$, the backward reaction rate becomes comparable to the rate of H$_2$D$^+$ + CO at temperatures of a few tens of K, and the reaction (\ref{H2D+}) becomes less effective for deuteration with increasing temperature. Another exchange reaction
\begin{equation}
    \rm CH_3^+ + HD \rightarrow CH_2D^+ + H_2 + \Delta E, \label{CH2D+}
\end{equation}
is exothermic by $\Delta E\sim 400$ K. Molecules whose formation pathways involve CH$_3^+$, such as hydrocarbons, can thus maintain relatively high D/H ratio at $T\sim$ several tens of K, compared with species mainly deuterated through reaction (\ref{H2D+}) \citep[e.g.,][]{Parise2009}. 
Evaluation of the exothermicity of these exchange reactions requires high-accuracy quantum chemical calculations of the zero-point energies of reactants and products \citep[e.g.][]{Nyman2019}.

\subsection{Oxygen and Nitrogen} \label{sec:O_N_frac}

%\textcolor{magenta}{Paola: We do not need to cite Riedel et al. 2023 here, as this paper deals with D/H ratio in methanol.  However, we should cite Sipila et al. 2023, as he included 15N (and 13C) fractionation rations in his code.}

Isotope fractionation of N and O can be caused by the isotope-selective photodissociation of N$_2$ and CO. When a molecule with high abundance is photodissociated mainly through line absorption rather than continuum absorption, it can attenuate UV radiation at specific wavelengths, which is called self-shielding. Their rare isotopomers, however, are not efficiently shielded by themselves.
For example, UV radiation that dissociates $^{12}$C$^{18}$O can penetrate deeper into clouds than that which dissociates $^{12}$C$^{16}$O \citep{vanDishoeck1988, Visser2009}. Isotope-selective photodissociation of CO has long been confirmed by observations of PDRs \citep[e.g.][]{Bally1982, Yamagishi2019}.
If water ice is efficiently formed in regions where $^{12}$C$^{16}$O is shielded but $^{12}$C$^{18}$O is not, the resulting water ice becomes enriched in $^{18}$O \citep{Yurimoto2004, Lyons2005}. Indeed, $^{18}$O-rich water is found in comet 67P/C-G (H$_2^{16}$O/H$_2^{18}$O=445$\pm 35$) \citep{Altwegg2019}.
While this mechanism was originally proposed to explain the oxygen isotope anomaly in meteorites, it is also important for molecular cloud chemistry.

Similarly, self-shielding is effective for N$_2$, which is one of the major reservoir of nitrogen \citep{Heays2014}. In regions where $^{14}$N$_2$ is self-shielded but $^{15}$N$^{14}$N is not, molecules formed from N atom become $^{15}$N-rich.
NH$_3$ ice is one of such species. In the formation stage of molecular clouds, $^{15}$N-rich NH$_3$ ice can be formed, which makes the ice phase $^{15}$N-rich and the gas phase $^{15}$N-poor \citep{FuruyaAikawa2018}. This mechanism can explain the high $^{14}$N/$^{15}$N ratio ($580-1000$) in N$_2$H$^+$ in dense cores \citep{Redaelli2018} and low  $^{14}$N/$^{15}$N ratio ($118-140$) in comets \citep{Rousselot2014, Altwegg2019} 
%In the formation stage of molecular clouds, $^{15}$N-rich NH$_3$ ice can be formed in regions where $^{14}$N$_2$ is self-shielded but $^{15}$N$^{14}$N is not. 
%This mechanism makes the ice phase $^{15}$N-rich and the gas phase $^{15}$N-poor \citep{FuruyaAikawa2018}, which can explain the high $^{14}$N/$^{15}$N ratio ($580-1000$) in N$_2$H$^+$ in dense cores \citep{Redaelli2018} and low  $^{14}$N/$^{15}$N ratio ($118-140$) in comets \citep{Rousselot2014, Altwegg2019} 
%\textcolor{blue}{and in protoplanetary discs \citep{Guzman2017,Hily-Blant2017,Rampinelli2025}} 
at least qualitatively. While several exothermic exchange reactions such as $^{15}$N$^+$ + N$_2$ $\rightarrow$ $^{14}$N$^+$ + N$^{15}$N were considered, they were found to possess a barrier in the entrance
channel and thus unlikely to proceed at 10 K \citep{Roueff2015}. 
\citet{Sipila2023} performed pseudo-time-dependent models of molecular clouds with tens of exchange reactions for $^{15}$N (and $^{13}$C) and found that the $^{14}$N/$^{15}$N ratios (e.g. HCN/HC$^{15}$N) tend to remain close to the elemental ratio within $\sim 10$ \%. 
%$^{14}$N/$^{15}$N ratios of molecules remain close to the elemental ratio within approximately $10 \%$ in models with these reactions \citep{Sipila2023}.

\subsection{Carbon}\label{sec:C_isotope}
Carbon isotope fractionation by the exothermic exchange reaction $^{13}$C$^+$ + $^{12}$CO $\rightarrow$ $^{12}$C$^+$ + $^{13}$CO + 35 K has long been investigated in theoretical models \citep{Watson1976,Langer1984}. While the $^{12}$C/$^{13}$C ratio of CO %(and of other molecules formed from CO, such as CH$_3$OH)
cannot deviate significantly from the elemental abundance ratio as long as CO is the dominant carbon reservoir, molecules formed from atomic C and C$^+$ become $^{13}$C-poor.
The trend is expected to extend to COMs \citep{Busch2025}; their $^{12}$C/$^{13}$C ratio is similar to the elemental abundance ratio if they are formed from CO (e.g. CH$_3$OH), while it becomes smaller if their carbon originates in C atom or C$^+$.
Direct C-atom addition reactions on ice surfaces, i.e., C + H$_2$O ice $\rightarrow$ H$_2$CO ice and C + CO ice $\rightarrow$ C$_2$O ice \citep{Molpeceres2021JPCL, Ferrero2023, Fedseev2022}, however, open pathways to form $^{13}$C-enriched COMs \citep{ichumura2024}.

Recent observations have found that CN, HCN, and HNC are enriched with $^{13}$C in prestellar and protostellar cores (\S \ref{sec_core_Cisotope}).
Several possible exchange reactions, such as $^{13}$C + C$_3$ $\rightarrow$ $^{12}$C + $^{13}$CC$_2$ + 27 K, have been proposed and are currently being investigated \citep{Langer1984, Colzi2020, Sipila2023}.

The abundance ratios among the $^{13}$C isotopomers can be a powerful probe for identifying molecular formation pathways. 
\citet{Takano1998} were the first to measure the $^{12}$C/$^{13}$C ratio in HC$_3$N by observing H$^{13}$CCN, HC$^{13}$CCN and HCC$^{13}$CN in TMC-1. Interestingly, they found H$^{13}$CCCN:HC$^{13}$CCN:HCC$^{13}$CN = 1.0:1.0:1.4, suggesting that HC$_3$N forms via the reaction C$_2$H$_2$ + CN, with CN enriched in $^{13}$C \citep[see also][]{Taniguchi2017b}. 
%In the case of HCC$^{13}$CN, a $^{13}$C fraction lower than the local ISM value was found \citep[$\sim$55;][]{Takano1998}, while for the other isotopologues, no significant $^{13}$C fractionation is apparent. 
%Similar conclusions have been reached by \cite{Araki2016}, who observed all the $^{13}$C isotopologues of HC$_3$N in the protostellar core L1527 with high sensitivity. 
Similarly, \citet{Sakai2010} found a column density ratio of C$^{13}$CH/$^{13}$CCH $=1.6\pm0.4$ in TMC-1. It indicates that a neutral-neutral reaction, C + CH$_2$, contributes significantly to the formation of CCH, rather than the electron recombination reactions of C$_2$H$_2^+$ and C$_2$H$_3^+$, in which the two carbon atoms are equivalent. %\citep[see also][]{Sakai2007, Sakai2013, Yoshida2015}.
Alternatively, $^{13}$CCH could be converted to C$^{13}$CH via the exchange reaction $^{13}$CCH + H $\rightarrow$ C$^{13}$CH + H + 8.1 K, if the reaction has no activation barrier. \citet{Furuya2011} also proposed a hydrogen-atom-catalyzed reaction, $^{13}$CCS + H $\rightarrow$ C$^{13}$CS + H + 17.4 K, to reproduce the observed ratio of C$^{13}$CS/$^{13}$CCS $=4.2\pm2.3$ in TMC-1. A quantum chemical study by \citet{Talbi2018} found that this reaction is barrierless.

%\cite{Taniguchi2016}, on the other hand, found no significant difference in $^{12}$C/$^{13}$C ratio among isomers of HC$_5$N in TMC-1 \citep[also confirmed by][]{Cernicharo2020}. It suggest that the formation route of HC$_5$N may proceed via ion-molecule formation (in contrast to HC$_3$N), involving relatively large hydrocarbon ions such as C$_5$H$_3^+$ reacting with atomic nitrogen. 
%$^{15}$N fractionation has also been measured in HC$_3$N and HC$_5$N in starless and pre-stellar cores. 
%\cite{Taniguchi2017} measured $^{14}$N/$^{15}$N = 344$\pm$53 and 257$\pm$54 in HC$_5$N and HC$_3$N, respectively towards TMC-1(CP), which again indicates different formation mechanisms for HC$_3$N and HC$_5$N. 

%\subsection{Molecular data for deriving isotopic ratios from observations}
%spectroscopic data  Oyama+23
%ND3/NHD2/NH3 Daniel+16

\section{Observations of Isotope fractionation in cores}
\subsection{Deuterium fractionation} %\label

The low temperatures ($<10$\,K) and high densities ($>10^5$\,cm$^{-3}$) in prestellar cores are ideal conditions for D enrichment (\S \ref{sec:D-enrichment}).
Since CO is the dominant reactant with H$_3^+$ and its deuterated forms, the catastrophic freeze-out of CO in prestellar cores (\S 3.2 in the main text) enhances D enrichment. The atomic D/H ratio in prestellar cores, for example, can approach unity \citep{Roberts2003,Walmsley2004}, i.e., more than four orders of magnitude higher than the elemental D/H ratio (Supplemental Table \ref{tab:isotope}). 

\subsubsection{Deuterated isotopologues of H$_3^+$}
The first strong (brightness temperature $\sim1$\,K) ground-state ortho-H$_2$D$^+$ ($o$-H$_2$D$^+$) line (372\,GHz) was observed toward the prestellar core L1544 with CSO by \cite{Caselli2003}. 
The same transition had previously been detected with JCMT in the cold outer envelope of a young stellar object NGC 1333 IRAS 4A \citep{Stark1999}, although with a much weaker line intensity.
This was followed by the first detection of para-D$_2$H$^+$ ($p$-D$_2$H$^+$) at 692\,GHz toward the prestellar core IRAS16293E \citep{Vastel2004,Parise2011} and several other observations toward nearby cold cores and clouds \citep{Caselli2008,Friesen2014,Bovino2021} as well as high-mass star-forming regions \citep{Pillai2012,RedaelliBovino2021,Sabatini2024}. Thanks to the high-frequency capabilities of APEX and SOFIA, ortho- and para- H$_2$D$^+$ lines at 1.4\,THz were detected towards the cold ($13-16$ K) envelope of the very young stellar object IRAS\,16293-2422 \citep{Brunken2014}. Observing ortho- and para-H$_2$D$^+$ allows a direct inference of the H$_2$ OPR, which regulates the deuterium fractionation (\S \ref{sec:D-enrichment}).
The derived H$_2$ OPR in the envelope of IRAS\,16293-2422 is $\sim 2 \times 10^{-4}$, which is low enough to allow efficient D enrichment.
The ground state transition of o-D$_2$H$^+$ (1.5\,THz) has also been detected toward IRAS16293-2422 with SOFIA by \cite{Harju2017}. 
The total (ortho+para) abundance of D$_2$H$^+$ is estimated to be comparable to that of H$_2$D$^+$. While it indicates abundant D$_3^+$, direct detection of D$_3^+$ remains
challenging, as the only observable transition is an absorption line at 5.3\,$\mu$m \citep{Goto2019}. With the shutdown of SOFIA and no FIR space telescope currently available, p-H$_2$D$^+$ and o-D$_2$H$^+$ can no longer be measured.

\subsubsection{Other molecules}
High D-fractions ($\sim 10$\% or higher) toward prestellar cores have been measured in N$_2$H$^+$ \citep{Caselli2002b,Crapsi2005,Pagani2007,Redaelli2019}, NH$_3$ \citep{Roueff2015,Crapsi2007,Harju2017b}, c-C$_3$H$_2$ \citep{Spezzano2013,Chantzos2018,Giers2022,Giers2023}, H$_2$CO \citep{Bacmann2003,ChaconTanarro2019}, H$_2$CS \citep{Spezzano2022}, HC$_3$N, CH$_3$CCH \citep{Giers2026} and CH$_3$OH \citep{Bizzocchi2014,ChaconTanarro2019,Lin2023b}, consistent with the D enrichment enhanced by CO freeze-out. The D-fraction in HCO$^+$ has been measured to be slightly below the 10\% level \citep{Caselli2002b,Redaelli2019}, probably due to the fact that both HCO$^+$ and DCO$^+$ mainly trace the lower density core envelope, as CO (their mother molecule) is highly depleted within the central regions, where the D-fraction peaks. 

While the direct measurements of molecular D/H ratios in ices are difficult (except for HDO/H$_2$O, see below), it can be inferred from the molecules desorbed from ice by shocks. 
\cite{Chahine2024} measured the D$_2$CO/HDCO abundance ratio along the outflow cavities of the Class 0 source NGC\,1333 IRAS\,4A, finding evidence of a decreasing D-fraction with distance from the central protostar. This trend is consistent with the radial decrease in D-fraction observed in prestellar cores \citep{Redaelli2019} and predicted by chemical modeling \citep{Aikawa2012,Sipila2015}.

%----- modified version by Yuri------------------
\subsubsection{Multiply deuterated molecules} \label{sec:multiD}

The relative abundances of multiply deuterated isotopologues (e.g., XD$_2$H, XH$_2$D, and XH$_3$) provide diagnostics of molecular formation pathways and the underlying deuterium fractionation reactions in molecular clouds.

Single-dish observations have detected NH$_3$, NH$_2$D, NHD$_2$, and ND$_3$ in several prestellar cores \citep{Roueff2005,Harju2017b}.
While the column density ratio of NH$_2$D/NH$_3$ varies among objects in the range $0.1-0.4$, it is higher than that of ND$_3$/NHD$_2$ ($\sim 0.02-0.06$). Considering uncertainties in the derivation of the column densities (e.g., excitation conditions), the ratio between NH$_2$D/NH$_3$ and ND$_3$/NHD$_2$ does not significantly deviate from the statistical expectation; i.e., NH$_3$ has three equivalent hydrogen positions that can be substituted by deuterium to form NH$_2$D. These observed ratios are also consistent with the chemical reaction network models of prestellar cores \citep{Harju2017, Aikawa2005}.

Higher values of doubly deuteration (XD$_2$/XHD) than mono-deuteration (XHD/XH$_2$) are observed in several molecules. Towards L1544, \cite{ChaconTanarro2019} obtained the column density ratios of HDCO/H$_2$CO$\simeq$0.03, D$_2$CO/HDCO$\simeq$1, and CH$_2$DOH/CH$_3$OH$\simeq$0.1.
\cite{Riedel2023} performed chemical network calculations using the density structure of L1544, and successfully reproduced the observed D/H ratios toward the center within a factor of 1.2.
Doubly deuterated methanol (CHD$_2$OH) was detected for the first time toward two prestellar cores \citep[H-MM1 and L694-2;][]{Lin2023b}. The CHD$_2$OH/CH$_2$DOH ratio (0.5-0.8) significantly exceeds the CH$_2$DOH/CH$_3$OH ratio (0.008-0.02).
While these high abundances of D$_2$CO and CHD$_2$OH could be due to the additional deuteration on grain surfaces (\S 2.4 and Figure 4 in the main text), a similarly high doubly-deuteration is found for H$_2$CS toward L1544: HDCS/H$_2$CS$\simeq$0.3 and D$_2$CS/HDCS$\simeq$1.0 \citep{Spezzano2022}.
Since H$_2$CS can also form on grain surfaces, laboratory experiments and quantum chemical calculations of H- and D-addition and abstraction reactions along its formation pathways --- similar to those performed for CH$_3$OH --- are highly desirable.
%\textcolor{red}{Yuri: the description on c-C3H2 is removed, since c-C3D2/c-C3HD (10\%) is similar to c-C3HD/ c-C3H2 (17\%) in Table 3 of Spezzano+22}.

%Interestingly, the ratio of D$_2$O/HDO is found to be higher than HDO/H$_2$O, which indicates that D$_2$O is mainly formed in dense cores, whereas H$_2$O is formed in an earlier phase, e.g., during molecular cloud formation \citep{Furuya2016}.

Water is another example. 
JWST recently detected HDO ice towards an intermediate-mass YSO (HOPS 370) and a massive YSO (IRAS 20126+4104).  The HDO/H$_2$O ice ratio is found to be similar to the highest reported gaseous HDO/H$_2$O ratios measured toward massive YSOs and the hot inner regions of isolated low-mass protostars ($(2-3)\times 10^{-3}$), suggesting that at least some of the gas HDO/H$_2$O ratios measured toward hot cores are representative of the HDO/H$_2$O ratios in ices \citep{Slavicinska_HDO_2025}.
\citet{Coutens2014} detected the D$_2$O $1_{1.0}-1_{0,1}$ transition toward the low-mass protostar NGC 1333 IRAS2A and derived D$_2$O/HDO ratio of $\sim (1.2\pm 0.5)\times 10^{-2}$, which is much higher than HDO/H$_2$O ratio of $\sim (1.7\pm 0.8)\times 10^{-3}$. It indicates that D$_2$O ice is mainly formed in cold dense cores with CO depletion, while H$_2$O ice is readily formed in an earlier phase, e.g., during molecular cloud formation \citep{Furuya2016}.

%{\bf MOVED FROM MAIN TEXT} 
%The detection of COMs in the gas phase along the bipolar outflows driven by protostars, at thousands of au from the central source, further supports their formation and storage in ices, since in these outer regions of dense cores it would take much longer than the shock timescale to synthesize COMs in the gas phase starting from simpler species \citep{Arce2008,Yamaguchi2012,Burkhardt2016,Lefloch2017}, with the exception of formamide \citep[NH$_2$CHO;][]{LopezSepulcre2024} and possibly acetaldehyde \citep[CH$_3$CHO;][]{Codella2020}.
%Interestingly, \cite{Chahine2024} measured the D$_2$CO/HDCO abundance ratio along the outflow cavities of the Class 0 source NGC\,1333 IRAS\,4A, finding evidence of a decreasing D-fraction with distance from the central protostar. This trend is consistent with the radial decrease in D-fraction observed in prestellar cores \citep{Redaelli2019} and predicted by chemical modeling \citep{Aikawa2012,Sipila2015}. Thus, \cite{Chahine2024} provide strong evidence that prestellar ices are released in the gas phase via sputtering along outflows, reinforcing the hypothesis that many COMs observed in these regions are directly injected into the gas by the shock processing of dust. 

\subsection{$^{15}$N fractionation}
While the elemental nitrogen isotope ratio remains uncertain (Supplemental Table \ref{tab:isotope}), observations of dense cores clearly show nitrogen isotope fractionation among molecules. In the following we adopt an elemental isotope ratio of $^{14}$N/$^{15}$N$\sim 400$. In the gas phase, NH$_3$, NH$_2$D and CN do not show significant fractionation \citep{Gerin2009,Lis2010,Daniel2013,Daniel2016,HilyBlant2013b}, whereas HCN and HNC are slightly $^{15}$N-rich, $^{14}$N/$^{15}$N$\sim 150 - 300$ \citep{Ikeda2002,HilyBlant2013a,Hily-Blant2020,Megalhaes2018}.
N$_2$H$^+$, on the other hand, is $^{15}$N-poor, with $^{14}$N/$^{15}$N $\sim$800 \citep{Bizzocchi2013,Redaelli2018}.
The combination of selective photodissociation of N$_2$ and $^{15}$N-rich NH$_3$ ice formation \citep[\S \ref{sec:O_N_frac},][]{FuruyaAikawa2018} explains these fractionation levels, at least qualitatively.

\citet{Colzi2019} showed that the variations in the N$_2$H$^+$ isotopic ratio observed toward the high-mass star-forming region IRAS 05358+3543 can be attributed to isotope-selective photodissociation of N$_2$ driven by the external interstellar UV field. This process produces higher ratios of N$_2$/N$^{15}$N in the outer envelope compared to the denser inner gas.
Similarly, the larger amount of $^{15}$N atom in regions exposed to the ISRF can explain the increased $^{15}$N fraction observed in HCN toward the south of L1544 \citep{Spezzano2022b}, if a significant fraction of HCN is produced by atomic nitrogen. 

With the aim of testing the model prediction on the $^{15}$N-rich ammonia ice, \cite{Redaelli2023} studied the ammonia $^{15}$N fractionation in L1544, the protostar IRAS\,4A, and the shocked region B1 along the protostellar outflow of L1157. If NH$_3$ ice is $^{15}$N-rich, large $^{15}$N fractions should be found in regions where icy mantles are either recently sublimated (near the protostar) or sputtered (in shocks).  Indeed, \cite{Redaelli2023} found a $^{14}$NH$_3$/$^{15}$NH$_3$ column density ratio of $\sim$200 toward the protostar, while no fractionation ($\sim$400) was measured toward L1544, in agreement with theory. Only upper limits could be obtained toward L1157-B1. 
Interestingly, the $^{15}$N fraction of NH$_3$ is about a factor of two larger than that measured in N$_2$H$^+$ \citep{Redaelli2018} in prestellar cores. This could be explained by the fact that, while N$_2$H$^+$ forms exclusively in the gas phase from $^{15}$N-poor N$_2$ (by reaction 2 in the main text), gas phase NH$_3$ is also produced from atomic N (which is relatively $^{15}$N-rich) on grains surfaces and then partially released in the gas phase by reactive desorption \citep[e.g.,][]{Sipila2019}. This could then imply that a significant fraction of nitrogen is still in atomic form in the core center and maintains the interstellar $^{14}$N/$^{15}$N value. 
That only a small fraction (a few percent) of N is in molecular form has been deduced by observations toward the starless core B68 \citep{Maret2006} and the prestellar core L183 \citep{Pagani2012}, as well as by astrochemical modeling \citep{Daranlot2012}.  
%{\bf YURI: The statements after "This could be" seems to contradict with the earlier statement that 14N/15N in NH3 is 400,i.e. ISM value in prestellar cores. Rather, 15N-poor N2H+ could be explained by the difference in dissociative recombination of 15NNH+ and N2H+ discussed below.} \textcolor{cyan}{PAOLA: I corrected the text. I hope now it is clearer.}

While \citet{FuruyaAikawa2018} predict that the gas-phase species are $^{15}$N-poor, observations of cores show an even higher $^{14}$N/$^{15}$N ratio for N$_2$H$^+$ than predicted in their model. To account for this additional fractionation, it has been proposed that the dissociative recombination of $^{15}$N isotopologues of N$_2$H$^+$ proceeds at a faster rate than that of the main isotopologue \citep{Loison2019,Hily-Blant2020}. In prestellar cores, where CO is frozen-out, N$_2$H$^+$ is mainly destroyed via dissociative recombination. Higher recombination rates of N$^{15}$NH$^+$ could produce N$_2$H$^+$/N$^{15}$NH$^+$ ratios as large as those observed. Although this scenario appears consistent with the observations of \cite{Redaelli2020}, who found higher $^{15}$N fractions in N$_2$H$^+$ toward protostellar objects, where CO is the main destruction partner of N$_2$H$^+$, definite confirmation awaits laboratory measurements of the dissociative recombination of $^{15}$N$^{14}$NH$^+$ and $^{14}$N$^{15}$NH$^+$ at low temperatures. 

$^{15}$N fractionation has also been used to probe formation pathways of cyanides. \cite{Taniguchi2017} measured $^{14}$N/$^{15}$N = 344$\pm$53 and 257$\pm$54 in HC$_5$N and HC$_3$N, respectively, towards the starless core TMC-1(CP) in Taurus.  From this and previous results, they deduced that the formation mechanisms of the two cyanopolyynes are different, with HC$_3$N forming from C$_2$H$_2$ + CN, while HC$_5$N is the product of ion-molecule reactions between hydrocarbon ions and nitrogen atoms. 
\cite{Hily-Blant2018} measured $^{14}$N/$^{15}$N in HC$_3$N towards L1544 and found values close to the ISM value ($\sim$400). Therefore, no $^{15}$N fractionation of HC$_3$N appears to be present at the higher densities of prestellar cores, similar to what was found in NH$_3$ by \cite{Redaelli2023}, thus suggesting that atomic N is important for the production of HC$_3$N.

A final note is that care must be taken when $^{14}$N/$^{15}$N ratios are deduced by assuming a fixed ISM $^{12}$C/$^{13}$C elemental abundance ratio ($\sim$68). This is typically done when observing HCN lines, as they are optically thick and self absorbed, so that the rarer isotopologue H$^{13}$CN is used instead. %But C-isotope can also be fractionated.
However, carbon isotopes themselves may also undergo isotopic fractionation.
\cite{Jensen2024} measured the HC$^{14}$N/HC$^{15}$N column density ratio toward six starless cores (including L1544 and L183) using the direct method (i.e., modeling the optically thick HCN lines with a non-LTE radiative transfer code) and the indirect method \citep[i.e., using the optically thin H$^{13}$CN and assuming $^{12}$C/$^{13}$C = 68 referring to][]{Milam2005}. The results were strikingly different, with the direct method providing significantly lower HC$^{14}$N/HC$^{15}$N column density ratios ($\sim$250) than the indirect method ($\sim$300-500). Similar direct analysis was carried out by \cite{Daniel2013} in Barnard\,B1b and \cite{Megalhaes2018} in L1498, who also found significant $^{15}$N fractionation of HCN. Knowledge on the physical structure of the source, as well as the rate coefficients of collisional excitation, is therefore important so that non-LTE radiative transfer methods can be applied to derive $^{15}$N fractionation
\citep[see also][who constructed the first time-dependent combined model for $^{13}$C and $^{15}$N fractionation inclusive of spin-state chemistry]{Sipila2023}. 
The same statement is valid for sources at later stages of evolution (protostars and protoplanetary disks), especially when attempting to investigate the chemical heritage, as discussed by \cite{Jensen2024}. 

\subsection{$^{13}$C fractionation}\label{sec_core_Cisotope}

%As mentioned in previous sections, the $^{12}$C/$^{13}$C isotopic ratio in the local interstellar medium is 68$\pm$15 \citep{Milam2005} and this value is typically assumed to derive the $^{14}$N/$^{15}$N (see Table\,\ref{tab:isotope}) isotopic ratio in abundant species, such as HCN, as the optically thinner lines of $^{13}$C isotopologues are preferentially observed as a proxy for the main species. However, 

In dark clouds, $^{13}$C fractionation of molecules is known to vary, mainly because of the exothermic exchange reaction $^{13}$C$^+$ + $^{12}$CO $\rightleftharpoons$ $^{13}$CO$^+$ + $^{12}$C$^+$ +  35\,K (\S 1.3). 
At low temperature, $^{13}$C is sequestered in CO molecules, and species mainly forming from atomic carbon (such as CN and  HCN), are expected to be $^{13}$C-poor. 
Theoretical models show that the $^{12}$C/$^{13}$C ratio in different molecules is highly time dependent and it can significantly change with the physical conditions, in particular number density, temperature, and cosmic-ray ionization rate \citep{Furuya2011, Roueff2015, Colzi2020}.
Besides the CO exchange reaction, \cite{Colzi2020} pointed out the importance of another exchange reaction: $^{13}$C + C$_3$ $\rightleftharpoons$ $^{12}$C + $^{13}$CC$_2$, which is exothermic by 27\,K. In regions where the C to CO conversion is not yet complete and in the central region of cores where CO is almost completely frozen, this reaction makes the molecules forming from atomic carbon $^{13}$C-rich.

%The $^{13}$C fractionation of molecules in dark clouds is known to vary, mainly because of the ion-neutral isotopic exchange reaction $^{13}$C$^+$ + $^{12}$CO $\rightleftharpoons$ $^{13}$CO$^+$ + $^{12}$C$^+$ + 35\,K (\S 1.3). 
%In cold regions, this reaction proceeds from left to right, therefore the $^{13}$C becomes sequestered in CO molecules, and species mainly forming from atomic carbon (such as CN and  HCN), are expected to have high $^{12}$C/$^{13}$C ratios.  
%In agreement with \cite{Roueff2015}, \cite{Colzi2020} show that the $^{12}$C/$^{13}$C ratio in different molecules is highly time dependent and it can significantly change with the physical conditions, in particular number density, temperature, and cosmic-ray ionization rate \citep[see also][for similar conclusions]{Furuya2011}. Besides the CO exchange reaction, \cite{Colzi2020} also pointed out another important reaction for $^{13}$C fractionation, the neutral-neutral reaction $^{13}$C + C$_3$ $\rightleftharpoons$ $^{12}$C + $^{13}$CC$_2$ + 27\,K. At the low temperatures of dark clouds, in regions where the C to CO conversion is not yet complete and in central zones of prestellar cores, where CO is almost completely frozen, this reaction maintains $^{12}$C/$^{13}$C ratios lower than 69 in molecules forming from atomic carbon. Another interesting result of \cite{Colzi2020} is that $^{12}$C/$^{13}$C ratios increase when the cosmic-ray ionization rate ($\zeta_{\rm H_2}$) increases, so measurements of the $^{13}$C fraction could be used to provide an independent estimate of $\zeta_{\rm H_2}$. 

High $^{13}$C fractions have been measured in CN, HCN and HNC toward the B1b cloud \citep[$^{12}$C/$^{13}$C $\sim$50, $\sim$30 and $\sim$20, respectively;][]{Daniel2013} and in HCN toward the starless core L1498 \citep[$\sim45$;][]{Megalhaes2018}. \cite{Spezzano2022b} measured the $^{12}$CN/$^{13}$CN column density ratio across L1544, finding values significantly higher than the elemental value of 69: 
between 130 and 190 across the core, although the differences are not significant, considering the uncertainties. Referring to the pseudo-time dependent models for various physical parameters by \cite{Sipila2023}, values around 130 are consistent with number densities around 10$^4$\,cm$^{-3}$ and chemical ages (starting with all elements, except for hydrogen, in atomic form) around one million years. Higher values (measured towards the core center and the north) would need higher densities and shorter chemical age. The higher $^{13}$C fractions measured in B1b and L1498 can be reproduced with number densities around 10$^4$\,cm$^{-3}$ and chemical ages between 10$^4$ and 10$^6$ years (see their Figure A.1). However, these model comparisons should be considered as qualitative, given that the physical structure of the source and its dynamical evolution are not taken into account in the pseudo-time dependent models.
%$\sim$130 toward the south, more exposed to the interstellar radiation field, and $\sim$190 toward the center and the northwest.\textcolor{red}{Referring to the pseudo-time dependent models for various physical parameters by \cite{Sipila2023}}, the L1544 results towards the south are consistent with number densities around 10$^4$\,cm$^{-3}$ and chemical ages (starting with all elements, except for hydrogen, in atomic form) around one million years, while the higher ratios toward the center and the northwest are consistent with model predictions for denser ($\sim$10$^6$\,cm$^{-3}$) material and younger chemical ages. The higher $^{13}$C fractions measured in B1b and L1498 can be reproduced with number densities around 10$^4$\,cm$^{-3}$ and chemical ages between 10$^4$ and 10$^6$ years (see their Figure A.1). However, these model comparisons should be considered as qualitative, given that the physical structure of the source and its dynamical evolution are not taken into account in the pseudo-time dependent models. 

Other $^{13}$C fractionation studies in dark clouds have targeted carbon-chain molecules, such as CCH, CCS, C$_3$S, C$_3$H$_2$, C$_4$H and cyanopolyynes (in particular HC$_3$N and HC$_5$N). In TMC-1, \cite{Takano1998} found H$^{13}$CCCN:HC$^{13}$CCN:HCC$^{13}$CN = 1.0:1.0:1.4. In the case of HCC$^{13}$CN, the $^{12}$C/$^{13}$C ratio lower than the local ISM value ($\sim$55) was found, while for the other isotopologues, no significant $^{13}$C fractionation is apparent. Similar conclusions have been reached by \cite{Araki2016}, who observed all the $^{13}$C isotopologues of HC$_3$N in the protostellar core L1527 with high sensitivity. 
\cite{Tercero2024} detected double-substituted isotopologues in TMC-1, as part of the high-sensitivity QUIJOTE survey carried out at the Yebes 40m telescope, also confirming the larger $^{13}$C fraction in HCC$^{13}$CN than the other isotopologues. However, they found similar $^{12}$C/$^{13}$C abundance ratios for DCCCN, pointing to additional chemical differentiation between HC$_3$N and DC$_3$N. 
In the case of HC$_5$N, \cite{Taniguchi2016} found no significant $^{13}$C fractionation \citep[also confirmed by][]{Cernicharo2020b}, suggesting that the formation route of HC$_5$N may proceed via ion-molecule formation (unlike in the case of HC$_3$N), involving relatively large hydrocarbon ions such as C$_5$H$_3^+$ reacting with atomic nitrogen. Interestingly, \cite{Cernicharo2024c} detected $^{13}$C isotopologues of CH$_2$CHCN, finding that CH$_2$CH$^{13}$CN has an increased abundance with respect to the other $^{13}$C isotopologues, as in the case of HC$_3$N. 

\cite{Sakai2007} found significant $^{13}$C dilution (i.e., high $^{12}$C/$^{13}$C) in CCS toward the dark cloud TMC-1 and the starless core L1521E. Anomalous $^{13}$C isotope abundances were then measured for C$_3$S and C$_4$H in TMC-1, with significant variations in the $^{12}$C/$^{13}$C ratio depending on the position of the $^{13}$C atom within the carbon chain molecule, ruling out some formation pathways such as $c-$C$_3$H$_3^+$ + S \citep{Sakai2013, Fuentetaja2025}.  Similar $^{13}$C dilutions have been measured in CCH toward the starless core L1521B and the prestellar core L183, although the $^{12}$C/$^{13}$C column density ratio is larger in L1521B by a factor of about 2 \citep[$\sim$250 vs. $\sim$100;][]{Taniguchi2019}. 
\citet{Fuentetaja2025}, on the other hand, found that the $^{12}$C/$^{13}$C ratio is practically the same as the Solar value for CS, HCS$^+$ and H$_2$CS in TMC-1.
%$^{13}$CCH was found to be less abundant than C$^{13}$CH, helping to constrain the formation route \citep[which involves hydrocarbons and C$^+$; see also][]{Sakai2010}. 
Finally, \cite{Yoshida2015} carried out a comprehensive observational study of $^{13}$C isotopologues of the cyclic C$_3$H$_2$ molecule (c-C$_3$H$_2$) toward the protostellar core L1527, using single dish telescopes, thus mainly tracing the cold envelope. They confirm the dilution of $^{13}$C species in carbon-chain molecules and found that the c-CC$^{13}$CH species exists more favorably than other $^{13}$C-substituted species.
This could be due to efficient isotopomer-exchange reactions, analogous to CCH \citep[$^{13}$CCH + H $\rightarrow$ C$^{13}$CH + H;][see also \S 1.3]{Sakai2010}. However, whether such H-mediated exchange reactions can proceed efficiently at low temperatures ($\sim 10$ K) without an activation barrier remains uncertain and requires further theoretical and experimental investigation.

Finally, the $^{12}$C/$^{13}$C ratios of CO$_2$ and CO ices have recently been measured in a quiescent region of the Chamaeleon I molecular cloud \citep{McClure2023} and in the cold envelopes of protostellar cores \citep{Brunken2024}. Considering the uncertainties in the band strengths, which vary depending on temperature and ice mixture composition, $^{12}$C/$^{13}$C ratios in protostellar core are consistent with, or slightly higher than, the local ISM value, whereas higher $^{12}$C/$^{13}$C ratios for CO ice ($99-184$) are found in Chamaeleon I.

\section{Chemical fractionation in the Galactic Center}
\label{sec:dfracGC}
%\textcolor{blue}{(Make the link with isotopic fractionation in the first sections)}

It is well known that isotopes provide diagnostics for stellar nucleosynthesis and hence, for the chemical enrichment and chemical history of galaxies \citep{Wilson1994, Viti2022}. Because the direct observation of isotopes in stellar spectra is extremely challenging, an alternative method consists of measuring isotopic ratios of gas-phase molecules in the ISM. %However, chemical fractionation in molecular clouds needs to be clearly understood in order to disentangle any possible chemical effect from the true isotopic ratio value, which is a fossil record of the stellar nucleosynthesis history of the Galaxy (see e.g. Roueff et al. 2015; Colzi et al. 2019).  
Since the kinetic temperatures of the gas are higher in Galactic Center GMCs with respect to clouds in the Galactic disk \citep[$T_{\rm kin}\sim70-150$ K vs. $\leq20$ K;][]{Ginsburg2018,Zeng2018, Colzi2024}, chemical fractionation processes are not expected to significantly affect molecular isotopic ratios, and therefore observations of these ratios offer unambiguous information about the stellar nucleosynthesis history and chemical enrichment in the center of galaxies including our own. 

The increasing interest for measuring isotopic ratios across the Galaxy  \citep{Milam2005, Adande2012, Zhang2015_isotope, Colzi2018, Yu2020, Yan2019, Yan2023}
has provided new information about the isotopic ratios toward GMCs in the Galactic Center. 
Early works established a low $^{12}$C/$^{13}$C isotopic ratio of $\sim$22 for the dense molecular gas in the Galactic Center, a factor of $>$3 lower than in the Solar neighborhood \citep[e.g.][]{Henkel1980, Henkel1985, Langer1990, Wilson1994}.
These low values have been confirmed recently by other groups using molecules such as CS, CH and even COMs \citep[see][]{Halfen2017, Humire2020, Yan2019, Yan2023, Jacob2020}. The isotopic ratios $^{18}$O/$^{17}$O and $^{14}$N/$^{15}$N also present significantly lower values in the Galactic Center than in molecular clouds in the Galactic disk \citep{Zhang2015_isotope, Chen2021}, although some discrepancies are found for NH$_3$ \citep{Guesten1985}. In contrast, the $^{32}$S/$^{34}$S ratio does not show noticeable differences between the CMZ and the inner disk ($R_{\rm GC}$$\leq$4.0 kpc), probably due to the fact that $^{34}$S is not a clean primary isotope of nucleosynthesis \citep{Humire2020, Yan2023}. The low CMZ values for the isotopic ratios are consistent with an inside-out formation scenario for our Galaxy, where the gas in the Galactic Center underwent an initial, intense chemical processing through stellar nucleosynthesis, likely produced by a higher rate of star formation and subsequent enrichment by massive stars. 

We note, however, that measurements of these ratios in the CMZ are challenging because lines of the typical molecules used to infer isotopic ratios such as CO, H$_2$CO, HCN, HNC, CS (and even their $^{13}$C, $^{15}$N, $^{18}$O, and $^{34}$S isotopologues) are largely optically thick. This implies that doubly-substituted molecular isotopologues need to be used for this analysis, which are not always available since they require high-sensitivity observations and hence, long telescope integration times \citep{Humire2020, Yan2023}. This was shown by \citet{Riquelme2010}, who found elevated $^{12}$C/$^{13}$C ratios ($>$40) toward high-altitude locations in the Galactic Center. This has been recently confirmed by \citet{Colzi2026} who, by analysing single and doubly-substituted isotopologues of HC$_3$N and HC$_5$N, have inferred similar $^{12}$C/$^{13}$C ratios of $\sim$45 toward the G+0.693 cloud. These higher isotopic ratios would suggest either accretion from less processed, lower metallicity gas from the halo \citep{Riquelme2010} or from the inner Galactic disk (Colzi et al. 2026), as predicted by hydrodynamical simulations of the Galaxy considering a Galactic bar potential \citep{Binney1991, Tress2020}. Additional high-sensitivity observations toward other GMCs in the Galactic Center are needed to confirm this hypothesis. 

Finally, in recent years, observations of deuterated species have attracted a lot of interest for Galactic Center studies. Since the kinetic temperatures of the gas in the CMZ are high (between 70 K and 150 K), no significant deuterium fractionation is expected in Galactic Center GMCs.  Observations of deuterated molecules in the Galactic Center are indeed very scarce \citep{Jacq1999, Lubowich2000, Polehampton2002}, with [D/H] ratios an order of magnitude lower than the primordial value (D/H of a few 10$^{-6}$ vs. 10$^{-5}$). However, these values were inferred indirectly using astrochemical models. Recent observations point to higher values. Using the high-sensitivity spectral surveys carried out toward G+0.693, \citet{Colzi2022a} analyzed the emission of deuterated species such as DCN, DNC, DCO$^{+}$, and N$_2$D$^+$ toward this cloud. These observations revealed the presence of a colder ($T_{\rm kin}$$\leq$30 K), denser ($n$(H$_2$)$\geq$5$\times$10$^4$ cm$^{-3}$), and narrower velocity component (FWHM$\sim 9$ km s$^{-1}$ vs. 20 km s$^{-1}$) in G+0.693, with an enhanced D/H ratio ($\geq$10$^{-4}$ vs. $\leq$6$\times$10$^{-5}$). This component is likely associated with gas on the verge of gravitational collapse and hence, under prestellar conditions. The analysis of the kinematics and spatial distribution of the dense gas probed by HC$_3$N, shows the existence of at least three fragments in G+0.693, kinematically differentiated and with higher H$_2$ volume densities and lower $T_{\rm kin}$ than the surrounding envelope \citep{Colzi2024}. G+0.693 is at a prestellar stage and it is hence on its way of forming a cluster of massive stars. This finding would be consistent with a scenario in which star formation in Sgr B2 (M), Sgr B2 (N) and G+0.693 occurs sequentially from Sgr B2(M) outwards \citep{Schmiedeke2016}. The study of the [D/H] ratio in galactic nuclei is thus a powerful tool to probe gas that will form the future generation of stars in external galaxies.

\section{Chemical inventory}
%Supplemental Tables 2 and 3 list the molecular column densities and abundances of all molecules and their isotopologues reported toward TMC-1 CP.
Supplemental Table 2 lists the molecular column densities and abundances of all molecules reported toward TMC-1 CP.
Molecular abundances are calculated considering a total H$_2$ column density of $1\times 10^{22}$ cm$^{-2}$.
The column density of a given species may vary by a factor of several depending on the assumed beam-filling factor and excitation temperature. The latter can be derived from rotational-diagram analysis when multiple transitions are available, but must otherwise be assumed.
Supplemental Table 3 lists the column densities and abundances of molecular isotopologues reported by the GOTHAM and QUIJOTE surveys toward TMC-1 CP. Earlier studies of molecular isotopologues in TMC-1 CP include \citet{Liszt2012, Sakai2013, Gratier2016, Majumdar2017} and \citet{Loison2020}; see also the references therein.
These studies are not included in Supplemental Table 3 because they often report only abundance ratios relative to the main isotopologues, rather than column densities of the isotopologues themselves.
Supplemental Table 4 lists the molecular column densities and abundances of all molecules reported toward G+0.693$-$0.027. Molecular abundances are calculated considering a total H$_2$ column density of $1.35\times 10^{23}$ cm$^{-2}$. 

%In Supplemental Table 2 and 4, molecular species are grouped by chemical family (e.g. hydrocarbons, N-bearing species, S-bearing species, and O-bearing species) and are listed in order of increasing molecular mass within each family. In Supplemental Table 3, molecular species are grouped according to the reference from which the column densities are adopted. }

%---TABLES BELOW ARE COPIED FROM Abundances-longtable on June20

\newpage
\begin{center}
%\begin{longtable}{lccc}
\begin{longtable}{lccp{6cm}}
\caption{Molecular column densities and abundances of all molecules reported toward TMC-1 CP. The molecular abundances have been calculated considering $N$(H$_2$)=$1\times10^{22}$ cm$^{-2}$.} \label{tab:long} \\

\hline \multicolumn{1}{c}{\textbf{Molecule}} & \multicolumn{1}{c}{\textbf{$N$(cm$^{-2}$)}} & \multicolumn{1}{c}{\textbf{Abundance}} & \multicolumn{1}{c}{\textbf{Reference}} \\ 
\hline 
\endfirsthead

\multicolumn{4}{c}%
{{\bfseries \tablename\ \thetable{} -- continued from previous page}} \\
\hline \multicolumn{1}{c}{\textbf{Molecule}} & \multicolumn{1}{c}{\textbf{$N$(cm$^{-2}$)}} & \multicolumn{1}{c}{\textbf{Abundance}} & \multicolumn{1}{c}{\textbf{Reference}} \\ \hline 
\endhead

\hline \multicolumn{4}{c}{{Continued on next page}} \\
\endfoot

\hline \hline
\endlastfoot

%\multicolumn{4}{c}{\bf C-bearing molecules} \\ \hline
%{\it 1 C atom} & & & \\ \hline
CH & $2.0\times 10^{14}$ & $2.0\times 10^{-8}$ & \citet{Agundez2013} \\
%{\it 2 C atoms} & & & \\ \hline
C$_2$H & $6.5\times 10^{14}$ & $6.5\times 10^{-8}$ & \citet{Sakai2010} \\
%{\it 3 C atoms} & & & \\ \hline
CH$_2$CCH & $8.7\times 10^{13}$ & $8.7\times 10^{-9}$ & \citet{Agundez2021a} \\
%C$_3$H & $6.0\times 10^{12}$ & $6.0\times 10^{-10}$ & Cernicharo et al. (2022c) \\
%l-C$_3$H & $6.0\times 10^{12}$ & $6.0\times 10^{-10}$ & \citet{Cernicharo2022c} \\
%c-C$_3$H & $1.2\times 10^{12}$ & $1.2\times 10^{-10}$ & \citet{Cernicharo2022c} \\
$l$-C$_3$H & $1.2\times 10^{13}$ & $1.2\times 10^{-9}$ & \citet{Agundez2026} \\
$c$-C$_3$H & $1.3\times 10^{13}$ & $1.3\times 10^{-9}$ & \citet{Agundez2026} \\
C$_3$H$^+$ & $2.4\times 10^{10}$ & $2.4\times 10^{-12}$ & \citet{Cernicharo2022a} \\
$c$-C$_3$H$_2$ & $1.2\times 10^{14}$ & $1.2\times 10^{-8}$ & \citet{Cernicharo2021f} \\
$l$-H$_2$C$_3$ & $(1.4-1.9)\times 10^{12}$ & $(1.4-1.9)\times 10^{-10}$ & \citet{cabezas2021e},  \citet{Xue2025} \\
%l-C$_3$H$_2$ & $1.41\times 10^{12}$ & $1.41\times 10^{-10}$ &  \citet{Xue2025} \\
%H$_2$C$_3$ & $1.9\times 10^{12}$ & $1.9\times 10^{-10}$ & Cabezas et al. (2021c) \\
CH$_3$CCH & $1.3\times 10^{14}$ & $1.3\times 10^{-8}$ & \citet{cabezas2021b} \\
CH$_2$CHCH$_3$ & $4.0\times 10^{13}$ & $4.0\times 10^{-9}$ & \citet{Marcelino2007} \\
H$_2$CCCH$^+$ & $7.0\times 10^{11}$ & $7.0\times 10^{-11}$ & \citet{Silva2023} \\
%{\it 4 C atoms} & & & \\ \hline
C$_4$H & $9.0\times 10^{13}$ & $9.0\times 10^{-9}$ & \citet{Agundez2023d} \\
C$_4$H$^-$ & $2.1\times 10^{10}$ & $2.1\times 10^{-12}$ & \citet{Agundez2023d} \\
$l$-H$_2$C$_4$ & $3.3\times 10^{12}$ & $3.3\times 10^{-10}$ & \citet{cabezas2021e} \\
CH$_2$CHCCH & $1.2\times 10^{13}$ & $1.2\times 10^{-9}$ & \citet{Cernicharo2021c} \\
CH$_3$CH$_2$CCH & $6.2\times 10^{11}$ & $6.2\times 10^{-11}$ &\citet{Cernicharo2024c} \\
%H$_2$C$_4$ & $3.3\times 10^{12}$ & $3.3\times 10^{-10}$ & Cabezas et al. (2021c) \\
%{\it 5 C atoms} & & & \\ \hline
HCCCH$_2$CCH & $5.0\times 10^{12}$ & $5.0\times 10^{-10}$ & \citet{Fuentetaja2024a} \\
CH$_3$C$_4$H & $1.3\times 10^{13}$ & $1.3\times 10^{-9}$ & \citet{Cabezas2025e} \\
%CH$_2$CCHCCH & $1.2\times 10^{13}$ & $1.2\times 10^{-9}$ & Cernicharo et al. (2021b) \\
$c$-C$_5$H & $9.0\times 10^{10}$ & $9.0\times 10^{-12}$ & \citet{Cabezas2022g} \\
$l$-C$_5$H & $1.3\times 10^{12}$ & $1.3\times 10^{-10}$ & \citet{Cabezas2022g} \\
C$_5$H$^+$ & $8.8\times 10^{10}$ & $8.8\times 10^{-12}$ & \citet{Cernicharo2022a} \\
%H$_2$C$_5$ & $1.8\times 10^{10}$ & $1.8\times 10^{-12}$ & Cabezas et al. (2021c) \\
$c$-C$_5$H$_6$ & $1.2\times 10^{13}$ & $1.2\times 10^{-9}$ & \citet{Cernicharo2021f} \\
$l$-H$_2$C$_5$ & $1.8\times 10^{10}$ & $1.8\times 10^{-12}$ & \citet{cabezas2021e} \\
$c$-C$_3$HCCH & $3.1\times 10^{11}$ & $3.1\times 10^{-11}$ & \citet{Cernicharo2021f} \\
%H$_2$C$_3$HCCH & $1.16\times 10^{13}$ & $1.16\times 10^{-9}$ & \citet{Xue2025} \\
H$_2$CCCHCCH & $1.2\times 10^{13}$ & $1.2\times 10^{-9}$ & \citet{Cernicharo2021d}, \citet{Xue2025} \\
%{\it 6 C atoms} & & & \\ \hline
HCCCHCCC & $1.3\times 10^{11}$ & $1.3\times 10^{-11}$ & \citet{Fuentetaja2022b} \\
C$_6$H & $6.0\times 10^{12}$ & $6.0\times 10^{-10}$ & \citet{Agundez2023d} \\
C$_6$H$^-$ & $1.5\times 10^{11}$ & $1.5\times 10^{-11}$ & \citet{Agundez2023d} \\
$l$-H$_2$C$_6$ & $8.0\times 10^{10}$ & $8.0\times 10^{-12}$ & \citet{cabezas2021e} \\
%H$_2$C$_6$ & $8.0\times 10^{10}$ & $8.0\times 10^{-12}$ & Cabezas et al. (2021c) \\
$o$-C$_6$H$_4$ & $5.0\times 10^{11}$ & $5.0\times 10^{-11}$ & \citet{Cernicharo2021h} \\
%{\it 7 C atoms} & & & \\ \hline
C$_7$H & $6.5\times 10^{10}$ & $6.5\times 10^{-12}$ & \citet{Cernicharo2022c} \\
$c$-C$_5$H$_4$CCH$_2$ & $2.7\times 10^{12}$ & $2.7\times 10^{-10}$ & \citet{Cernicharo2022c} \\
CH$_2$CCHC$_4$H & $2.2\times 10^{12}$ & $2.2\times 10^{-10}$ & \citet{Fuentetaja2022a} \\
CH$_3$C$_6$H & $7.0\times 10^{11}$ & $7.0\times 10^{-11}$ & \citet{Fuentetaja2022a} \\
1-$c$-C$_5$H$_5$CCH & $1.4\times 10^{12}$ & $1.4\times 10^{-10}$ & \citet{Cernicharo2021i} \\
2-$c$-C$_5$H$_5$CCH & $2.0\times 10^{12}$ & $2.0\times 10^{-10}$ & \citet{Cernicharo2021i} \\
%{\it 8 C atoms} & & & \\ \hline
C$_8$H & $5.0\times 10^{11}$ & $5.0\times 10^{-11}$ & \citet{Agundez2023d} \\
C$_8$H$^-$ & $2.0\times 10^{10}$ & $2.0\times 10^{-12}$ & \citet{Agundez2023d} \\
$c$-C$_6$H$_5$CCH & $3.0\times 10^{12}$ & $3.0\times 10^{-10}$ & \citet{Loru2023} \\
%{\it 9 C atoms} & & & \\ \hline
$c$-C$_9$H$_8$ & $1.6\times 10^{13}$ & $1.6\times 10^{-9}$ & \citet{Cernicharo2021f} \\
%{\it 10 C atoms} & & & \\ \hline
C$_{10}$H & $2.0 \times 10^{11}$$^{a}$ & $2.0\times 10^{-11}$$^{a}$ & \citet{Remijan2023} \\
C$_{10}$H$^-$ & $(1.8-40)\times 10^{10}$ & $(1.8-40)\times 10^{-12}$ & \citet{Remijan2023,Xue2025} \\
%{\it 13 C atoms} & & & \\ \hline
$c$-C$_{11}$H$_{8}$ & $6.0\times 10^{12}$ & $6.0\times 10^{-10}$ & \citet{Fuentetaja2026} \\
$c$-C$_{13}$H$_{10}$ & $2.8\times 10^{13}$ & $2.8\times 10^{-9}$ & \citet{Cabezas2025d} \\
%\multicolumn{4}{c}{\bf N-bearing molecules} \\ \hline
%NH$_3$ & $2.99\times 10^{14}$ & $2.99\times 10^{-8}$ & \citet{Xue2025} \\
NH$_3$ & $3.0\times 10^{14}$ & $3.0\times 10^{-8}$ & \citet{Xue2025} \\N$_2$H$^+$ & $2.8\times 10^{12}$ & $2.8\times 10^{-10}$ & \citet{Agundez2013} \\
%{\it 1 C atom} & & & \\ \hline
HNC & $2.6\times 10^{14}$ & $2.6\times 10^{-8}$ & \citet{Agundez2013} \\
HCN & $1.1\times 10^{14}$ & $1.1\times 10^{-8}$ & \citet{Agundez2013} \\
CN & $7.4\times 10^{12}$ & $7.4\times 10^{-10}$ & \citet{Agundez2013} \\
HCNH$^+$ & $4.2\times 10^{13}$ & $4.2\times 10^{-9}$ & \citet{Bop2024} \\
%H$_2$CN & $1.5\times 10^{11}$ & $1.5\times 10^{-11}$ & \citet{Agundez2013} \\
H$_2$CN & $1.5\times 10^{11}$ & $1.5\times 10^{-11}$ & \citet{Ohishi1994} \\
%CH$_2$CN & $1.5\times 10^{13}$ & $1.5\times 10^{-9}$ & Cabezas et al. (2021a) \\
CNCN & $8.0\times 10^{11}$ & $8.0\times 10^{-11}$ & \citet{Agundez2023a} \\
CH$_2$CN & $(1.5-1.9)\times 10^{13}$ & $(1.5-1.9)\times 10^{-9}$ & \citet{cabezas2021b}, \citet{Xue2025} \\
CH$_3$CN & $4.7\times 10^{12}$ & $4.7\times 10^{-10}$ & \citet{cabezas2021b} \\
CH$_3$NC & $2.7\times 10^{11}$ & $2.7\times 10^{-11}$ & \citet{Xue2025} \\
%{\it 2 C atoms} & & & \\ \hline
HCCN & $4.4\times 10^{11}$ & $4.4\times 10^{-11}$ & \citet{Cernicharo2021c} \\
NCCNH$^+$ & $6.8\times 10^{10}$ & $6.8\times 10^{-12}$ & \citet{Xue2025} \\ 
%NCCHO & $3.51\times 10^{11}$ & $3.51\times 10^{-11}$ & Cabezas et al. (2025b) \\
%{\it 3 C atoms} & & & \\ \hline
CH$_2$(CN)$_2$ & $1.8\times 10^{11}$ & $1.8\times 10^{-11}$ & \citet{Agundez2024} \\
HC$_3$N & $1.9\times 10^{14}$ & $1.9\times 10^{-8}$ & \citet{Tercero2024} \\
HC$_3$N$^+$ & $6.0\times 10^{10}$ & $6.0\times 10^{-12}$ & \citet{Cabezas2024b} \\
HCCNC & $3.4\times 10^{12}$ & $3.4\times 10^{-10}$ & \citet{Cernicharo2024b} \\
HNCCC & $5.8\times 10^{11}$ & $5.8\times 10^{-11}$ & \citet{Cernicharo2024b}, \citet{Xue2025} \\
%HNC$_3$ & $4.8\times 10^{12}$ & $4.8\times 10^{-10}$ & \citet{Xue2025} \\
H$_2$CCCN & $2.5\times 10^{11}$ & $2.5\times 10^{-11}$ & \citet{Cabezas2023} \\
HCCNCH$^+$ & $3.0\times 10^{10}$ & $3.0\times 10^{-12}$ & \citet{Agundez2022} \\
HC$_3$NH$^+$ & $8.6\times 10^{11}$ & $8.6\times 10^{-11}$ & \citet{Cabezas2022d} \\
CH$_2$CHCN & $6.2\times 10^{12}$ & $6.2\times 10^{-10}$ & \citet{Cernicharo2024c} \\
CH$_3$CH$_2$CN & $1.3\times 10^{11}$ & $1.3\times 10^{-11}$ & \citet{Cernicharo2024c} \\
C$_3$N & $1.8\times 10^{13}$ & $1.8\times 10^{-9}$ & \citet{Cernicharo2020a} \\
C$_3$N$^-$ & $6.4\times 10^{10}$ & $6.4\times 10^{-12}$ & \citet{Agundez2023d} \\
%{\it 4 C atoms} & & & \\ \hline
CH$_2$C$_3$N & $1.6\times 10^{11}$ & $1.6\times 10^{-11}$ & \citet{Cabezas2021h} \\
CH$_3$C$_3$N & $1.8\times 10^{12}$ & $1.8\times 10^{-10}$ & \citet{Fuentetaja2024a} \\
HC$_3$HCN & $2.2\times 10^{11}$ & $2.2\times 10^{-11}$ & \citet{Cabezas2025a} \\
Z-NCCHCHCN & $5.1\times 10^{10}$ & $5.1\times 10^{-12}$ & \citet{Agundez2024} \\
%HCCCH$_2$CN & $2.78\times 10^{12}$ & $2.78\times 10^{-10}$ & \citet{Fuentetaja2024a} \\
HCCCH$_2$CN & $2.8\times 10^{12}$ & $2.8\times 10^{-10}$ & \citet{Fuentetaja2024a} \\
%CH$_2$CCHCN & $2.67\times 10^{12}$ & $2.67\times 10^{-10}$ & \citet{Fuentetaja2024a} \\
CH$_2$CCHCN & $2.7\times 10^{12}$ & $2.7\times 10^{-10}$ & \citet{Fuentetaja2024a} \\
NC$_4$NH$^+$ & $1.1\times 10^{10}$ & $1.1\times 10^{-12}$ & \citet{Agundez2023a} \\
\textit{trans}-CH$_3$CHCHCN & $5.0\times 10^{10}$ & $5.0\times 10^{-12}$ & \citet{Cernicharo2022b} \\
\textit{cis}-CH$_3$CHCHCN & $1.3\times 10^{11}$ & $1.3\times 10^{-11}$ & \citet{Cernicharo2022b} \\
CH$_2$C(CH$_3$)CN & $1.0\times 10^{11}$ & $1.0\times 10^{-11}$ & \citet{Cernicharo2022b} \\
$g$-CH$_2$CHCH$_2$CN & $8.0\times 10^{10}$ & $8.0\times 10^{-12}$ & \citet{Cernicharo2022b} \\
$c$-CH$_2$CHCH$_2$CN & $7.0\times 10^{10}$ & $7.0\times 10^{-12}$ & \citet{Cernicharo2022b} \\
HC$_4$N & $3.7\times 10^{11}$ & $3.7\times 10^{-11}$ & \citet{Cernicharo2021c} \\
%{\it 5 C atoms} & & & \\ \hline
C$_2$H$_3$C$_3$N & $8.3\times 10^{10}$ & $8.3\times 10^{-12}$ & \citet{Xue2025} \\
HNC$_5$ & $1.3\times 10^{10}$ & $1.3\times 10^{-12}$ & \citet{Fuentetaja2024b} \\
E-HCCCHCHCN & $1.7\times 10^{11}$ & $1.7\times 10^{-11}$ & \citet{Agundez2024} \\
HC$_4$NC & $3.0\times 10^{11}$ & $3.0\times 10^{-11}$ & \citet{Cernicharo2020b} \\
C$_5$N & $4.8\times 10^{11}$ & $4.8\times 10^{-11}$ & \citet{Cernicharo2024d} \\
%C$_5$N$^-$ & $1.188\times 10^{11}$ & $1.188\times 10^{-11}$ & \citet{Cernicharo2024d} \\
C$_5$N$^-$ & $1.2\times 10^{11}$ & $1.2\times 10^{-11}$ & \citet{Cernicharo2024d} \\
HC$_5$N & $6.6\times 10^{13}$ & $6.6\times 10^{-9}$ & \citet{Fuentetaja2024b} \\
HC$_5$N$^+$ & $9.9\times 10^{10}$ & $9.9\times 10^{-12}$ & \citet{Cernicharo2024d} \\
HC$_5$NH$^+$ & $2.7\times 10^{11}$ & $2.7\times 10^{-11}$ & \citet{Cabezas2022d} \\
E-1-C$_4$H$_5$CN & $4.0\times 10^{10}$$^{b}$ & $4.0\times 10^{-12}$$^{b}$ & \citet{Xue2025} \\
2-C$_4$H$_5$CN & $3.1\times 10^{10}$$^{a}$ & $3.1\times 10^{-12}$$^{a}$ &  \citet{Agundez2025b}\\
%{\it 6 C atoms} & & & \\ \hline
CH$_2$CCHC$_3$N & $1.2\times 10^{11}$ & $1.2\times 10^{-11}$ & \citet{Fuentetaja2022a} \\
CH$_3$C$_5$N & $9.5\times 10^{10}$ & $9.5\times 10^{-12}$ & \citet{Fuentetaja2022a} \\
1-$c$-C$_5$H$_5$CN & $(3.1-8.3)\times 10^{11}$ & $(3.1-8.3)\times 10^{-11}$ & \citet{Lee_Changala_2021}, \citet{Cernicharo2021i} \\
2-$c$-C$_5$H$_5$CN & $(1.3-1.9)\times 10^{11}$ & $(1.3-1.9)\times 10^{-11}$ & \citet{Lee_Changala_2021}, \citet{Cernicharo2021i} \\
%{\it 7 C atoms} & & & \\ \hline
C$_7$N$^-$ & $5.0\times 10^{10}$ & $5.0\times 10^{-12}$ & \citet{Cernicharo2023b} \\
HC$_7$N & $2.1\times 10^{13}$ & $2.1\times 10^{-9}$ & \citet{Cabezas2022d} \\
HC$_7$N$^+$ & $2.3\times 10^{10}$ & $2.3\times 10^{-12}$ & \citet{Cernicharo2024d} \\
HC$_7$NH$^+$ & $5.5\times 10^{10}$ & $5.5\times 10^{-12}$ & \citet{Cabezas2022d} \\
$c$-C$_6$H$_5$CN & $1.2\times 10^{12}$ & $1.2\times 10^{-10}$ & \citet{Loru2023} \\
%{\it 8 C atoms} & & & \\ \hline
%CH$_3$C$_7$N & $5.75\times 10^{10}$ & $5.75\times 10^{-12}$ & \citet{Xue2025} \\
CH$_3$C$_7$N & $5.8\times 10^{10}$ & $5.8\times 10^{-12}$ & \citet{Xue2025} \\
%{\it 9 C atoms} & & & \\ \hline
%HC$_9$N & $4.32\times 10^{12}$ & $4.32\times 10^{-10}$ & \citet{Xue2025} \\
HC$_9$N & $4.3\times 10^{12}$ & $4.3\times 10^{-10}$ & \citet{Xue2025} \\
%{\it 10 C atoms} & & & \\ \hline
2-C$_9$H$_7$CN & $1.5\times 10^{11}$ & $1.5\times 10^{-11}$ & \citet{Xue2025} \\ 
%{\it 11 C atoms} & & & \\ \hline
HC$_{11}$N & $1.0\times 10^{11}$ & $1.0\times 10^{-11}$ & \citet{Xue2025} \\
1-C$_{10}$H$_7$CN & $(5.5-8.0)\times 10^{11}$ & $(5.5-8.0)\times 10^{-11}$ & \citet{Cernicharo2024f}, \citet{Xue2025} \\
2-C$_{10}$H$_7$CN & $(5.3-5.5)\times 10^{11}$ & $(5.3-5.5)\times 10^{-11}$ & \citet{Cernicharo2024f}, \citet{Xue2025} \\
%1-C$_{10}$H$_7$CN & $5.5\times 10^{11}$ & $5.5\times 10^{-11}$ & Cernicharo et al. (2024f) \\
%2-C$_{10}$H$_7$CN & $5.5\times 10^{11}$ & $5.5\times 10^{-11}$ & Cernicharo et al. (2024f) \\ \hline
%{\it 13 C atoms} & & & \\ \hline
1-C$_{12}$H$_7$CN & $(1.0-1.1)\times 10^{12}$ & $(1.0-1.1)\times 10^{-10}$ & \citet{Cernicharo2024f}, \citet{Xue2025} \\
3-C$_{12}$H$_7$CN & $7.0\times 10^{11}$ & $7.0\times 10^{-11}$ & \citet{Cernicharo2026} \\
4-C$_{12}$H$_7$CN & $5.0\times 10^{11}$ & $5.0\times 10^{-11}$ & \citet{Cernicharo2026} \\
5-C$_{12}$H$_7$CN & $(0.8-1.0)\times 10^{12}$ & $(0.8-1.0)\times 10^{-10}$ & \citet{Cernicharo2024f}, \citet{Xue2025} \\
%1-C$_{12}$H$_7$CN & $9.5\times 10^{11}$ & $9.5\times 10^{-11}$ & Cernicharo et al. (2024) \\
%5-C$_{12}$H$_7$CN & $9.5\times 10^{11}$ & $9.5\times 10^{-11}$ & Cernicharo et al. (2024) \\ \hline
%{\it 17 C atoms} & & & \\ \hline
1-C$_{16}$H$_9$CN & $9.1\times 10^{11}$ & $9.1\times 10^{-11}$ & \citet{Xue2025} \\
2-C$_{16}$H$_9$CN & $5.7\times 10^{11}$ & $5.7\times 10^{-11}$ & \citet{Xue2025} \\
4-C$_{16}$H$_9$CN & $9.2\times 10^{11}$ & $9.2\times 10^{-11}$ & \citet{Xue2025} \\ 
%{\it 25 C atoms} & & & \\ \hline
%C$_{24}$H$_{11}$CN & $2.69\times 10^{12}$ & $2.69\times 10^{-10}$ & \citet{Wenzel2025b} \\
C$_{24}$H$_{11}$CN & $2.7\times 10^{12}$ & $2.7\times 10^{-10}$ & \citet{Wenzel2025b} \\
%\multicolumn{4}{c}{\bf S-bearing molecules} \\ \hline
HSO & $7.0\times 10^{10}$ & $7\times 10^{-12}$ & \citet{Marcelino2023} \\
SO & $3.3\times 10^{13}$ & $3.3\times 10^{-9}$ & \citet{Xue2025} \\
HS$_2$ & $5.7\times 10^{11}$ & $5.7\times 10^{-11}$ & \citet{Esplugues2025} \\
%NS & $8\times 10^{12}$ & $8\times 10^{-10}$ & \citet{Agundez2013} \\
NS & $1.7\times 10^{12}$ & $1.7\times 10^{-10}$ & \citet{Cernicharo2018} \\
NS$^+$ & $5.2\times 10^{10}$ & $5.2\times 10^{-12}$ & \citet{Cernicharo2018} \\
H$_2$S & $3.1\times 10^{13}$ & $3.1\times 10^{-9}$ & \citet{Navarro2020} \\
SO$_2$ & $3\times 10^{12}$ & $3\times 10^{-10}$ & \citet{Agundez2013,Cernicharo2011} \\
%{\it 1 C atom} & & & \\ \hline
HCNS & $9.0\times 10^{9}$ & $9\times 10^{-13}$ & \citet{Cernicharo2024a} \\
HNCS & $3.2\times 10^{11}$ & $3.2\times 10^{-11}$ & \citet{Cernicharo2024a} \\
HSCN & $8.3\times 10^{11}$ & $8.3\times 10^{-11}$ & \citet{Cernicharo2024a} \\
NCS & $9.5\times 10^{11}$ & $9.5\times 10^{-11}$ & \citet{Cernicharo2024a} \\
%HCS & $5.51\times 10^{12}$ & $5.51\times 10^{-10}$ & \citet{Fuentetaja2022b} \\
HCS & $5.5\times 10^{12}$ & $5.5\times 10^{-10}$ & \citet{Fuentetaja2022b} \\
HCS$^+$ & $1\times 10^{13}$ & $1\times 10^{-9}$ & \citet{Cernicharo2021e} \\
H$_2$CS & $4.7\times 10^{13}$ & $4.7\times 10^{-9}$ & \citet{Cernicharo2021a} \\
CS & $3.5\times 10^{14}$ & $3.5\times 10^{-8}$ & \citet{Cernicharo2021e} \\
HSC & $1.3\times 10^{11}$ & $1.3\times 10^{-11}$ & \citet{Cernicharo2021e} \\
OCS & $2.2\times 10^{13}$ & $2.2\times 10^{-9}$ & \citet{Agundez2013} \\
CH$_3$SH & $1.7\times 10^{12}$ & $1.7\times 10^{-10}$ & \citet{Agundez2025} \\
%{\it 2 C atoms} & & & \\ \hline
CH$_3$CHS & $9.8\times 10^{10}$ & $9.8\times 10^{-12}$ & \citet{Agundez2025} \\
%NCCHS & $1.3\times 10^{12}$ & $1.3\times 10^{-10}$ & Cernicharo et al. (2021c) \\
H$_2$CCS & $7.8\times 10^{11}$ & $7.8\times 10^{-11}$ & \citet{Cabezas2024a} \\
%HCCS & $6.84\times 10^{11}$ & $6.84\times 10^{-11}$ & \citet{Fuentetaja2022b} \\
HCCS & $6.8\times 10^{11}$ & $6.8\times 10^{-11}$ & \citet{Fuentetaja2022b} \\
HCCS$^+$ & $1.1\times 10^{12}$ & $1.1\times 10^{-10}$ & \citet{Cabezas2022b} \\
CCS & $5.5\times 10^{13}$ & $5.5\times 10^{-9}$ & \citet{Cernicharo2021e} \\
HCSCN & $1.3\times 10^{12}$ & $1.3\times 10^{-10}$ & \citet{Cernicharo2021g} \\
%{\it 3 C atoms} & & & \\ \hline
CH$_2$CHCHS & $4.4\times 10^{10}$ & $4.4\times 10^{-12}$ & \citet{Cabezas2025b} \\
%HCCCHS & $3.2\times 10^{11}$ & $3.2\times 10^{-11}$ & Cernicharo et al. (2021c) \\
CCCS & $1.3\times 10^{13}$ & $1.3\times 10^{-9}$ & \citet{Cernicharo2021e} \\
HC$_3$S & $1.5\times 10^{11}$ & $1.5\times 10^{-11}$ & \citet{Cernicharo2024e} \\
HC$_3$S$^+$ & $2\times 10^{11}$ & $2\times 10^{-11}$ & \citet{Cernicharo2021a} \\
NC$_3$S & $1.4\times 10^{11}$ & $1.4\times 10^{-11}$ & \citet{Cernicharo2024e} \\
NCCHCS & $1.2\times 10^{11}$ & $1.2\times 10^{-11}$ & \citet{Cabezas2024a} \\
HCSCCH & $3.2\times 10^{11}$ & $3.2\times 10^{-11}$ & \citet{Cernicharo2021g} \\
$c$-H$_2$C$_3$S & $4.9\times 10^{10}$ & $4.9\times 10^{-12}$ & \citet{Xue2025} \\ 
$l$-H$_2$C$_3$S & $3.7\times 10^{11}$ & $3.7\times 10^{-11}$ & \citet{Cernicharo2021e} \\
%{\it 4 C atoms} & & & \\ \hline
HCCCCS & $9.5\times 10^{10}$ & $9.5\times 10^{-12}$ & \citet{Fuentetaja2022b} \\
C$_4$S & $3.8\times 10^{10}$ & $3.8\times 10^{-12}$ & \citet{Cernicharo2021e} \\
%{\it 5 C atoms} & & & \\ \hline
C$_5$S & $5\times 10^{10}$ & $5\times 10^{-12}$ & \citet{Cernicharo2021e} \\
%\multicolumn{4}{c}{\bf O-bearing molecules} \\ \hline
OH & $3\times 10^{15}$ & $3\times 10^{-7}$ & \citet{Agundez2013} \\
O$_2$ & $\leq7.7\times 10^{14}$ & $\leq7.7\times 10^{-8}$ & \citet{Agundez2013} \\
H$_2$O & $\leq7.0\times 10^{14}$ & $\leq7.0\times 10^{-8}$ & \citet{Agundez2013} \\
NO & $2.7\times 10^{14}$ & $2.7\times 10^{-8}$ & \citet{Agundez2013} \\
%{\it 1 C atom} & & & \\ \hline
CO & $1.7\times 10^{18}$ & $1.7\times 10^{-4}$ & \citet{Agundez2013} \\
HCO & $1.1\times 10^{12}$ & $1.1\times 10^{-10}$ & \citet{Cernicharo2021j} \\
HCO$^+$ & $9.3\times 10^{13}$ & $9.3\times 10^{-9}$ & \citet{Agundez2013} \\
H$_2$CO & $5\times 10^{14}$ & $5\times 10^{-8}$ & \citet{Agundez2013} \\
HOCO$^+$ & $3.1\times 10^{11}$ & $3.1\times 10^{-11}$ & \citet{Xue2025} \\
%t-HCOOH & $7.3\times 10^{11}$ & $7.3\times 10^{-11}$ & \citet{Xue2025} \\
\textit{trans}-HCOOH & $9.3\times 10^{11}$ & $9.3\times 10^{-11}$ & \citet{Molpeceres2025b} \\
\textit{cis}-HCOOH & $5.3\times 10^{10}$ & $5.3\times 10^{-12}$ & \citet{Molpeceres2025b} \\
CH$_3$OH & $3.2\times 10^{13}$ & $3.2\times 10^{-9}$ & \citet{Agundez2013} \\
CH$_3$O & $\leq1\times 10^{12}$ & $\leq1\times 10^{-10}$ & \citet{Agundez2013} \\
HNCO & $1.1\times 10^{13}$ & $1.1\times 10^{-9}$ & \citet{Cernicharo2024a} \\
HCNO & $7.8\times 10^{10}$ & $7.8\times 10^{-12}$ & \citet{Cernicharo2024a} \\
HOCN & $1.5\times 10^{11}$ & $1.5\times 10^{-11}$ & \citet{Cernicharo2024a} \\
NCO & $7.4\times 10^{11}$ & $7.4\times 10^{-11}$ & \citet{Cernicharo2024a} \\
H$_2$NCO$^+$ & $1.8\times 10^{10}$ & $1.8\times 10^{-12}$ & \citet{Cernicharo2024a} \\
%{\it 2 C atoms} & & & \\ \hline
C$_2$H$_5$OH & $1.1\times 10^{12}$ & $1.1\times 10^{-10}$ & \citet{Agundez2023c} \\
H$_2$CCO & $1.4\times 10^{13}$ & $1.4\times 10^{-9}$ & \citet{Fuentetaja2023} \\
CCO & $7.5\times 10^{11}$ & $7.5\times 10^{-11}$ & \citet{Cernicharo2021j} \\
HC$_2$O & $7.7\times 10^{11}$ & $7.7\times 10^{-11}$ & \citet{Cernicharo2021j} \\
CH$_3$CHO & $3.5\times 10^{12}$ & $3.5\times 10^{-10}$ & \citet{Cabezas2025b} \\
C$_2$H$_3$OH & $2.5\times 10^{12}$ & $2.5\times 10^{-10}$ & \citet{Agundez2021c} \\
CH$_3$OCHO & $1.1\times 10^{12}$ & $1.1\times 10^{-10}$ & \citet{Agundez2021c} \\
CH$_3$OCH$_3$ & $2.5\times 10^{12}$ & $2.5\times 10^{-10}$ & \citet{Agundez2021c} \\
CH$_3$CO$^+$ & $3.2\times 10^{11}$ & $3.2\times 10^{-11}$ & \citet{Cernicharo2021b} \\
HCOCN & $3.5\times 10^{11}$ & $3.5\times 10^{-11}$ & \citet{Cernicharo2021g} \\
%{\it 3 C atoms} & & & \\ \hline
CH$_3$COCH$_3$ & $1.4\times 10^{11}$ & $1.4\times 10^{-11}$ & \citet{Agundez2023c} \\
C$_2$H$_3$CHO & $2.2\times 10^{11}$ & $2.2\times 10^{-11}$ & \citet{Agundez2021c} \\
C$_2$H$_5$CHO & $1.9\times 10^{11}$ & $1.9\times 10^{-11}$ & \citet{Agundez2023c} \\
CH$_3$CHCO & $1.5\times 10^{11}$ & $1.5\times 10^{-11}$ & \citet{Fuentetaja2023} \\
HCCCO & $1.3\times 10^{11}$ & $1.3\times 10^{-11}$ & \citet{Cernicharo2021j} \\
C$_3$O & $1.2\times 10^{12}$ & $1.2\times 10^{-10}$ & \citet{Cernicharo2021j} \\
%HCCCHO & $9.0\times 10^{12}$ & $9.0\times 10^{-10}$ & \citet{Xue2025} \\
HCCCHO & $(9.0-90)\times 10^{11}$ & $(9.0-90)\times 10^{-11}$ & \citet{Xue2025, Esplugues2026} \\
%HCOCCH & $1.5\times 10^{12}$ & $1.5\times 10^{-10}$ & Cernicharo et al. (2021h) \\
HC$_3$O$^+$ & $2.1\times 10^{11}$ & $2.1\times 10^{-11}$ & \citet{Cernicharo2020c} \\
$c$-H$_2$C$_3$O & $5.2\times 10^{11}$ & $5.2\times 10^{-11}$ & \citet{Xue2025} \\ 
$l$-H$_2$C$_3$O & $3.7\times 10^{10}$ & $3.7\times 10^{-12}$ & \citet{Esplugues2026} \\ 
%{\it 5 C atoms} & & & \\ \hline
C$_5$O & $1.5\times 10^{10}$ & $1.5\times 10^{-12}$ & \citet{Cernicharo2021j} \\
HC$_5$O & $(1.4-1.7)\times 10^{12}$ & $(1.4-1.7)\times 10^{-10}$ & \citet{mcguire2017}, \citet{Cernicharo2021j} \\
%{\it 7 C atoms} & & & \\ \hline
HC$_7$O & $6.5\times 10^{11}$ & $6.5\times 10^{-11}$ & \citet{Cernicharo2021j} \\
\end{longtable}
\end{center}
\vspace{-2.5\baselineskip}
$^{a}$ Tentative detection. $^{b}$ See also \citet{Agundez2025b}

\vspace{1cm}

\begin{center}
\begin{longtable}{lccc}
\caption{Molecular column densities and abundances of molecular isotopologues reported toward TMC-1 CP by the GOTHAM and QUIJOTE surveys. The molecular abundances have been calculated considering $N$(H$_2$)=$1\times 10^{22}$ cm$^{-2}$.} \label{tab:long} \\

\hline \multicolumn{1}{c}{\textbf{Molecule}} & \multicolumn{1}{c}{\textbf{$N$(cm$^{-2}$)}} & \multicolumn{1}{c}{\textbf{Abundance}} & \multicolumn{1}{c}{\textbf{Reference}} \\ 
\hline 
\endfirsthead

\multicolumn{4}{c}%
{{\bfseries \tablename\ \thetable{} -- continued from previous page}} \\
\hline \multicolumn{1}{c}{\textbf{Molecule}} & \multicolumn{1}{c}{\textbf{$N$(cm$^{-2}$)}} & \multicolumn{1}{c}{\textbf{Abundance}} & \multicolumn{1}{c}{\textbf{Reference}} \\ \hline 
\endhead

\hline \multicolumn{4}{c}{{Continued on next page}} \\
\endfoot

\hline \hline
\endlastfoot

$^{13}$CH$_3$CCCCH & $1.4\times 10^{11}$ & $1.4\times 10^{-11}$ & \citet{Cabezas2025e} \\
CH$_3$$^{13}$CCCCH & $1.4\times 10^{11}$ & $1.4\times 10^{-11}$ & \citet{Cabezas2025e} \\
CH$_3$C$^{13}$CCCH & $1.4\times 10^{11}$ & $1.4\times 10^{-11}$ & \citet{Cabezas2025e} \\
CH$_3$CC$^{13}$CCH & $1.4\times 10^{11}$ & $1.4\times 10^{-11}$ & \citet{Cabezas2025e} \\
CH$_3$CCC$^{13}$CH & $1.4\times 10^{11}$ & $1.4\times 10^{-11}$ & \citet{Cabezas2025e} \\
CH$_3$CCCCD & $2.4\times 10^{11}$ & $2.4\times 10^{-11}$ & \citet{Cabezas2025e} \\
CH$_2$DCCCCH & $5.5\times 10^{11}$ & $5.5\times 10^{-11}$ & \citet{Cabezas2025e} \\
$^{13}$CH$_2$CHCN & $6.2\times 10^{10}$ & $6.2\times 10^{-12}$ & \citet{Cernicharo2024c} \\
CH$_2$$^{13}$CHCN & $5.9\times 10^{10}$ & $5.9\times 10^{-12}$ & \citet{Cernicharo2024c}  \\
CH$_2$CH$^{13}$CN & $8.1\times 10^{10}$ & $8.1\times 10^{-12}$ & \citet{Cernicharo2024c} \\
CH$_2$CDCN & $4.7\times 10^{10}$ & $4.7\times 10^{-12}$ & \citet{Cernicharo2024c} \\
\textit{trans}-CHDCHCN & $4.5\times 10^{10}$ & $4.5\times 10^{-12}$ & \citet{Cernicharo2024c}  \\
\textit{cis}-CHDCHCN & $4.9\times 10^{10}$ & $4.9\times 10^{-12}$ & \citet{Cernicharo2024c}  \\
CH$_2$CHC$^{15}$N & $2.2\times 10^{10}$ & $2.2\times 10^{-12}$ & \citet{Cernicharo2024c}  \\
H$^{13}$CCNC & $3.7\times 10^{10}$ & $3.7\times 10^{-12}$ & \citet{Cernicharo2024b}  \\
HC$^{13}$CNC & $3.7\times 10^{10}$ & $3.7\times 10^{-12}$ & \citet{Cernicharo2024b} \\
HCCN$^{13}$C & $3.4\times 10^{10}$ & $3.4\times 10^{-12}$ & \citet{Cernicharo2024b} \\
HCC$^{15}$NC & $1.4\times 10^{10}$ & $1.4\times 10^{-12}$ & \citet{Cernicharo2024b} \\
DCCNC & $1.1\times 10^{11}$ & $1.1\times 10^{-11}$ & \citet{Cernicharo2024b} \\
HN$^{13}$CCC & $6.2\times 10^{9}$ & $6.2\times 10^{-13}$ & \citet{Cernicharo2024b} \\
HNC$^{13}$CC & $6.2\times 10^{9}$ & $6.2\times 10^{-13}$ & \citet{Cernicharo2024b} \\
HNC$^{13}$C & $6.6\times 10^{9}$ & $6.6\times 10^{-13}$ & \citet{Cernicharo2024b} \\
DNCCC & $1.1\times 10^{10}$ & $1.1\times 10^{-12}$ & \citet{Cernicharo2024b} \\
DCCCN & $3.3\times 10^{12}$ & $3.3\times 10^{-10}$ & \citet{Tercero2024} \\
H$^{13}$CCCN & $1.8\times 10^{12}$ & $1.8\times 10^{-10}$ & \citet{Tercero2024} \\
HC$^{13}$CCN & $2.0\times 10^{12}$ & $2.0\times 10^{-10}$ & \citet{Tercero2024} \\
HCC$^{13}$CN & $2.8\times 10^{12}$ & $2.8\times 10^{-10}$ & \citet{Tercero2024} \\
HCCC$^{15}$N & $6.0\times 10^{11}$ & $6.0\times 10^{-11}$ & \citet{Tercero2024} \\
D$^{13}$CCCN & $3.2\times 10^{10}$ & $3.2\times 10^{-12}$ & \citet{Tercero2024} \\
DC$^{13}$CCN & $2.5\times 10^{10}$ & $2.5\times 10^{-12}$ & \citet{Tercero2024} \\
DCC$^{13}$CN & $3.5\times 10^{10}$ & $3.5\times 10^{-12}$ & \citet{Tercero2024} \\
DCCC$^{15}$N & $1.0\times 10^{10}$ & $1.0\times 10^{-12}$ & \citet{Tercero2024} \\
H$^{13}$C$^{13}$CCN & $1.9\times 10^{10}$ & $1.9\times 10^{-12}$ & \citet{Tercero2024} \\
H$^{13}$CC$^{13}$CN & $2.3\times 10^{10}$ & $2.3\times 10^{-12}$ & \citet{Tercero2024} \\
HC$^{13}$C$^{13}$CN & $2.5\times 10^{10}$ & $2.5\times 10^{-12}$ & \citet{Tercero2024} \\
HCC$^{13}$C$^{15}$N & $1.1\times 10^{10}$ & $1.1\times 10^{-12}$ & \citet{Tercero2024} \\
HC$^{13}$CC$^{15}$N & $8.0\times 10^{9}$ & $8.0\times 10^{-13}$ & \citet{Tercero2024} \\
H$^{34}$SCN & $2.1\times 10^{10}$ & $2.1\times 10^{-12}$ & \citet{Cernicharo2024a} \\
DSCN & $2.1\times 10^{10}$ & $2.1\times 10^{-12}$ & \citet{Cernicharo2024a} \\
H$^{15}$NCO & $3.3\times 10^{10}$ & $3.3\times 10^{-12}$ & \citet{Cernicharo2024a} \\
HN$^{13}$CO & $2.1\times 10^{11}$ & $2.1\times 10^{-11}$ & \citet{Cernicharo2024a} \\
DNCO & $2.5\times 10^{11}$ & $2.5\times 10^{-11}$ & \citet{Cernicharo2024a} \\
DCNO & $3.1\times 10^{9}$ & $3.1\times 10^{-13}$ & \citet{Cernicharo2024a} \\
$c$-HDC$_3$O & $2.6\times 10^{10}$ & $2.6\times 10^{-12}$ & \citet{Esplugues2026} \\
$c$-H$_2$$^{13}$CCCO & $1.4\times 10^{10}$ & $1.4\times 10^{-12}$ & \citet{Esplugues2026} \\
CH$_2$DC$_4$H & $5.5\times 10^{11}$ & $5.5\times 10^{-11}$ & \citet{Cabezas2022c} \\
CH$_2$DC$_3$N & $8.0\times 10^{10}$ & $8.0\times 10^{-12}$ & \citet{Cabezas2021f} \\
C$^{34}$S & $1.5\times 10^{13}$ & $1.5\times 10^{-9}$ & \citet{Cernicharo2021a} \\
$^{13}$C$^{34}$S & $1.5\times 10^{11}$ & $1.5\times 10^{-11}$ & \citet{Cernicharo2021a} \\
H$_2$C$^{34}$S & $1.8\times 10^{12}$ & $1.8\times 10^{-10}$ & \citet{Cernicharo2021a} \\
CC$^{34}$S & $5.0\times 10^{12}$ & $5.0\times 10^{-10}$ & \citet{Cernicharo2021a} \\
CCC$^{34}$S & $5.3\times 10^{11}$ & $5.3\times 10^{-11}$ & \citet{Cernicharo2021a} \\
HC$^{34}$S$^+$ & $7.1\times 10^{11}$ & $7.1\times 10^{-11}$ & \citet{Cernicharo2021a} \\
HDCCN & $7.4\times 10^{11}$ & $7.4\times 10^{-11}$ & \citet{cabezas2021b} \\
H$^{13}$CCCCCN & $1.8\times 10^{12}$ & $1.8\times 10^{-10}$ & \citet{Cernicharo2020b} \\
HC$^{13}$CCCCN & $2.1\times 10^{12}$ & $2.1\times 10^{-10}$ & \citet{Cernicharo2020b} \\
HCC$^{13}$CCCN & $2.0\times 10^{12}$ & $2.0\times 10^{-10}$ & \citet{Cernicharo2020b} \\
HCCC$^{13}$CCN & $1.7\times 10^{12}$ & $1.7\times 10^{-10}$ & \citet{Cernicharo2020b} \\
HCCCC$^{13}$CN & $2.1\times 10^{12}$ & $2.1\times 10^{-10}$ & \citet{Cernicharo2020b} \\
HCCCCC$^{15}$N & $4.6\times 10^{11}$ & $4.6\times 10^{-11}$ & \citet{Cernicharo2020b} \\
DCCCCCN & $2.2\times 10^{12}$ & $2.2\times 10^{-10}$ & \citet{Cernicharo2020b} \\
$^{13}$CCCCH & $6.2\times 10^{11}$ & $6.2\times 10^{-11}$ & \citet{Xue2025} \\
C$^{13}$CCCH & $1.1\times 10^{12}$ & $1.1\times 10^{-10}$ & \citet{Xue2025} \\
CC$^{13}$CCH & $1.5\times 10^{12}$ & $1.5\times 10^{-10}$ & \citet{Xue2025} \\
CCC$^{13}$CH & $1.2\times 10^{12}$ & $1.2\times 10^{-10}$ & \citet{Xue2025} \\
C$^{13}$CS & $2.3\times 10^{12}$ & $2.3\times 10^{-10}$ & \citet{Xue2025} \\
H$^{13}$CC$_6$N & $2.3\times 10^{11}$ & $2.3\times 10^{-11}$ & \citet{Xue2025} \\
HC$^{13}$CC$_5$N & $2.8\times 10^{11}$ & $2.8\times 10^{-11}$ & \citet{Xue2025} \\
HC$_2$$^{13}$CC$_4$N & $3.1\times 10^{11}$ & $3.1\times 10^{-11}$ & \citet{Xue2025} \\
HC$_3$$^{13}$CC$_3$N & $1.5\times 10^{11}$ & $1.5\times 10^{-11}$ & \citet{Xue2025} \\
HC$_4$$^{13}$CC$_2$N & $1.5\times 10^{11}$ & $1.5\times 10^{-11}$ & \citet{Xue2025} \\
HC$_5$$^{13}$CCN & $1.7\times 10^{11}$ & $1.7\times 10^{-11}$ & \citet{Xue2025} \\
HC$_6$$^{13}$CN & $2.2\times 10^{11}$ & $2.2\times 10^{-11}$ & \citet{Xue2025} \\
HDCS & $6.2\times 10^{12}$ & $6.2\times 10^{-10}$ & \citet{Xue2025} \\
C$_4$D & $1.3\times 10^{12}$ & $1.3\times 10^{-10}$ & \citet{Xue2025} \\
DC$_3$N & $3.8\times 10^{12}$ & $3.8\times 10^{-10}$ & \citet{Xue2025} \\
DC$_5$N & $8.8\times 10^{11}$ & $8.8\times 10^{-11}$ & \citet{Xue2025} \\
DC$_7$N & $1.6\times 10^{11}$ & $1.6\times 10^{-11}$ & \citet{Xue2025} \\
\end{longtable}
\end{center}

\begin{center}
\begin{longtable}{lccc}
\caption{Molecular column densities and abundances of all molecules reported toward G+0.693. The molecular abundances have been calculated considering $N$(H$_2$)=1.35$\times$10$^{23}$ cm$^{-2}$.} \label{tab:long} \\

\hline \multicolumn{1}{c}{\textbf{Molecule}} & \multicolumn{1}{c}{\textbf{$N$(cm$^{-2}$)}} & \multicolumn{1}{c}{\textbf{Abundance}} & \multicolumn{1}{c}{\textbf{Reference}} \\ 
\hline 
\endfirsthead

\multicolumn{4}{c}%
{{\bfseries \tablename\ \thetable{} -- continued from previous page}} \\
\hline \multicolumn{1}{c}{\textbf{Molecule}} & \multicolumn{1}{c}{\textbf{$N$(cm$^{-2}$)}} & \multicolumn{1}{c}{\textbf{Abundance}} & \multicolumn{1}{c}{\textbf{Reference}} \\ \hline 
\endhead

\hline \multicolumn{4}{c}{{Continued on next page}} \\
\endfoot

\hline \hline
\endlastfoot
%\multicolumn{4}{c}{\bf C-bearing molecules} \\ \hline
CCH & $5.3\times 10^{15}$ & $3.9\times 10^{-8}$ & \citet{Bizzocchi2020} \\
CH$_3$CCH & $1.7\times 10^{15}$ & $1.3\times 10^{-8}$ & \citet{Fatima2023} \\
CH$_2$CHCH$_3$ & $8.2\times 10^{14}$ & $6.1\times 10^{-9}$ & \citet{Fatima2023} \\
(CH$_3$)$_2$CCH$_2$ & $2.3\times 10^{14}$ & $1.7\times 10^{-9}$ & \citet{Fatima2023} \\
%\multicolumn{4}{c}{\bf N-bearing molecules} \\ \hline
H$_2$CNCN & $2.9\times 10^{12}$ & $2.2\times 10^{-11}$ & \citet{SanAndres2024} \\
N$_2$H$^+$ & $5.8\times 10^{15}$ & $4.3\times 10^{-8}$ & \citet{Colzi2022a} \\
CN & $2.0\times 10^{15}$ & $1.5\times 10^{-8}$ & \citet{Rivilla2019} \\
HCN & $3.1\times 10^{16}$ & $2.3\times 10^{-7}$ & \citet{Colzi2022a} \\
HNC & $5.0\times 10^{15}$ & $3.7\times 10^{-8}$ & \citet{Colzi2022a} \\
H$_2$NC & $4.2\times 10^{12}$ & $3.1\times 10^{-11}$ & \citet{Sanandres2023} \\
H$_2$CN & $9.2\times 10^{12}$ & $6.8\times 10^{-11}$ & \citet{Sanandres2023} \\
H$_2$CCN & $2.2\times 10^{14}$ & $1.6\times 10^{-9}$ & \citet{Zeng2018} \\
HNCN & $1.2\times 10^{13}$ & $8.9\times 10^{-11}$ & \citet{Rivilla2021b} \\
CH$_3$CN & $1.2\times 10^{14}$ & $8.5\times 10^{-10}$ & \citet{Zeng2018} \\
C$_2$H$_3$CN & $9.0\times 10^{13}$ & $6.7\times 10^{-10}$ & \citet{Zeng2018} \\
C$_2$H$_5$CN & $4.1\times 10^{13}$ & $3.0\times 10^{-10}$ & \citet{Zeng2018} \\
C$_3$N & $4.3\times 10^{13}$ & $3.2\times 10^{-10}$ & \citet{Zeng2018} \\
HC$_3$N & $7.1\times 10^{14}$ & $5.3\times 10^{-9}$ & \citet{Zeng2018} \\
HC$_5$N & $2.6\times 10^{14}$ & $1.9\times 10^{-9}$ & \citet{Zeng2018} \\
HC$_7$N & $1.5\times 10^{13}$ & $1.1\times 10^{-10}$ & \citet{Zeng2018} \\
HNCCC & $2.0\times 10^{12}$ & $1.5\times 10^{-11}$ & \citet{Zeng2018} \\
HCCNC & $2.3\times 10^{13}$ & $1.7\times 10^{-10}$ & \citet{Zeng2018} \\
CH$_3$C$_3$N & $4.5\times 10^{13}$ & $3.3\times 10^{-10}$ & \citet{Zeng2018} \\
CH$_2$CCHCN & $2.3\times 10^{13}$ & $1.7\times 10^{-10}$ &  \citet{Rivilla2022c} \\
HCCCH$_2$CN & $1.8\times 10^{13}$ & $1.3\times 10^{-10}$ & \citet{Rivilla2022c} \\
CH$_3$CCCN & $1.4\times 10^{13}$ & $1.0\times 10^{-10}$ & \citet{Rivilla2022c} \\
HNCO & $3.2\times 10^{15}$ & $2.4\times 10^{-8}$ & \citet{Sanznovo2024b}  \\
HOCN & $2.1\times 10^{13}$ & $1.6\times 10^{-10}$ & \citet{Sanznovo2024b} \\
HCOCN & $7.6\times 10^{12}$ & $5.6\times 10^{-11}$ & \citet{Rivilla2022c} \\
HOCH$_2$CN & $8.0\times 10^{12}$ & $5.9\times 10^{-11}$ & \citet{Rivilla2022c} \\
H$_2$NCO$^+$ & $1.7\times 10^{12}$ & $1.3\times 10^{-11}$ & \citet{Rodriguez2021b} \\
NH$_2$CHO & $2.5\times 10^{14}$ & $1.9\times 10^{-9}$ & \citet{Zeng2023} \\
CH$_3$C(O)NH$_2$ & $1.2\times 10^{14}$ & $8.5\times 10^{-10}$ & \citet{Zeng2023} \\
\textit{trans}-CH$_3$NHCHO & $4.3\times 10^{13}$ & $3.2\times 10^{-10}$ & \citet{Zeng2023} \\
\textit{cis}-CH$_3$NHCHO & $1.5\times 10^{13}$ & $1.1\times 10^{-10}$ & \citet{Zeng2025} \\
NH$_2$C(O)NH$_2$ & $6.3\times 10^{12}$ & $4.7\times 10^{-11}$ & \citet{Jimenezserra2020} \\
NH$_2$C(O)CH$_2$OH & $7.4\times 10^{12}$ & $5.5\times 10^{-11}$ & \citet{Rivilla2023} \\
CH$_3$NCO & $6.6\times 10^{13}$ & $4.9\times 10^{-10}$ & \citet{Zeng2018} \\
C$_2$H$_5$NCO & $8.1\times 10^{12}$ & $6.0\times 10^{-11}$ & \citet{Rodriguez2021b} \\
NH$_2$CN & $3.1\times 10^{14}$ & $2.3\times 10^{-9}$ & \citet{Rivilla2021b} \\
CH$_2$NH & $5.4\times 10^{14}$ & $4.0\times 10^{-9}$ & \citet{Zeng2018} \\
CH$_3$NH$_2$ & $3.0\times 10^{15}$ & $2.2\times 10^{-8}$ & \citet{Zeng2018} \\
CH$_2$CHCHNH & $1.5\times 10^{13}$ & $1.1\times 10^{-10}$ & \citet{alberton2023} \\
HNCHCN & $9.2\times 10^{13}$ & $6.8\times 10^{-10}$ & \citet{SanAndres2024} \\
HCCCHNH & $2.4\times 10^{13}$ & $1.8\times 10^{-10}$ & \citet{Bizzocchi2020} \\
NCCHNH & $2.3\times 10^{14}$ & $1.7\times 10^{-9}$ & \citet{Bizzocchi2020} \\
NCCNH$^+$ & $1.9\times 10^{11}$ & $1.4\times 10^{-12}$ & \citet{Rivilla2019} \\
C$_2$H$_3$NH$_2$ & $4.5\times 10^{13}$ & $3.3\times 10^{-10}$ & \citet{Zeng2021} \\
C$_2$H$_5$NH$_2$ & $2.5\times 10^{13}$ & $1.9\times 10^{-10}$ & \citet{Zeng2021} \\
NO$^+$ & $3.9\times 10^{13}$ & $2.9\times 10^{-10}$ & \citet{Rivilla2022b}  \\
NO & $1.6\times 10^{16}$ & $1.2\times 10^{-7}$ & \citet{Rivilla2022b} \\
HNO & $5.8\times 10^{13}$ & $4.3\times 10^{-10}$ & \citet{rivilla2020b} \\
N$_2$O & $4.8\times 10^{14}$ & $3.6\times 10^{-9}$ & \citet{rivilla2020b} \\
NH$_2$OH & $2.8\times 10^{13}$ & $2.1\times 10^{-10}$ & \citet{rivilla2020b} \\
NH$_2$CH$_2$CH$_2$OH & $1.5\times 10^{13}$ & $1.1\times 10^{-10}$ & \citet{rivilla2021a} \\
%\multicolumn{4}{c}{\bf S-bearing molecules} \\ \hline
CS & $6.2\times 10^{14}$ & $4.6\times 10^{-9}$ & \citet{Requena2006} \\
CH$_3$SH & $6.5\times 10^{14}$ & $4.8\times 10^{-9}$ & \citet{rodriguez2021a} \\
$g$-CH$_3$CH$_2$SH & $4.0\times 10^{13}$ & $3.0\times 10^{-10}$ & \citet{rodriguez2021a} \\
CH$_3$SCH$_3$ & $2.6\times 10^{13}$ & $1.9\times 10^{-10}$ & \citet{SanzNovo2025a} \\
$c$-C$_6$H$_6$S & $5.6\times 10^{12}$ & $4.1\times 10^{-11}$ & \citet{Araki2025} \\
SO & $3.0\times 10^{15}$ & $2.2\times 10^{-8}$ & \citet{Sanznovo2024b} \\
SO$_2$ & $3.8\times 10^{14}$ & $2.8\times 10^{-9}$ & \citet{Sanznovo2024b} \\
SO$^+$ & $1.3\times 10^{13}$ & $9.9\times 10^{-11}$ & \citet{Rivilla2022b} \\
OCS & $3.6\times 10^{15}$ & $2.7\times 10^{-8}$ & \citet{Sanznovo2024Aa} \\
HOCS$^+$ & $9.0\times 10^{12}$ & $6.7\times 10^{-11}$ & \citet{Sanznovo2024Aa} \\
HCOSH & $1.6\times 10^{13}$ & $1.2\times 10^{-10}$ & \citet{rodriguez2021a} \\
NS & $2.8\times 10^{14}$ & $2.1\times 10^{-9}$ & \citet{Sanznovo2024b} \\
HNCS & $6.2\times 10^{13}$ & $4.6\times 10^{-10}$ & \citet{Sanznovo2024b} \\
HNSO & $8.0\times 10^{13}$ & $5.9\times 10^{-10}$ & \citet{Sanznovo2024b} \\
%\multicolumn{4}{c}{\textbf{O-bearing molecules}} \\ \hline
CO & $7.4\times 10^{18}$ & $5.5\times 10^{-5}$ & \citet{Polehampton2007} \\
HCO$^+$ & $7.6\times 10^{15}$ & $5.7\times 10^{-8}$ & \citet{Colzi2022a} \\
HOCO$^+$ & $2.8\times 10^{14}$ & $2.1\times 10^{-9}$ & \citet{Sanznovo2024Aa} \\
HCCO & $5.0\times 10^{12}$ & $3.7\times 10^{-11}$ & \citet{rivilla2021a} \\
H$_2$CCO & $2.9\times 10^{14}$ & $2.2\times 10^{-9}$ & \citet{rivilla2021a} \\
H$_2$CO & $4.2\times 10^{14}$ & $3.1\times 10^{-9}$ & \citet{Jimenezserra2020} \\
H$_2$COH$^+$ & $9.8\times 10^{13}$ & $7.3\times 10^{-10}$ & \citet{Requena2008} \\
HCOOH & $2.0\times 10^{14}$ & $1.5\times 10^{-9}$ & \citet{Sanznovo2023} \\
HOCOOH & $6.4\times 10^{12}$ & $4.7\times 10^{-11}$ & \citet{Sanznovo2023}\\
CH$_3$COOH & $4.2\times 10^{13}$ & $3.1\times 10^{-10}$ & \citet{Sanznovo2023} \\
CH$_3$OH & $1.5\times 10^{16}$ & $1.1\times 10^{-7}$ & \citet{rodriguez2021a} \\
C$_2$H$_5$OH & $6.2\times 10^{14}$ & $4.6\times 10^{-9}$ & \citet{rodriguez2021a} \\
$n$-C$_3$H$_7$OH & $8.9\times 10^{13}$ & $6.6\times 10^{-10}$ & \citet{Jimenezserra2022} \\
H$_2$C=CHOH & $1.2\times 10^{14}$ & $9.2\times 10^{-10}$ & \citet{Jimenezserra2022} \\
(CHOH)$_2$ & $1.8\times 10^{13}$ & $1.3\times 10^{-10}$ & \citet{Rivilla2022a} \\
CH$_3$OCH$_3$ & $7.8\times 10^{14}$ & $5.8\times 10^{-9}$ & \citet{SanzNovo2025a} \\
CH$_3$CHO & $5.0\times 10^{14}$ & $3.7\times 10^{-9}$ & \citet{SanzNovo2022} \\
HCCCHO & $3.2\times 10^{13}$ & $2.4\times 10^{-10}$ & \citet{Bizzocchi2020} \\
HOCH$_2$CHO & $9.3\times 10^{13}$ & $6.9\times 10^{-10}$ & \citet{Rivilla2022a} \\
$s$-C$_2$H$_5$CHO & $7.4\times 10^{13}$ & $5.5\times 10^{-10}$ & \citet{SanzNovo2022} \\
CH$_3$OCHO & $1.9\times 10^{15}$ & $1.4\times 10^{-8}$ & \citet{Requena2008} \\
$c$-C$_2$H$_4$O & $1.2\times 10^{14}$ & $9.1\times 10^{-10}$ & \citet{Requena2008} \\
CH$_2$CHCHO & $3.7\times 10^{13}$ & $2.7\times 10^{-10}$ & \citet{Requena2008} \\
CH$_3$CH$_2$CHO & $7.4\times 10^{13}$ & $5.5\times 10^{-10}$ & \citet{SanzNovo2022} \\
HOCH$_2$CH$_2$CHO & $8.6\times 10^{12}$ & $6.4\times 10^{-11}$ & \citet{Fried2025} \\
CH$_3$C(O)OCH$_3$ & $2.2\times 10^{14}$ & $1.6\times 10^{-9}$ & \citet{sanznovo2026} \\
$a$-CH$_3$CH$_2$OC(O)H & $1.8\times 10^{13}$ & $1.3\times 10^{-10}$ & \citet{sanznovo2026} \\
$g$-CH$_3$CH$_2$OC(O)H & $1.9\times 10^{13}$ & $1.4\times 10^{-10}$ & \citet{sanznovo2026} \\
CH$_3$C(O)CH$_2$OH & $2.1\times 10^{13}$ & $1.6\times 10^{-10}$ & \citet{sanznovo2026} \\
CH$_3$CH(OH)C(O)H & $1.1\times 10^{13}$ & $8.1\times 10^{-11}$ & \citet{sanznovo2026} \\
HO(CH$_2$)$_2$C(O)H & $8.6\times 10^{12}$ & $6.4\times 10^{-11}$ & \citet{sanznovo2026} \\
CH$_3$OCH$_2$C(O)H & $3.3\times 10^{12}$ & $2.4\times 10^{-11}$ & \citet{sanznovo2026} \\
\textit{aGg’}-(CH$_2$OH)$_2$ & $2.4\times 10^{14}$ & $1.7\times 10^{-9}$ & \citet{Jimenezserra2026} \\
HOCH$_2$CH(OH)COCH$_2$OH & $8.7\times 10^{13}$ & $6.4\times 10^{-10}$ & \citet{Jimenezserra2026} \\
%\multicolumn{4}{c}{\textbf{Metal-bearing molecules}} \\ \hline
SiC$_2$ & $1.0\times 10^{13}$ & $7.6\times 10^{-11}$ & \citet{Massalkhi2023} \\
NaS & $5.0\times 10^{10}$ & $3.7\times 10^{-13}$ & \citet{Reymontejo2024} \\
MgS & $6.0\times 10^{10}$ & $4.4\times 10^{-13}$ & \citet{Reymontejo2024} \\
SiS & $5.3\times 10^{13}$ & $3.9\times 10^{-10}$ & \citet{Massalkhi2023} \\
CaO & $2.0\times 10^{10}$ & $1.5\times 10^{-13}$ & \citet{Reymontejo2024} \\
SiO & $7.2\times 10^{14}$ & $5.3\times 10^{-9}$ & \citet{Massalkhi2023} \\
PO & $4.9\times 10^{12}$ & $3.6\times 10^{-11}$ & \citet{Rivilla2022b} \\
PO$^+$ & $6.0\times 10^{11}$ & $4.4\times 10^{-12}$ & \citet{Rivilla2022b} \\
\end{longtable}
\end{center}

\bibliography{ref}